\documentclass{aa}  

\usepackage{graphicx}
\usepackage{txfonts}
\usepackage{enumerate}
\usepackage{paralist}
\usepackage[section]{placeins}
\bibpunct{(}{)}{;}{a}{}{,} 

\begin{document}

   \title{Radio continuum properties of luminous infrared galaxies}
\titlerunning{Radio Continuum Properties of LIRGs}
   \subtitle{Identifying the presence of an AGN in the radio}

   \author{E. Vardoulaki
          \inst{1}\fnmsep\thanks{email: eleniv@physics.uoc.gr},  
        V. Charmandaris
        \inst{1}\fnmsep\inst{2}\fnmsep\inst{3},
         E. J. Murphy
         \inst{4},
          T. Diaz-Santos
       \inst{4,5},
         L. Armus
         \inst{4},
           A. S. Evans
           \inst{6},
           J. Mazzarella
           \inst{4},
           G. C. Privon
           \inst{6},
           S. Stierwalt
           \inst{6},
           L. Barcos-Mu\~{n}oz
          \inst{6}
           \\
          }
\authorrunning{Vardoulaki et al.}

   \institute{Department of Physics, University of Crete, GR-71003, Heraklion, Greece 
        \and
                Institute for Astronomy, Astrophysics, Space Applications \& Remote Sensing, National Observatory of Athens, GR-15236, Athens, Greece
       \and 
                 Chercheur Associ\'e, Observatoire de Paris, F-75014,  Paris, France
       \and
                Infrared Processing \& Analysis Center, MS 100-22, California Institute of Technology, Pasadena, CA, USA 
      \and
      Nucleo de Astronomia de la Facultad de Ingenieria, Universidad Diego Portales, Av. Ejercito Libertador 441, Santiago, Chile
      \and
              Department of Astronomy, University of Virginia, Charlottesville, VA, USA
             }

   \date{Received ; accepted }

 
  \abstract
   {Luminous infrared galaxies (LIRGs) are systems enshrouded in dust, which absorbs most of their optical/UV emission and radiates it again in the mid- and far-infrared. Radio observations are largely unaffected by dust obscuration, enabling us to study the central regions of LIRGs in an unbiased manner. }
   {The main goal of this project is to examine how the radio properties of local LIRGs relate to their infrared spectral characteristics. Here we present an analysis of the radio continuum properties of a subset of the Great Observatories All-sky LIRG Survey (GOALS), which consists of 202 nearby systems ($z < $0.088). Our radio sample consists of 35 systems, containing 46 individual galaxies, that were observed at both 1.49 and 8.44 GHz with the VLA with a resolution of about 1 arcsec (FWHM). The aim of the project is to use the radio imagery to probe the central kpc of these LIRGs in search of active galactic nuclei (AGN).}
   {We used the archival data\thanks{The VLA images used in this study are available in electronic form
at the CDS via anonymous ftp to cdsarc.u-strasbg.fr (130.79.128.5) or via http://cdsweb.u-strasbg.fr/cgi-bin/qcat?J/A+A/} at 1.49 and 8.44 GHz to create radio-spectral-index maps using the standard relation between flux density $S_{\nu}$ and frequency $\nu$, $S_{\nu} \sim \nu^{-\alpha}$, where $\alpha$ is the radio spectral index. By studying the spatial variations in $\alpha$, we classified the objects as radio-AGN, radio-SB, and AGN/SB (a mixture). We identified the presence of an active nucleus using the radio morphology, deviations from the radio/infrared correlation, and spatially resolved spectral index maps, and then correlated this to the usual mid-infrared ([NeV]/[NeII] and [OIV]/[NeII] line ratios and equivalent width of the 6.2 $\rm \mu$m PAH feature) and optical (BPT diagram) AGN diagnostics.}
   {We find that 21 out of the 46 objects in our sample ($\sim$ 45\%) are radio-AGN, 9 out of the 46 ($\sim$ 20\%) are classified as starbursts (SB) based on the radio analysis, and 16 ($\sim$ 35\%) are AGN/SB. After comparing to other AGN diagnostics we find 3 objects out of the 46 ($\sim$ 7\%) that are identified as AGN based on the radio analysis, but are not classified as such based on the mid-infrared and optical AGN diagnostics presented in this study. }
   {}


   \maketitle
%

\section{Introduction}

The all-sky survey of the Infrared Astronomical Satellite \citep[IRAS][]{neugebauer84} revealed the existence in the local universe of the so-called Luminous and Ultraluminous Infrared Galaxies (U/LIRGs). This is a class of galaxies with an infrared luminosity (L$_{\rm IR}$; 8 $\rm \mu$m $< \lambda <$ 1000 $\rm \mu$m) between 10$^{11}$ L$_{\odot}$ and 10$^{12}$ L$_{\odot}$ for the LIRGs and over 10$^{12}$ L$_{\odot}$ for the ULIRGs. Detailed analysis of the properties of these systems has revealed that they contain large quantities of dust, which absorbs over 90\% of their optical/UV emission (for the most luminous systems) and re-radiates it in the infrared \citep{armus09}. Recent advances in space infrared surveys by ISO and Spitzer have also demonstrated that LIRGs compose a significant fraction ($\geq$ 50\%) of the cosmic infrared background and dominate the star-formation activity at $z \sim$ 1 \cite[e.g.][]{elbaz02, lefloch05, caputi07, magnelli09, murphy09}. Furthermore, the dominant population of the cosmic X-ray background \cite[e.g.][]{shanks91, hasinger05} are Seyfert galaxies at $z \sim$ 0.7, whose local analogues are also LIRGs. It is thus evident that studying these intriguing objects is extremely important owing to their key role in the galaxy evolution scheme at high and low redshifts.

Over the past 20 years multi-wavelength studies have revealed that unlike local ULIRGs, which are nearly always associated with the last stages of violent gas-rich major mergers \citep{armus87, sanders88}, only a low percentage of LIRGs are interacting systems and they often form most of their stars in their extended disks. Recently, \cite{stierwalt13} have shown that only $\sim$ 18\% of LIRGs, compared to $\sim$ 90\% of ULIRGs, are associated with late type mergers based on analysis of the Spitzer/IRAC images. These numbers change to $\sim$ 19\% in LIRGs and $\sim$ 81\% in ULIRGs, based on HST classification of merger stages \citep[][not all objects have an associated HST merger stage]{haan11}. In total, $\sim$ 68\% of LIRGs and $\sim$ 95\% of ULIRGs are merging systems based on IRAC merger stages \citep{stierwalt13}, while from HST merger stages we get $\sim$ 91\% of LIRGs and $\sim$ 93\% of ULIRGs associated with a merger \citep{haan11}. 

The difference in physical size of the area responsible for the energy production results in observable differences in their spectral energy distributions, especially in the mid-IR. Spectroscopic studies, in particular in the infrared range, which is less affected by extinction, reveal a great diversity in the 5-30 $\rm \mu$m mid-IR spectral features \citep{tanio10b, tanio11}. The emission from polycyclic aromatic hydrocarbons (PAH), tracers of photodissociation regions and the gas heating via star formation, is clearly seen in LIRGs and has been used extensively to estimate their star formation rate in the local universe \cite[i.e.][]{calzetti10}. However, as the infrared luminosity of the systems increases and reaches the ULIRG range, PAH emission becomes weaker \citep{desai07}. We also see that in LIRGs \citep{stierwalt14}. In addition, in LIRGs the average size of the area producing the mid-IR continuum emission is larger than in ULIRGs, and it appears that there is a correlation between the compactness of the mid-IR continuum emission and the IRAS far-IR colours of the source \cite[e.g.][]{tanio10b}. The typical size of the mid-IR emitting region in most LIRGs is of the order of 500 pc (FWHM), while the estimated upper limit for the core size in ULIRGs in the mid-IR is $<$ 1 kpc and can be below 200 pc \citep[FWHM;][]{soifer00}.
  
The actual nature of the dominant energy source in the central regions of LIRGs and ULIRGs, whether massive star formation or radiation from an accreting supermassive black hole in an active nucleus (AGN), has been a subject of intense study over the years. However, since by virtue of their high extinction the usual optical diagnostic methods could not be easily applied in these systems, observations in longer wavelengths were necessary. It has been shown that $\sim$ 20\% of LIRGs host an AGN detected in the mid-IR, and its presence becomes more prevalent as the IR luminosity increases \cite[e.g.][]{armus07, petric11, inami13, stierwalt13}. The actual contribution of this AGN to the bolometric luminosity of a galaxy is a challenging problem. Detailed knowledge of each source and observations across the large portion of its spectrum are often required \citep{veilleux09}. In some very dust-obscured sources, the very central regions are hidden from direct view at wavelengths shorter than $\sim$ 10 $\rm \mu$m \citep[e.g.][]{laurent00, spoon04}. At radio wavelengths one can overcome the effects of dust obscuration and study the centres of these objects in great detail. Thus we can identify signs of AGN activity, such as weak jets \citep[e.g.][]{lonsdale03}, or observe more extended structures not associated with AGN but typical of star-forming regions \citep{condon91}.

In this paper we use a radio-analysis technique to the radio continuum observations of \cite{condon90, condon91} in order to identify the presence of AGN. As discussed earlier, LIRG nuclei are enshrouded in dust, making them optically thick in the UV/optical, and often even in the mid-IR. Even though X-ray photons originating in an AGN penetrate the dust efficiently, nuclei with hydrogen column densities N(H) $>10^{24}$ cm$^{-2}$ become optically thick to X-ray emission. On the other hand, radio observations are largely unaffected by dust obscuration, enabling us to study the central regions of LIRGs in an unbiased manner. The high spatial resolution ($\sim$ arcsec) of VLA radio observations already in hand allows us to investigate the presence of an active nucleus in U/LIRGs via the identification of weak jets \cite[e.g.][]{lonsdale03} or spatial variation of the radio spectra index.

Even though radio observations of LIRGs have been in the literature for more than 20 years \citep{condon90, condon91}, with our analysis we aim to obtain additional information about the nature of these objects. \cite{condon91} presented a classification of these objects based on the integrated radio spectral index, calculated between 1.4 and 8.44 GHz. But using the integrated radio spectral index of an object to distinguish between an AGN and a nuclear SB can be challenging, since both exhibit flat and steep radio spectral indices. Here we present a method that takes advantage of the data available at 1.49 and 8.44 GHz. By studying the variations in the radio spectral index spatially, we aim to distinguish between the contribution from the AGN and the SB in the radio (Sec.~\ref{sec:analysis}).

Furthermore, we compare our radio findings to known mid-IR and optical AGN diagnostics. 
Current AGN diagnostics include a combination of emission-line diagnostics such as [O III]/H$\beta$ and [Ne III]/[O II] ratios \cite[BPT diagram][]{baldwin81, kewley06}, mid-IR colours and lines \cite[e.g.][]{genzel98, laurent00, lacy04} and hard X-ray emission. Diagnostics based on Spitzer mid-IR observations of the GOALS sample reveal that $\sim$20\% host an AGN, but only in 10\% the AGN dominates the mid-IR emission from the galaxy \citep{petric11}. These estimates rely either on the presence of dust emission heated to near sublimation temperatures ($\sim$ 1000 K) due to the UV and hard X-ray photons from the central accretion disk, or on the effects these photons have on the interstellar gas. More specifically, PAH emission from AGN is known to be suppressed \cite[see][]{wu09}, while emission lines such as [Ne V]14.3/23.2 $\rm \mu$m and [O IV]25.9 $\rm \mu$m, with ionisation potential of 97 and 55 eV respectively, are observed under the same circumstances \citep{tommasin10}. We wish to examine how these diagnostic-tools vary in LIRGs for which evidence of an AGN is established via their radio emission. 
   
We use J2000.0 positions and the convention for all spectral indices, $\alpha$, that flux density of a radio source $S_{\nu} \sim \nu^{-\alpha}$, where $\nu$ is the observing frequency. In Sec.~\ref{sec:sample} we present the sample used in the analysis (Sec.~\ref{sec:analysis}). In Sec.~\ref{sec:results} we present our results of the radio analysis and our discussion on the comparison to mid-IR and optical AGN diagnostics. We also give an upper limit for the size of the compact starburst in the radio. In Sec.~\ref{sec:conc} we present our conclusions. In the Appendix we give an object-by-object analysis and the radio-spectral-index maps and histograms for all the objects of our sample.


\section{Sample}
\label{sec:sample}

This study is part of the Great Observatories All-sky LIRG Survey \citep[GOALS;][]{armus09}, a local ($z <$ 0.088) infrared-selected unbiased sample of 181 LIRG and 21 ULIRG systems, made up from 293 individual galaxies. GOALS was selected from the IRAS Revised Bright Galaxy Sample (RBGS), a flux-limited sample of all extragalactic objects brighter than 5.24 Jy at 60 $\rm \mu$m, covering the entire sky surveyed by IRAS at Galactic latitudes $|b| >$ 5${\degr}$. The GOALS sample is ideal not only in its completeness and sample size, but also in the proximity and brightness of the galaxies. The galaxies span the full range of nuclear spectral types (type-1 and type-2 AGN, LINERs and starbursts) and interaction stages (major mergers, minor mergers and isolated galaxies). 

The multi-wavelength coverage of the GOALS sample includes Spitzer/IRAC imaging at 3.6, 4.5, 5.8 and 8.0 $\rm \mu$m as well as Spitzer/MIPS at 24, 70 and 160 $\rm \mu$m. Spitzer/IRS 5-37 $\rm \mu$m spectra are available for the nuclei of all 293 galaxies. High resolution optical and near-IR Hubble Space Telescope imaging is available for 88 galaxies with L$_{\rm IR} >$ 10$^{11.4}$ L$_{\odot}$ in addition to far-UV data for 25 objects, making GOALS one of the most complete local sample imaged by Hubble. Complete tables showing all available data can be found at http://goals.ipac.caltech.edu/.

Here we present a new analysis of archival radio continuum maps available for the GOALS galaxies. For 151 LIRG systems, radio maps were obtained with the Very Large Array (VLA) at 1.4 GHz \citep{condon90}. In particular, 57 objects were observed in A configuration (1$\farcs$5, 1$\farcs$8, 2$\farcs$1 and 2$\farcs$4 resolution), 97 in B configuration (5$\arcsec$, 6$\arcsec$, 7$\arcsec$ and 8$\arcsec$ resolution), 103 in C configuration (15$\arcsec$, 18$\arcsec$, 21$\arcsec$ and 24$\arcsec$ resolution) and 6 in D configuration (48$\arcsec$, 54$\arcsec$ and 60$\arcsec$ resolution). Objects that contain diffuse features, such as extended disks, require low-resolution configurations in order to recover that faint diffuse emission, while more luminous (L$_{\rm IR} > 10^{11}$ L$_{\odot}$) infrared sources are associated with smaller ($\sim$ 1 kpc linear) radio sizes that are resolvable only at A configuration \citep{condon90}. Additionally, 40 objects with log$(L_{\rm IR} / L_{\odot}) \geq$ 11.25 have been observed at 8.44 GHz in A configuration \citep[0.25-0.50 arcsec resolution;][]{condon91}. This luminosity cut adopted by \cite{condon91} might favour objects containing an AGN. Details about the data reduction can be found in the referenced papers. 

For the purpose of our analysis we require the best resolution maps at 1.49 GHz, which is given by A configuration at the VLA. After cross-correlating the list of objects observed at A array 1.49 and at 8.44 GHz, and rejecting maps with: (a) astrometry issues (MCG -03-04-014); (b) missing maps (NGC 1614, NGC 5257 S); (c) objects that were not observed at the highest resolution at 1.49 GHz (NGC 0034 N, NGC 0695, NGC 6286, NGC 7469); (d) no radio components associated with the galaxy (UGC 02369 N, VV 340a S), we are left with 35 systems or 46 individual objects that were observed at both 1.49 and 8.44 GHz in A configuration. We need objects to be observed at both radio frequencies in order to construct radio spectral index maps, using the standard relation between flux density and frequency given in the Introduction.

\section{Analysis of radio data}
\label{sec:analysis}

   \begin{figure}[!ht]
   \centering
   \resizebox{\hsize}{!}{
   \includegraphics{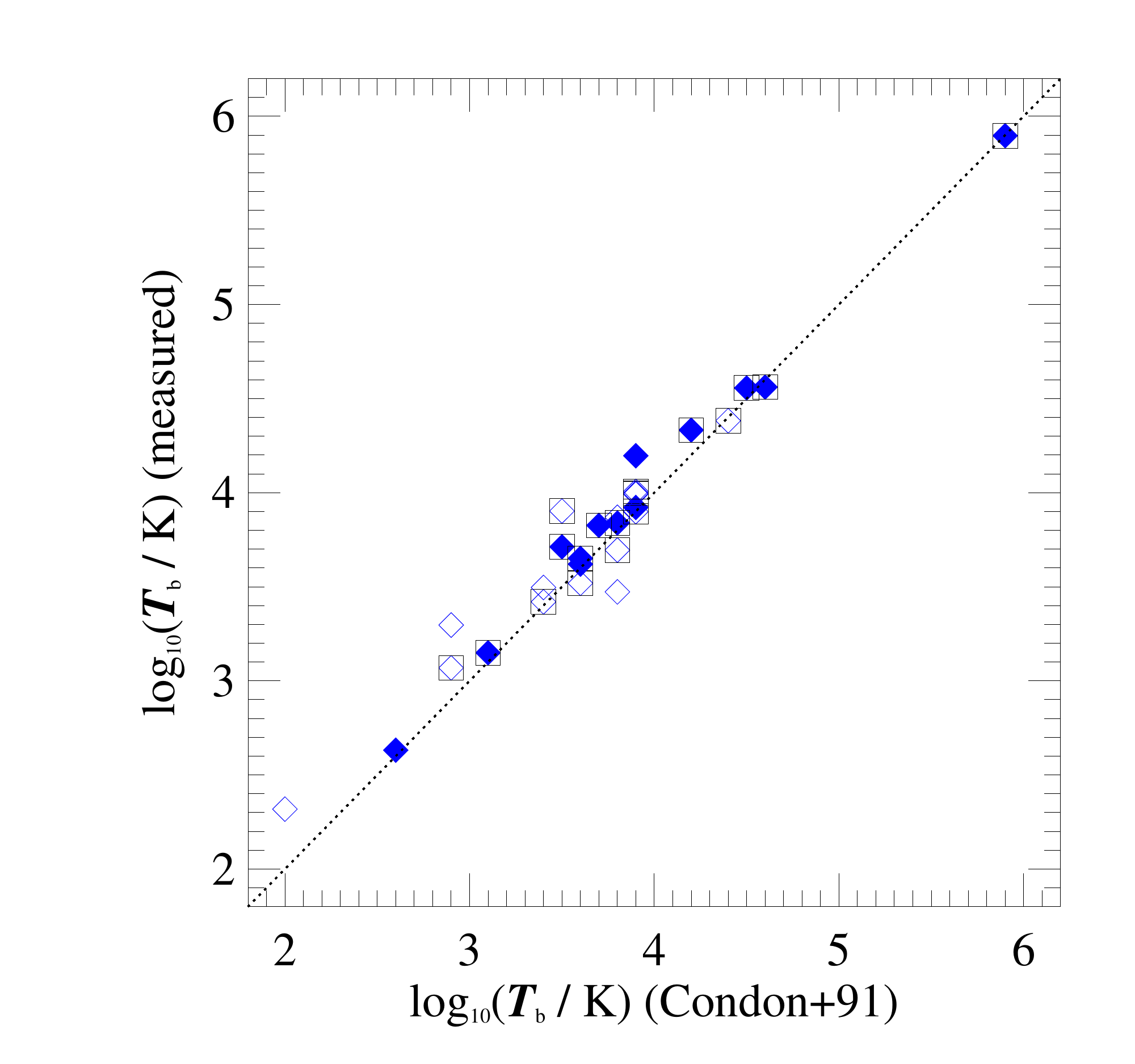}}
      \caption{Brightness temperature $T_{\rm b}$ (in K) as measured from the 8.44 GHz maps, versus brightness temperature from \cite{condon91}. Not all of our objects have a measurement in \cite{condon91}, thus are omitted from the plot. Symbols: blue diamonds denote the common objects, filled blue diamonds denote objects classified as radio-AGN (see Table~\ref{table:agn}), and the black squares denote objects that are classified as mid-IR AGN based on \cite{petric11}. The dotted line is the one-to-one relation. The relative mean dispersion is 2.5\%.
              }
         \label{tbcon}
   \end{figure}

Using the 1.49 and 8.44 GHz images we construct radio spectral index maps ($\alpha$-maps) to probe the origin of the emission arising from the central kpc of our LIRG sample. The available data at 1.49 and 8.44 GHz have different resolutions. In order to be compared, they need to be convolved to the same resolution. Thus we change the resolution of high-resolution 8.44-GHz map to the one of the lower-resolution 1.49-GHz map, and convolve the 8.44 GHz map to the Point Spread Function (PSF) of the 1.49 GHz map, after deconvolving with the PSF of the 8.44 GHz map. We finally make sure that the flux in the maps is conserved after the image manipulation.

To ensure that there are no misalignments in the final image after convolution, we checked that the maximum-flux pixel in the original and final image of each object have the same R.A. and Dec. We also checked whether the original 1.49- and 8.44-GHz images have the maximum-flux pixel at the same coordinates. In most cases the difference is negligible, and can be ignored. In a couple of cases though (NGC 0034 S, IC 1623A/B), the difference is larger than the pixel size of the 1.49 GHz map ($>$ 0.4 arcsec). We believe the shift is real and not a resolution effect, and could be attributed to the existence of an AGN. Studies \citep[e.g.][]{lobanov98} have shown that the peak of core emission gets closer to the black-hole (BH) as we observe at higher frequencies. This is obvious in NGC 0034 (see $\alpha$-map, Fig.~\ref{amaps}), where the 8.44 GHz maximum-flux position is matching the bulge position, whereas the 1.49 GHz maximum-flux position is further out (see notes on the objects in Appendix~\ref{sec:notes} for further discussion).

After assuring the maps are aligned and have the same resolution, we divide them based on the relation between flux density and frequency given above, to obtain the radio spectral index maps. Only flux densities above 3 $\sigma$ in both images are used in the calculation of $\alpha$. The resulting $\alpha$-maps are shown in Fig.~\ref{amaps}. The errors in $\alpha$ were calculated using standard error propagation. Since the synthesised beam-size of the 8.44 GHz observations is much less than the one at 1.49 GHz, the area observed at 8.44 GHz is much smaller than at 1.49 GHz. As a result, even after convolving the 8.44 GHz to the resolution of the 1.49 GHz maps, the resulting $\alpha$-maps can be smaller than the size of the source at 1.49 GHz due to lack of information from the smoothed 8.44 GHz images (e.g. IC 1623A/B).

Radio and infrared properties of our LIRGs are presented in Table~\ref{data}. Fluxes at 1.49 and 8.44 GHz were measured from the original maps by applying a 3$\sigma$ flux cut. We wish to mention that at 8.44 GHz there is an obvious difference between our measurements and the ones presented in \cite{condon91}. The possible explanation for that is the 3 $\sigma$ flux cut we apply, that could be excluding some diffuse emission from the source, while in some cases \cite{condon91} only report the peak flux density at 8.44 GHz. 

As part of the radio analysis of our objects, we also classify them based on their radio structure at both frequencies available (1.49 \& 8.44 GHz) following the scheme: compact (COM) for objects whose contours are circular around the maximum-flux position; extended (EXT) for objects whose contours are not circular around the maximum-flux position and extend outwards; diffuse (DF) for objects whose radio structure appears diffuse; double (DB) for objects that show a double radio structure; and complex (CMX) for objects that show multiple radio components. These classifications are presented in Table~\ref{data}, and aid in our understanding of whether the object contains an AGN or not. 

By studying the variation in the value of $\alpha$ in the radio-spectral-index maps we aim to identify which objects have an AGN component (see Sec.~\ref{sec:radioAGNclass}). Furthermore, by comparing our radio-based result with those from known mid-IR AGN diagnostic tools (see Sec.~\ref{sec:mid-IR}), we are looking for evidence of buried AGN identified in the radio that are not detected in the mid-IR.

We also construct histograms ($\alpha$-hists), shown in Fig.~\ref{amaps}-$Right$. The $\alpha$-hists are based on the $\alpha$-maps and give the values of $\alpha$ at each pixel ($top$ panel). The $middle$ panel is the $\alpha$+error histogram, where the $\alpha$-map and error map (Fig.~\ref{amaps}-$Middle$) were added. The $bottom$ panel shows the $\alpha$-error histogram, produced after subtracting the error map (Fig.~\ref{amaps}-$Middle$) from the $\alpha$-map. These histograms are necessary in the interpretation of the analysis and in order to classify the objects in our sample. The distribution shown reflects the number of values of the radio spectral index, from flat to steep, within the area occupied by the source in the $\alpha$-maps. Whenever we refer to distance from the centre we mean the distance from the position in the map that corresponds to the maximum flux density at 1.49 GHz, as given in Table~\ref{data}.

\begin{sidewaystable*}
\caption{Radio and Infrared data for the 35 LIRG systems of our sample.}\label{data}
\centering
\begin{tabular}{llllllcclrccl}
\hline\hline             \\
\multicolumn{1}{c}{Name} & \multicolumn{1}{c}{R.A.} & \multicolumn{1}{c}{Dec.} & \multicolumn{1}{c}{$ S_{\rm 1.49 GHz}$} & \multicolumn{1}{c}{$S_{\rm 8.44 GHz}$} & 
\multicolumn{1}{c}{$\alpha^{\rm 8.44GHz}_{\rm 1.49 GHz}$} & $\log L_{\rm 1.4 GHz}$ &
$\log L_{\rm IR}$ & $q_{\rm IR}$ & IR8 & \multicolumn{2}{c}{Radio Structure}&  \multicolumn{1}{l}{Alternative}\\\\ 
\hline
 & \multicolumn{1}{c}{(J2000)} & \multicolumn{1}{c}{(J2000)} & \multicolumn{1}{c}{(mJy)} & \multicolumn{1}{c}{(mJy)} & & ($\rm W~Hz^{-1}$) & ($\rm L_{\sun}$) &
 & & {\small 1.49 GHz} & {\small 8.44 GHz} & \multicolumn{1}{l}{Name} \\
\hline
\multicolumn{1}{c}{(1)} & \multicolumn{1}{c}{(2)} & \multicolumn{1}{c}{(3)} & \multicolumn{1}{c}{(4)} & \multicolumn{1}{c}{(5)} & \multicolumn{1}{c}{(6)} & (7) & (8) & (9) & (10) & (11) & (12) & \multicolumn{1}{l}{(13)} \\
\hline
\\
  {\small (1) NGC 0034 (S)}&       $00^{\rm h} 11^{\rm m} 06\fs5$ & $-12{\degr} 06^{\rm m} 26^{\rm s}$ &   56.0$\pm 0.3$ & 15.21$\pm 0.01$ &  0.748$\pm$0.002 & 16.68 & 11.431 & 2.75 & 11.901 & {\tiny COM?} & {\tiny EXT?} & {\tiny NGC 0017} \\
  {\small (2) IC 1623A/B} &       $01^{\rm h} 07^{\rm m} 47\fs5$ & $-17{\degr} 30^{\rm m} 24^{\rm s}$ &    201.27$\pm 0.09$ & 13.491$\pm 0.003$ & 1.558$\pm$0.001& 17.25 & 11.706 & 2.45 & 9.521&{\tiny CMX} & {\tiny CMX} & {\tiny Arp 236}\\
  {\small (3) CGCG 436-030} &       $01^{\rm h}  20^{\rm m} 02\fs6$ & $ ~~14{\degr}  21^{\rm m} 42^{\rm s}$ &  41.4$\pm 0.2$ & 13.337$\pm 0.009$ &  0.654$\pm$0.002 & 16.96 & 11.686 & 2.72& 11.464& {\tiny COM?} & {\tiny CMX }& {\tiny MCG }\\
  & & & & & & & & & & & &{\tiny +02-04-025}\\
  {\small (4) IRAS F01364-1042} &  $01^{\rm h}  38^{\rm m} 52\fs8$ & $ -10{\degr} 27^{\rm m}  12^{\rm s}$ &  15.9$\pm 0.1$ & 8.35$\pm 0.02$ & 0.371$\pm$0.006 & 16.94 & 11.848 & 2.90 & 55.498& {\tiny COM?} & {\tiny COM?} & {\tiny 2MASX J0138} \\
  & & & & & & & & & & & &{\tiny 5289-1027113}\\
  {\small (5) IRAS F01417+1651} &  $01^{\rm h} 44^{\rm m} 30\fs5$ & $  ~~17{\degr}  06^{\rm m} 08^{\rm s}$ &  38.4$\pm 0.5$ & 19.95$\pm 0.05$ & 0.378$\pm$0.003 & 16.83 & 11.639 & 2.81 & 36.755&{\tiny COM} & {\tiny COM} & {\tiny III Zw 035} \\
  {\small (6) UGC 02369 (S)} &  $ 02^{\rm h} 54^{\rm m}  01\fs8$ & $ ~~14{\degr} 58^{\rm m} 15^{\rm s}$ &   41.9$\pm 0.2$ & 8.917$\pm 0.007$ &  0.892$\pm$0.002 & 16.98 & 11.66 & 2.68 & 14.333&{\tiny CMX} & {\tiny CMX} \\
 {\small (7) IRAS F03359+1523} &    $03^{\rm h} 38^{\rm m} 47\fs0$ & $ ~~15{\degr} 32^{\rm m}      53^{\rm s}$ &  17.4$\pm 0.1$ & 7.46$\pm 0.01$ & 0.490$\pm$0.005 & 16.70 & 11.545 & 2.85 & 21.425&{\tiny EXT} & {\tiny CMX} \\
  {\small (8) ESO 550-IG025 (N)} &       $04^{\rm h}  21^{\rm m} 19\fs9$ & $-18{\degr} 48^{\rm m} 39^{\rm s}$ &   15.79$\pm 0.06$ & 6.68$\pm 0.01$ &  0.495$\pm$0.005 & 16.57 & 11.133 & 2.56 & ---&{\tiny COM} & {\tiny DB} & {\tiny MCG}\\
  & & & & & & & & & & & &{\tiny -03-12-002}\\
  {\small (9) ESO 550-IG025 (S)} &       $04^{\rm h} 21^{\rm m} 20\fs0$ & $-18{\degr} 48^{\rm m} 56^{\rm s}$  & 12.44$\pm 0.07$ & 3.024$\pm 0.003$ & 0.815$\pm$0.005 & 16.47 & 11.268 & 2.80 &--- &{\tiny COM} & {\tiny DF} & {\tiny MCG }\\
  & & & & & & & & & & & &{\tiny -03-12-002}\\
 {\small (10) IRAS F05189-2524} &  $ 05^{\rm h} 21^{\rm m}       01\fs4$ & $-25{\degr} 21^{\rm m} 45^{\rm s}$ &   26.9$\pm 0.2$ & 10.97$\pm 0.02$ &  0.518$\pm$0.005 & 17.07 & 12.163 & 3.09 & 10.660&{\tiny COM} & {\tiny EXT?} & {\tiny 2MASX J0521}\\
 & & & & & & & & & & & &{\tiny 0136-2521450}\\
  {\small (11) NGC 2623} &    $ 08^{\rm h} 38^{\rm m} 24\fs0$ & $ ~~25{\degr}      45^{\rm m}      16^{\rm s}$&   95.6$\pm 0.7$ & 34.63$\pm 0.03$ &  0.585$\pm$0.001 & 16.92 & 11.597 & 2.68 & 29.703&{\tiny COM} & {\tiny COM?} &{\tiny Arp 243} \\
  {\small (12) IRAS F08572+3915} &  $ 09^{\rm h} 00^{\rm m} 25\fs3$ & $ ~~39{\degr} 03^{\rm m}  53^{\rm s}$   &  5.4$\pm 0.1$ & 4.07$\pm 0.02$ & 0.17$\pm$0.04 & 16.69 & 12.163 & 3.48 & 4.949&{\tiny COM} & {\tiny COM} \\
  {\small (13) UGC 04881 (NE)} & $09^{\rm h} 15^{\rm m} 55\fs5$ & $~~44{\degr} 19^{\rm m} 58^{\rm s}$ &19.8$\pm 0.2$ & 7.41$\pm 0.02$ &   0.566$\pm$0.004 & 16.89 & 11.578 & 2.69 & 21.656&{\tiny COM?} & {\tiny COM?} & {\tiny MCG }\\
 & & & & & & & & & & & &{\tiny +08-17-065}\\
  {\small (14) UGC 04881 (SW)} &       $09^{\rm h} 15^{\rm m} 54\fs7$ & $~~44{\degr} 19^{\rm m} 51^{\rm s}$ & 8.00$\pm 0.07$ & 1.102$\pm 0.004$ & 1.143$\pm$0.004 & 16.50 & 11.243 & 2.74 & 21.656&{\tiny DF} & {\tiny DF} & {\tiny Arp 055}\\
  {\small (15) UGC 05101} &     $ 09^{\rm h} 35^{\rm m} 51\fs6$ & $      ~~61{\degr} 21^{\rm m} 11^{\rm s}$&   133.9$\pm 0.6$ & 51.05$\pm 0.06$ & 0.556$\pm$0.001 & 17.72 & 12.015 & 2.30 & 17.085&{\tiny COM?} & {\tiny EXT} & {\tiny MCG }\\
  & & & & & & & & & & & &{\tiny +10-14-026}\\
  {\small (16) IRAS F10173+0828} &  $10^{\rm h} 20^{\rm m} 00\fs2$ & $~~08{\degr} 13^{\rm m}  33^{\rm s}$ & 7.9$\pm 0.1$ & 5.21$\pm 0.04$ & 0.24$\pm$0.01 & 16.70 & 11.861 & 3.16 & 102.227&{\tiny COM} & {\tiny COM} & {\tiny 2MASX J1020}\\
  & & & & & & & & & & & &{\tiny 0023+0813342}\\
  {\small (17) IRAS F10565+2448} & $ 10^{\rm h} 59^{\rm m} 18\fs1$ & $ ~~24{\degr}  32^{\rm m}  34^{\rm s}$ &  44.9$\pm 0.4$ & 12.70$\pm 0.01$ & 0.728$\pm$0.002 & 17.34 & 12.083 & 2.75 & 18.245&{\tiny COM} & {\tiny EXT} & {\tiny 2MASX J1059}\\
  & & & & & & & & & & & &{\tiny 1815+2432343}\\
  {\small (18) MCG +07-23-019} &$ 11^{\rm h} 03^{\rm m}  53\fs9$ & $~~40{\degr} 50^{\rm m}  59^{\rm s}$ &   28.2$\pm 0.1$ & 10.27$\pm 0.01$ &  0.582$\pm$0.003 & 16.94 & 11.668 & 2.73 & 14.881&{\tiny EXT} & {\tiny EXT} & {\tiny Arp 148}\\
  {\small (19) UGC 06436 (NW)} & $11^{\rm h} 25^{\rm m} 45\fs0$ & $~~14{\degr} 40^{\rm m} 35^{\rm s}$ & 13.9$\pm 0.1$ & 3.354$\pm 0.005$ &  0.822$\pm$0.007 & 16.63 & 11.456 & 2.83 & 11.430&{\tiny COM?} & {\tiny COM?} & {\tiny IRAS F11231}  \\
  & & & & & & & & & & & &{\tiny +1456}\\
  {\small (20) UGC 06436 (SE)} &    $11^{\rm h} 25^{\rm m} 49\fs5$ & $~~14{\degr} 40^{\rm m} 06^{\rm s}$ &    6.09$\pm 0.09$ & 1.207$\pm 0.005$ & 0.933$\pm$0.007 & 16.27 & 11.178 & 2.91 & &{\tiny COM?} & {\tiny COM?} & {\tiny IC 2810}\\
  {\small (21) NGC 3690 (E)}&   $11^{\rm h} 28^{\rm m} 33\fs6$ & $~~58{\degr} 33^{\rm m} 46^{\rm s}$ & 271.7$\pm 0.2$ &  89.94$\pm 0.06$ & 0.637$\pm$0.001 & 16.93 & 11.668 & 2.74 & 12.707&{\tiny CMX} & {\tiny CMX} & {\tiny IC 694}\\
  {\small (22) NGC 3690 (W)}&   $11^{\rm h} 28^{\rm m} 30\fs9$ & $~~58{\degr} 33^{\rm m} 40^{\rm s}$ &   118.1$\pm 0.2$ &   23.32$\pm 0.01$ & 0.935$\pm$0.001 & 16.56 & 11.578 & 3.01 & 12.707&{\tiny CMX} & {\tiny CMX} & {\tiny IC 694}\\
 
\hline
\end{tabular}
\tablefoot{ Columns: (1) Name of the galaxy; systems are marked with the relative position to each other (e.g.\ N, S, E, W); {\bf the numbers before the object names are used in Fig.~\ref{bpt}}. (2) \& (3) Right Ascension (J2000) \& Declination (J2000), respectively, at the radio position with maximum flux density at 1.49 GHz. (4) \& (5) Integrated flux density at 1.49 \& 8.44 GHz, respectively, of the galaxy where flux above 3$\sigma$ is being considered in the calculation. We also give the error on the measurement. (6) Radio spectral index calculated between 1.49 and 8.44 GHz. (7) The radio luminosity at 1.49 GHz calculated using the standard relation between flux and luminosity; the values of luminosity distance are taken from \cite{tanio10b}. (8) The total infrared luminosity calculated from 8 to 1000 $\rm \mu m$ for each galaxy \citep[see][]{tanio10b}. (9) The parameter $q_{\rm IR}$ = $\log_{10} [(S_{\rm IR(8-1000 \mu m)}/3.75\times10^{12} \rm W~m^{-2}$)/($S_{\rm 1.4 GHz}/\rm W~m^{-2}~Hz^{-1}$)] \citep{helou85}. (10) The ratio of total IR luminosity to the luminosity at 8 $\mu \rm m$ \cite[IR8; e.g.][]{elbaz11}. (11) \& (12) The radio structure at 1.49 \& 8.44 GHz, respectively: COM for compact; EXT for extended; CMX for complex, DB for double; DBDB for double double; and DF for diffuse; a '?' indicates uncertainty in the classification. (13) Gives an alternative name for the galaxy.}
\end{sidewaystable*}

\addtocounter{table}{-1}

\begin{sidewaystable*}
\caption{(Continued)}\label{data}
\centering
\begin{tabular}{llllllcclrccl}
\hline\hline             \\
\multicolumn{1}{c}{Name} & \multicolumn{1}{c}{R.A.} & \multicolumn{1}{c}{Dec.} & \multicolumn{1}{c}{$ S_{\rm 1.49 GHz}$} & \multicolumn{1}{c}{$S_{\rm 8.44 GHz}$} & 
 \multicolumn{1}{c}{$\alpha^{\rm 8.44GHz}_{\rm 1.49 GHz}$} & $\log L_{\rm 1.4 GHz}$ &
$\log L_{\rm IR}$ & $q_{\rm IR}$ & IR8& \multicolumn{2}{c}{Radio Structure} &  \multicolumn{1}{l}{Alternative} \\\\
\hline
 & \multicolumn{1}{c}{(J2000)} & \multicolumn{1}{c}{(J2000)} & \multicolumn{1}{c}{(mJy)} & \multicolumn{1}{c}{(mJy)} & & ($\rm W~Hz^{-1}$) & ($\rm L_{\sun}$) &
 & &{\small 1.49 GHz} & {\small 8.44 GHz} & \multicolumn{1}{l}{Name}\\
\hline
\multicolumn{1}{c}{(1)} & \multicolumn{1}{c}{(2)} & \multicolumn{1}{c}{(3)} & \multicolumn{1}{c}{(4)} & \multicolumn{1}{c}{(5)} &  \multicolumn{1}{c}{(6)} & (7) & (8) & (9) & (10) & (11) & (12) & \multicolumn{1}{l}{(13)}\\
\hline
\\

 {\small (23) IRAS F12112+0305 (NE)}&   $ 12^{\rm h} 13^{\rm m} 46\fs0$ & $ ~~02{\degr}  48^{\rm m} 41^{\rm s}$ & 18.3$\pm 0.3$& 8.06$\pm 0.02$ & 0.473$\pm$0.009  & 17.43 & 12.123 &  2.69 &  29.811$^{\rm t}$ &{\tiny COM} & {\tiny COM} \\
 {\small (24) IRAS F12112+0305 (SW)}&    $ 12^{\rm h} 13^{\rm m} 45\fs9$ & $  ~~02{\degr}  48^{\rm m} 38^{\rm s}$ &  4.6$\pm 0.1$ & 1.67$\pm 0.01$ &   0.586$\pm$0.009 & 16.84 & 11.990 & 3.15 & 29.811$^{\rm t}$&{\tiny COM} & {\tiny COM} \\
  {\small (25) UGC 08058} &  $12^{\rm h}  56^{\rm m}  14\fs2$ & $  ~~56{\degr} 52^{\rm m} 24^{\rm s}$ &  239.9$\pm 1.4$ & 266.5$\pm 0.5$ & -0.060$\pm$0.001 & 18.04 & 12.571 & 2.53 & 8.272&{\tiny COM} & {\tiny COM?}& {\tiny Mrk 231} \\
  {\small (26) VV 250a (NW)}&  $13^{\rm h}  15^{\rm m}  30\fs7$ & $  ~~62{\degr} 07^{\rm m} 45^{\rm s}$ &  6.7$\pm 0.1$  & 1.475$\pm 0.008$ &  0.87$\pm$0.02 & 16.22 & 11.115 & 2.89 & 10.514&{\tiny COM} & {\tiny CMX} &{\tiny UGC 08335 (W)} \\
  {\small (27) VV 250a (SE)}&  $13^{\rm h}  15^{\rm m}  34\fs9$ & $  ~~62{\degr} 07^{\rm m} 28^{\rm s}$&  40.9$\pm 0.4$ &  32.67$\pm 0.02$ &  0.13$\pm$0.02 & 17.01 & 11.709 & 2.70 & 10.514&{\tiny COM} & {\tiny CMX} &{\tiny UGC 08335 (E)}\\
  {\small (28) UGC 08387} &  $ 13^{\rm h}      20^{\rm m}    35\fs3$ & $ ~~34{\degr}  08^{\rm m}  21^{\rm s}$ & 100.5$\pm 0.5$ & 33.93$\pm 0.01$  & 0.626$\pm$0.001 & 17.17 & 11.731 & 2.56 & 15.105&{\tiny COM?} & {\tiny CMX} & {\tiny Arp 193}\\
  {\small (29) NGC 5256 (NE)}&  $ 13^{\rm h} 38^{\rm m} 17\fs7$ & $ ~~48{\degr}  16^{\rm m}  41^{\rm s}$ &  25.8$\pm 0.2$ & 7.88$\pm 0.01$ & 0.684$\pm$0.001 & 17.16 &  11.132 & 2.41 & 10.371&{\tiny COM} & {\tiny COM} & {\tiny Mrk 266 (NE)}  \\
  {\small (30) NGC 5256 (SW)}&  $ 13^{\rm h} 38^{\rm m} 17\fs2$ & $ ~~48{\degr}  16^{\rm m}  32^{\rm s}$ &  31.9$\pm 0.2$ & 8.603$\pm 0.003$  & 0.757$\pm$0.001 &  16.82 &  11.355 & 2.54 & --- &{\tiny EXT} & {\tiny EXT}  & {\tiny Mrk 266 (SW)} \\
{\small (31) NGC 5256 (C)}& $ 13^{\rm h} 38^{\rm m} 17\fs5$ & $ ~~48{\degr}  16^{\rm m}  35^{\rm s}$ &  38.3$\pm 0.2$ & 3.789$\pm 0.003$  & 1.334$\pm$0.001 &  16.50 & --- & --- & --- &{\tiny EXT} & {\tiny EXT} & {\tiny Mrk 266 (C)}  \\
 {\small (32) NGC 5257 (NW)}&  $ 13^{\rm h} 39^{\rm m} 52\fs9$ & $ ~~00{\degr}  50^{\rm m}  24^{\rm s}$ & 6.27$\pm 0.04$ & 1.115$\pm 0.002$ & 0.99$\pm$0.05 & 15.95 & 11.312 & 3.36 & 6.159&{\tiny CMX} & {\tiny CMX} & {\tiny Arp 240} \\
 {\small (33) UGC 08696} &     $ 13^{\rm h} 44^{\rm m} 42\fs$1 & $  ~~55{\degr} 53^{\rm m}13^{\rm s}$ & 129.5$\pm 0.6$ & 43.75$\pm 0.06$ & 0.725$\pm$0.001 & 17.68 & 12.209 & 2.53 & 27.983&{\tiny COM?} & {\tiny CMX} & {\tiny Mrk 273}\\
 {\small (34) IRAS F14348-1447 (NE)}&    $14^{\rm h}  37^{\rm m}  38\fs3$ & $-15{\degr}  00^{\rm m}  21^{\rm s}$ & 12.5$\pm 0.1$& 3.55$\pm 0.01$  & 0.727$\pm$0.003 & 17.39 & 11.950 & 2.56 & 24.887$^{\rm t}$&{\tiny COM} & {\tiny COM} \\
 {\small (35) IRAS F14348-1447 (SW)}&    $14^{\rm h}  37^{\rm m}  38\fs2$ & $-15{\degr}  00^{\rm m}  24^{\rm s}$ & 20.28$\pm 0.1$& 6.05$\pm 0.01$  & 0.697$\pm$0.003 & 17.59 & 12.193 & 2.60 & 24.887$^{\rm t}$&{\tiny COM} &{\tiny COM} \\
 {\small (36) VV 340a (N)}& $ 14^{\rm h}  57^{\rm m} 00\fs6$ & $ ~~24{\degr} 37^{\rm m} 02^{\rm s}$ & 55.4$\pm 0.1$ & 3.099$\pm 0.005$ & 1.66$\pm$0.01 & 17.23 & 11.658 & 2.43& 6.958& {\tiny DBDB} & {\tiny DB} & {\tiny UGC 09618}\\
 {\small (37) VV 705 (N)}& $ 15^{\rm h}  18^{\rm m} 06\fs1$ & $ ~~42{\degr} 44^{\rm m} 44^{\rm s}$ &   26.4$\pm 0.2$  & 8.23$\pm 0.01$ & 0.673$\pm$0.002 & 17.04 & 11.849 & 2.81 & 14.589&{\tiny COM?} & {\tiny COM?} & {\tiny I ZW 107}\\
 {\small (38) VV 705 (S)}&  $ 15^{\rm h}  18^{\rm m} 06\fs3$ & $ ~~42{\degr} 44^{\rm m} 38^{\rm s}$  &   18.0$\pm 0.1$ & 2.178$\pm 0.008$ & 1.217$\pm$0.002 & 16.88 & 11.120 &  2.24 & &{\tiny COM?} & {\tiny COM?} & {\tiny I ZW 107}\\
 {\small (39) IRAS F15250+3608} & $15^{\rm h} 26^{\rm m} 59\fs4$ & $~~35{\degr} 58^{\rm m} 36^{\rm s}$ &  12.1$\pm 0.2$& 10.34$\pm 0.06$& 0.09$\pm$0.01 & 17.00 & 12.081 & 3.08 & 11.320&{\tiny COM} & {\tiny COM} \\
 {\small (40) UGC 09913} &    $15^{\rm h} 34^{\rm m} 57\fs2$ & $ ~~23{\degr} 30^{\rm m}11^{\rm s}$ & 303.5$\pm 1.9$& 147.09$\pm 0.09$& 0.417$\pm$0.001 & 17.46 & 12.275 & 2.82 & 72.130& {\tiny COM?} & {\tiny DB} & {\tiny Arp 220} \\
 {\small (41) NGC 6090} &  $ 16^{\rm h} 11^{\rm m}  40\fs8$ & $ ~~52{\degr} 27^{\rm m} 26^{\rm s}$ & 45.24$\pm 0.06$ & 0.513$\pm 0.004$ &  2.583$\pm$0.003 & 17.02& 11.580 & 2.56& 8.525&{\tiny DB} & {\tiny DB} & {\tiny I ZW 135}\\
 {\small (42) IRAS F17132+5313 (NE)} & $ 17^{\rm h} 14^{\rm m} 20\fs3$ & $ ~~53{\degr} 10^{\rm m} 31^{\rm s}$ &  5.92$\pm 0.1$ & 2.31$\pm 0.01$ & 1.065$\pm$0.009 & 17.00 & 11.876 & 2.88 & 11.440$^{\rm t}$&{\tiny EXT} & {\tiny DF} \\
 {\small (43) IRAS F17132+5313 (SW)} & $ 17^{\rm h} 14^{\rm m} 19\fs7$ & $ ~~53{\degr} 10^{\rm m} 28^{\rm s}$ & 14.71$\pm 0.1$ & 2.320$\pm 0.004$ & 0.542$\pm$0.009 & 16.60 & 11.206 & 2.60 & 11.440$^{\rm t}$&{\tiny COM} & {\tiny COM} \\
 {\small (44) IRAS F22491-1808} & $ 22^{\rm h} 51^{\rm m} 49\fs3$ & $ -17{\degr} 52^{\rm m} 23^{\rm s}$ & 5.37$\pm 0.05$ & 2.82$\pm 0.01$ &  0.37$\pm$0.01 & 16.93 & 12.201 & 3.27 & 33.912&{\tiny COM?} & {\tiny COM?} \\
 {\small (45) IC 5298} &  $  23^{\rm h} 16^{\rm m}  00\fs7$ & $  ~~25{\degr} 33^{\rm m} 23^{\rm s}$ & 25.5$\pm 0.1$ & 7.079$\pm 0.009$ &   0.740$\pm$0.004 & 16.65 & 11.600 & 2.95 & 15.843&{\tiny COM} & {\tiny CMX} & {\tiny ZW 475.065}\\
 {\small (46) Mrk 0331 (N)} &  $  23^{\rm h} 51^{\rm m}  26\fs7$ & $  ~~20{\degr} 35^{\rm m} 10^{\rm s}$ & 63.7$\pm 0.2$ & 15.743$\pm 0.008$ & 0.806$\pm$0.002 & 16.69 & 11.497 & 2.81 & 11.549&{\tiny COM?} & {\tiny CMX} & {\tiny UGC 12812 (E)}\\
\hline
\end{tabular}
\tablefoot{The base-10 logarithm of the total infrared luminosity of IRAS F14348-1447 is 12.390. Here we give the fractional contribution of the two components to the total infrared luminosity. The fractional contribution was calculated by doing photometry on the 8 and 24 $\mu \rm m$ maps of the object. Similarly for  IRASF12112+0305 $\rm log_{10}(L_{\rm IR})$ = 12.363 and for IRAS F17132+5313 $\rm log_{10}(L_{\rm IR})$ = 11.961. A (t) next to the values of IR8 stands for total measurement for the system.
}
\end{sidewaystable*}

\begin{table}
\caption{Brightness Temperature measurements}             
\label{tab:Tb}      
\centering                          
\begin{tabular}{l l l l r}        
\hline\hline                 
\multicolumn{1}{c}{name} & \multicolumn{1}{c}{log$_{10}$} & a  & b  & \multicolumn{1}{c}{P.A.}\\    
& \multicolumn{1}{c}{($T_{\rm b}$/K)}& ('') & ('') & \multicolumn{1}{c}{(deg.)} \\ 

\hline                        
\multicolumn{1}{c}{(1)} & \multicolumn{1}{l}{(2)}  & \multicolumn{1}{l}{(3)} & 
\multicolumn{1}{l}{(4)} & \multicolumn{1}{c}{(5)} \\
\hline
NGC 0034 (S) & 3.15 & 0.39 & 0.32 &   12.2 \\
IC 1623A/B & 3.75 & 0.21 & 0.13 &   -4.8 \\
CGCG 436-030 & 3.07 & 0.43 & 0.31 &  -12.4 \\
IRAS F01364-1042 & 3.69 & 0.14 & 0.14 &  -14.3 \\
IRAS F01417+1651 & 3.99 & 0.17 & 0.13 &  -32.2 \\
UGC 02369 (S) & 3.71 & 0.14 & 0.14 &  -14.4 \\
IRAS F03359+1523 & 3.87 & 0.13 & 0.09 &   31.5 \\
ESO 550-IG025 (S) & 2.41 & 0.82 & 0.37 &   27.5 \\
ESO 550-IG025 (N) & 3.47 & 0.12 & 0.10 &    2.7 \\
IRAS F05189-2524 & 3.52 & 0.17 & 0.18 &   -2.0 \\
NGC 2623 & 3.65 & 0.38 & 0.23 &   -4.41 \\
IRAS F08572+3915 & 3.92 & 0.09 & 0.07 &  -12.2 \\
UGC 04881 (NE) & 3.42 & 0.21 & 0.15 &  -12.5 \\
UGC 04881 (SW) & 1.66 & 0.80 & 0.33 &    9.9 \\
UGC 05101 & 4.56 & 0.14 & 0.11 &    5.2 \\
IRAS F10173+0828 & 4.33 & 0.07 & 0.04 &  -11.1 \\
IRAS F10565+2448 & 2.98 & 0.47 & 0.33 &  -24.2 \\
MCG +07-23-019 & 3.62 & 0.23 & 0.10 &   -2.1 \\
UGC 06436 (NW) & 2.24 & 0.68 & 0.30 &    6.7 \\
UGC 06436 (SE) & 2.04 & 0.56 & 0.18 &  -34.2 \\
NGC 3690 (E) & 4.00 & 0.37 & 0.28 &  -34.7 \\
NGC 3690 (W) & 3.90 & 0.19 & 0.18 &  -14.7 \\
IRAS F12112+0305 (NE) & 4.40 & 0.12 & 0.10 &-0.1 \\
IRAS F12112+0305 (SW) & 3.61 & 0.18 & 0.08 &-0.1 \\
UGC 08058 & 5.90 & 0.07 & 0.06 &  -14.6 \\
VV 250a (NW) & 2.36 & 0.49 & 0.02 &  -19.5 \\
VV 250a (SE) & 3.30 & 0.54 & 0.35 &  -18.1\\
UGC 08387 & 2.93 & 1.46 & 0.26 & -143.2 \\
NGC 5256 (NE) & 3.48 & 0.25 & 0.19 & -131.4 \\
NGC 5256 (SW) & 2.20 & 0.70 & 2.03 & 0.2 \\
NGC 5256 (C) & 2.19 & 0.69 & 1.22 & 0.8 \\
NGC 5257 & 2.32 & 0.26 & 0.25 &  -25.3 \\
UGC 08696 & 3.84 & 0.32 & 0.18 &   25.7 \\
IRAS F14348-1447 (NE) & 3.13 & 0.22 & 0.28 &0.0 \\
IRAS F14348-1447 (SW) & 3.15 & 0.20 & 0.30 & 0.0 \\
VV 340a (N) & 3.19 & 0.18 & 0.12 & -120.4 \\
VV 705 (N) & 3.50 & 0.20 & 0.15 &  -53.3 \\
VV 705 (S) & 2.63 & 0.28 & 0.21 &  -27.0 \\
IRAS F15250+3608 & 4.56 & 0.06 & 0.06 &  106.3 \\
UGC 09913 & 4.38 & 0.21 & 0.15 &    5.4 \\
NGC 6090 & 2.62 & 0.19 & 0.04 &    4.3 \\
IRAS F17132+5313 (NE) & 2.07 & 0.61 & 1.83 &-3.0 \\
IRAS F17132+5313 (SW) & 4.08 & 0.15 & 0.13 &-0.3 \\
IRAS F22491-1808 & 3.90 & 0.07 & 0.06 &   -2.0 \\
IC 5298 & 3.83 & 0.13 & 0.09 &  -30.7 \\
Mrk 0331 (N)& 4.20 & 0.12 & 0.09 &   -5.9 \\

 \hline                                   
\end{tabular}
\tablefoot{The brightness temperature was measured directly from the 8.44 GHz radio maps using the relation stated in \cite{condon91}, log[(S/$\Omega$)(mJy/arcsec$^{2}$)] $\approx$ log[T$_{\rm b}$(K)] - 1.3, where $\Omega$ = 3$\pi~ \times$ a $\times$ b/(8 $\times$ ln(2)) and S is the flux density, a the major axis and b the minor axis in arcsec. The values are given in Column 2. We used the function GAUSS2DFIT of IDL to fit ellipses to the radio components. This gives us the deconvolved major and minor axes, shown in Columns 3 \& 4 respectively, and the position angle PA of the ellipse from the x-axis in degrees, counter-clockwise, shown in Column 5. 
}
\end{table}

\cite{condon91} used the brightness temperature $T_{\rm b}$ to distinguish between SB and AGN, where an AGN would have   $T_{\rm b}  {\sim} 10^{5}$ K, but they do not provide brightness temperatures for all of the objects in our sample. Here we provide measurements of the brightness temperature $T_{\rm b}$ for all individual objects in our sample, using the relation given in \cite{condon91}, log[($S/\Omega$)(mJy/arcsec$^{2}$)] $\approx$ log[$T_{\rm b}$(K)] - 1.3, where $S$ is the flux density and $\Omega$ = 3$\pi \times$ a $\times$ b/(8 $\times$ ln(2)), where a is the major axis and b the minor axis both in arcsec. We use the function GAUSS2DFIT in IDL to fit an ellipse to each radio component at 8.44 GHz and calculate the deconvolved major and minor axes and the corresponding position angle PA. All values are given in Table~\ref{tab:Tb}. The differences from the values of $T_{\rm b}$ reported in \cite{condon91} are due to the difference in the flux densities we measured at 8.44 GHz from the ones in literature (see Fig.~\ref{tbcon}): the relative mean dispersion between measured values and those from the literature is 2.5\%.

\subsection{Interpretation of radio analysis}
\label{sec:radioAGNclass}

\begin{figure*}[!ht]
    \resizebox{\hsize}{!}
            {\includegraphics{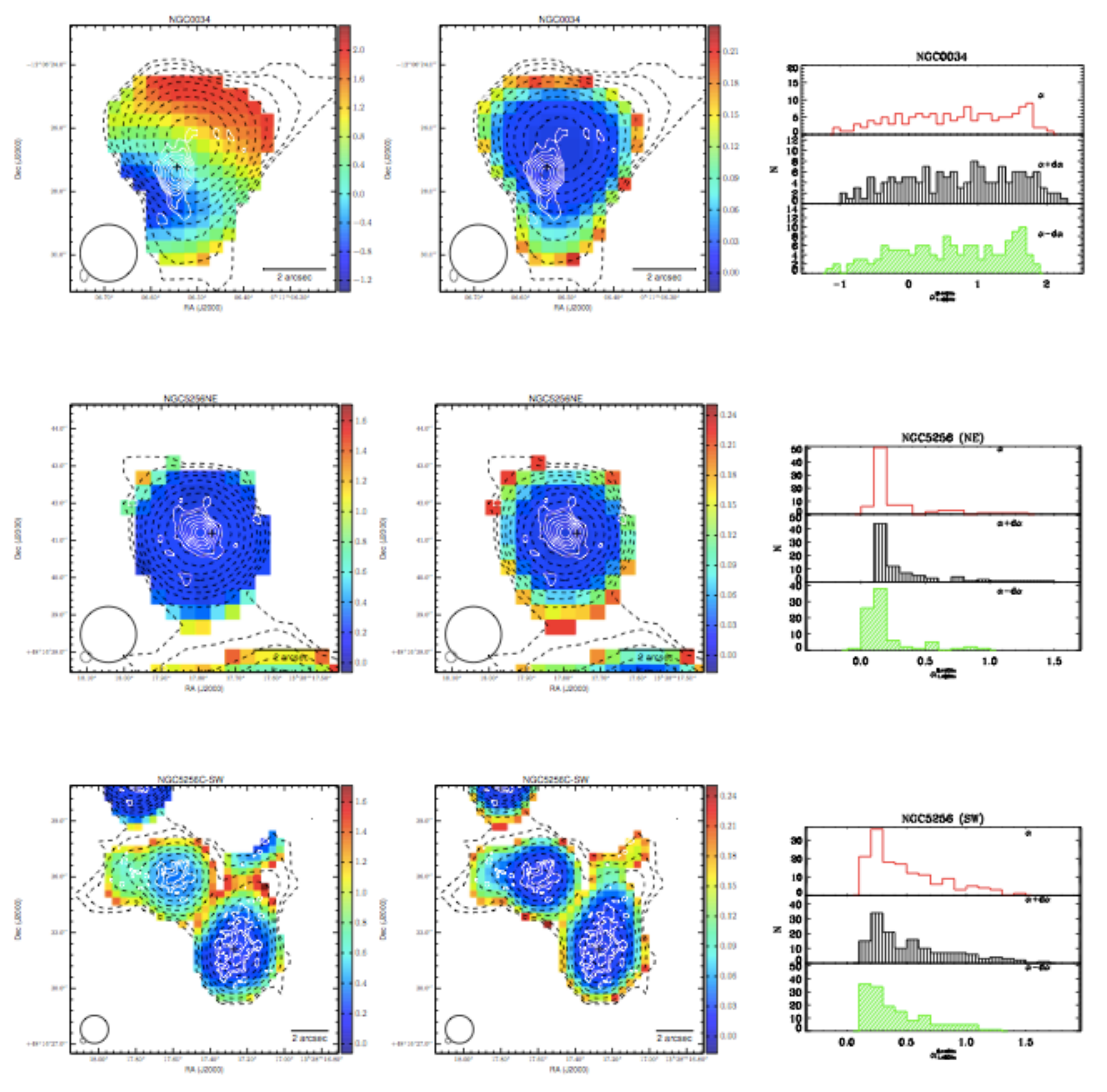}
            }
   
   \caption{Each row corresponds to an object from our sample. ($Left$) Radio-spectral-index maps ($\alpha$-maps) are represented with the colour-map; the colour bar on the right shows the radio spectral index values from low (flat $\alpha$) to high values (steep $\alpha$). Overlaid, the radio contours at 1.49 GHz (black dashed) and at 8.44 GHz (white solid), where the contour levels scale as log$_{10}$ of the flux density from the lowest flux value (3 $\sigma$) to the highest.  
($Middle$) The error map for the radio spectral index per pixel value. The colour bar on the right shows the range of values. Contours as on the $Left$. In both plots, on the bottom left corner we give the beam of the 1.49 GHz map as black circle, and the beam of the 8.44 GHz map as grey circle/ellipse. On the bottom right we give a scale-bar of 2 arcsec, as visual aid. 
($Right$) Histograms of the $\alpha$-maps: ($Top$) Histograms of the radio spectral index, based on Fig.~\ref{amaps}-$Left$, that give the values of $\alpha$ at each pixel. ($Middle$) Histograms where the errors on $\alpha$ (Fig.~\ref{amaps}-$Middle$) have been added. ($Bottom$) Histograms where the errors on $\alpha$ have been subtracted. The rest of the figure is located in the Appendix.
     }
              \label{amaps}%
    \end{figure*}

One of the goals of this study is to unveil buried AGN in the centre of local LIRGs by identifying variations in the radio-spectral-index maps (see Fig.~\ref{amaps}) created using the radio continuum maps at 1.49 and at 8.44 GHz. AGN or SB classification based on integrated values of the radio spectral index is inconclusive, since both can display similar values.

In the case of radio-loud radio sources that display radio jets, separating SB and AGN is more straight-forward, since these structures can only be created by powerful AGN. In the case of radio-quiet radio sources though, separating the two contributions is challenging. The LIRGs in our sample have been known to contain a nuclear SB that is identified in radio via synchrotron emission and low brightness temperatures \citep{condon91}; only powerful AGN have $ T_{\rm b} \gtrsim 10^{5}$-$10^{6}$ K \citep[see also][]{clemens08}. At the same time, some of them are known to have an AGN contribution \cite[see][for the AGN percentage]{vega08, petric11, alonso13}. Still, AGN and SB can both give a steep radio spectral index \citep[$\sim$ 0.8; e.g.][]{krolik99, odea98, niklas97, mao14}, a signature of synchrotron emission, which complicates the analysis. Therefore, a measure of the integrated radio spectral index does not provide enough information to distinguish between the two power sources.

Another challenging task is to separate synchrotron from thermal free-free emission. As illustrated by \cite{condon91}, free-free emission from HII regions displays a flat radio spectral index ($\alpha \sim$ 0.1), and at frequencies below 10 GHz, synchrotron emission dominates free-free emission. The $\alpha$-maps, created from data below this wavelength range, will help us to discern between these two cases.

Due to the unified AGN scheme \citep[e.g.][]{antonucci93}, we have to also take into account orientation effects. In radio-loud AGN, depending on the viewing angle we can get a compact radio source or a more extended radio structure \cite[FRI or FRII or in-between, e.g.][]{fr74}. In the former case, this can occur when the jet is pointing towards the observer and there is constant flow of material causing a relativistic effect known as Doppler boosting. The resulting integrated radio spectral index from this source would be inverted, .i.e we get negative values of $\alpha$. In the case of FRI or FRII radio sources, where one can distinguish the jet and lobes, especially in the case of FRII, the picture changes. Material gets transferred through the jet to the lobes further out form the core. The hot-spots created display flat radio spectral indices ($\alpha <$ 0.5) due to the constant supply by the jet. At areas, though, of the lobe further out from the hot-spot, where material has been ejected earlier and cooled down due to loss of energy \citep{carilli96}, the radio spectral index is steeper in the hot-spots \cite[but see][for an alternative physical picture]{blundell01}. This is a signature of synchrotron emission and the phenomenon is called spectral ageing. To summarise, in an AGN, unless there is constant supply of newly ejected material, there is rapid loss of energy, causing synchrotron ageing that is seen as steepening in the values of the radio spectral index further out from the injection point.

The starburst phase in LIRGs is known to have a short synchrotron lifetime \citep[e.g.][]{clemens08}. Also, compact, nuclear starbursts should have flat spectra due to optically-thick free-free absorption \citep{condon91, clemens08, clemens10, leroy11, murphy13, murphy13b}. If the SB is situated in an HII region, there may be some steepening at the edges of the radio source, i.e. spectral ageing, as cosmic-ray (CR) electrons are escaping the system and cool down. Still, we do not expect the values of $\alpha$ to get steeper than $\sim$ 0.83 $\pm$ 0.13 \citep{niklas97}. We adopt a 2 $\sigma$ uncertainty based on this range ($\alpha$ = 1.1) for our radio interpretation below.

By studying the $\alpha$-histograms (see Fig.~\ref{amaps}-$Right$) we find that the objects in study fall in the following categories, after taking into account the uncertainties:
\begin{enumerate}
      \item flat $\alpha$ values with steep tail (29/46);
      \item flat $\alpha$ values with small dispersion (7/46);
      \item steep $\alpha$ values (5/46);
      \item inverted $\alpha$ values in the centre (4/46); 
      \item large range of $\alpha$ values, from flat to steep (3/46);
      \item large range with core shift (2/46).
   \end{enumerate}

We classify the LIRGs in our sample based on the following three categories, after taking into account the way the values of the radio spectral index vary with distance from the maximum-flux position in the radio, as seen in the $\alpha$-maps, the radio structure at both 1.49 \& 8.44 GHz, the $\alpha$-histograms, and the uncertainties in $\alpha$: 
  \begin{enumerate}
      \item {\bf radio-AGN} when
		\begin{inparaenum}[\itshape i\upshape)]
		\item $\alpha$ has negative values (inverted) due to synchrotron self-absorption from a jet pointing towards the observer;
		\item $\alpha$ is flat in the centre, displaying a steep tail suggesting synchrotron ageing, and the values of $\alpha$ get larger than 1.1; 
		\item the $\alpha$-map has steep $\alpha$ values that exceed 1.1;
		\item there is a shift in the core radio emission between 1.49 \& 8.44 GHz (see NGC 0034 N \& IC 1623A/B in Appendix~\ref{sec:notes}).
		\end{inparaenum}
    \item {\bf radio-SB} when
    		\begin{inparaenum}[\itshape i\upshape)]
    		\item $\alpha$ is flat across the distribution, and the brightness temperature is less than $10^{5}$-$10^{6}$ K; 
    		\item $\alpha$ is flat in the centre, displaying a steep tail suggesting synchrotron ageing due to CR electrons, but $\alpha$ does not exceed the value of 1.1;
		\item $\alpha$ has a steep distribution ($\sim$ 0.8) but does not exceed the value of 1.1.
        		\end{inparaenum}
    \item {\bf radio-AGN/SB} when we cannot distinguish between the two, and when due to the uncertainties in $\alpha$ it is not clear that the steepening in the radio spectral index exceeds the theoretical value of 1.1. This category can contain an AGN, or a SB or a mixture of both.
 \end{enumerate}

\clearpage

\section{Results and Disscussion}
\label{sec:results}

The radio classification, which is based on the interpretation of our radio analysis (see Sec.~\ref{sec:radioAGNclass}) is presented in Table~\ref{table:agn}. For a better understanding of our results, notes on individual objects are presented in the Appendix~\ref{sec:notes}, where we explain in detail the characteristics of each object and the way we classified them. Furthermore, all the $\alpha$-maps are available online at http://users.physics.uoc.gr/$\sim$eleniv/radiogoals.html.

Based on the above classification and the radio properties presented in Table~\ref{data}, we find that from the 46 individual LIRGs, 21 are radio-AGN, 9 are radio-SB and 16 are radio-AGN/SB.

\subsection{The infrared-radio correlation at low redshifts}
\label{sec:ir-radio_corr}

Luminous infrared galaxies are known for their tight relation between radio and infrared emission \cite[e.g.][]{helou85, bell03, murphy09, ivison10}. This is expressed by the parameter $q_{\rm IR}$ = $\log_{10} [(F_{\rm IR(8-1000 \mu m)}/3.75\times10^{12} \rm W~m^{-2}$)/($S_{\rm 1.4 GHz}/\rm W~m^{-2}~Hz^{-1}$)], a measure of the star-formation-rate of a galaxy, where $F_{\rm IR(8-1000 \mu m)}$ is the total flux in the infrared band, from 8 to 1000 $\mu \rm m$ in W m$^{-2}$ and $S_{\rm 1.4 GHz}$ is the flux density at 1.4 GHz in W m$^{-2}$ Hz$^{-1}$.

\begin{figure*}[!ht]  

    \resizebox{\hsize}{!}
            {\includegraphics{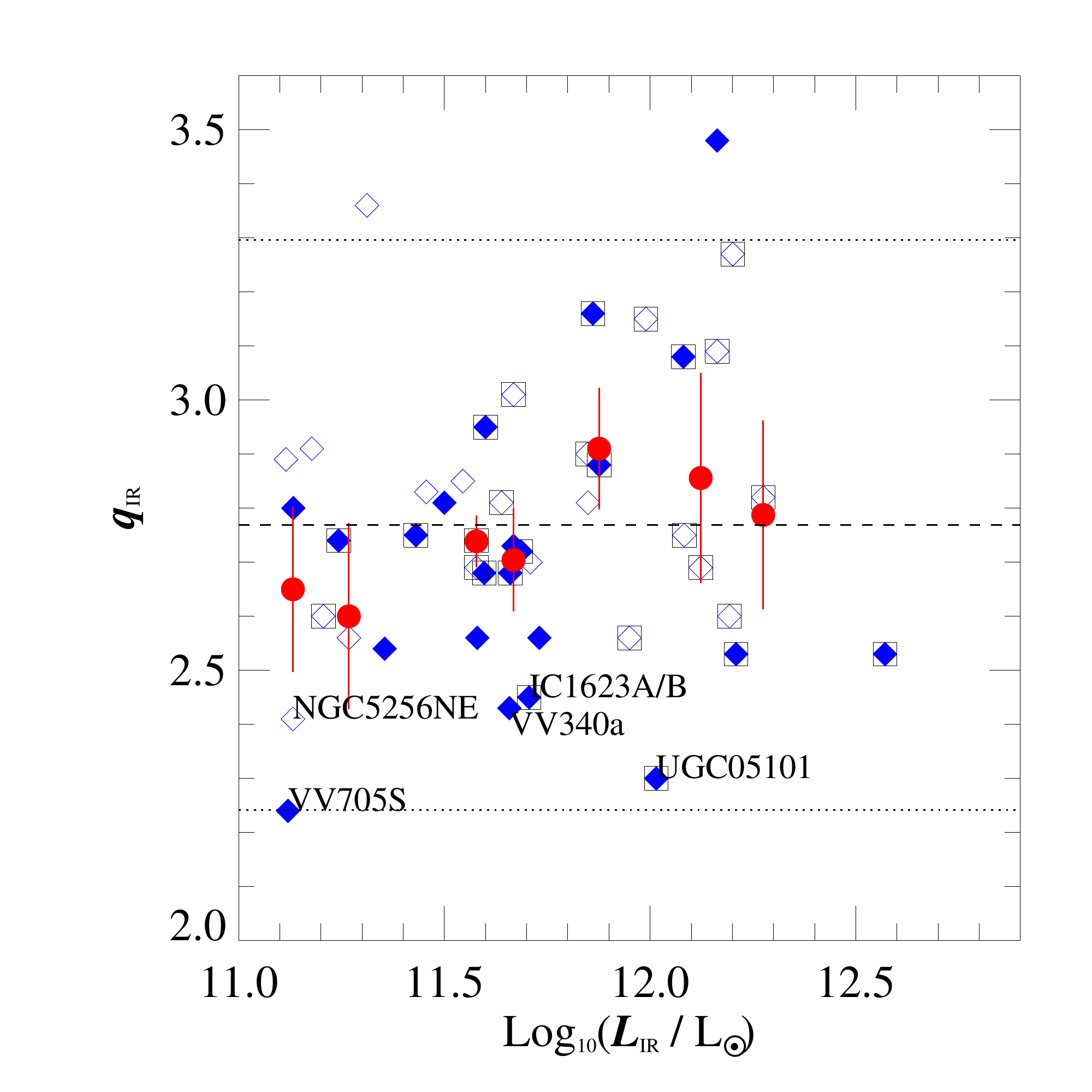}
            \includegraphics{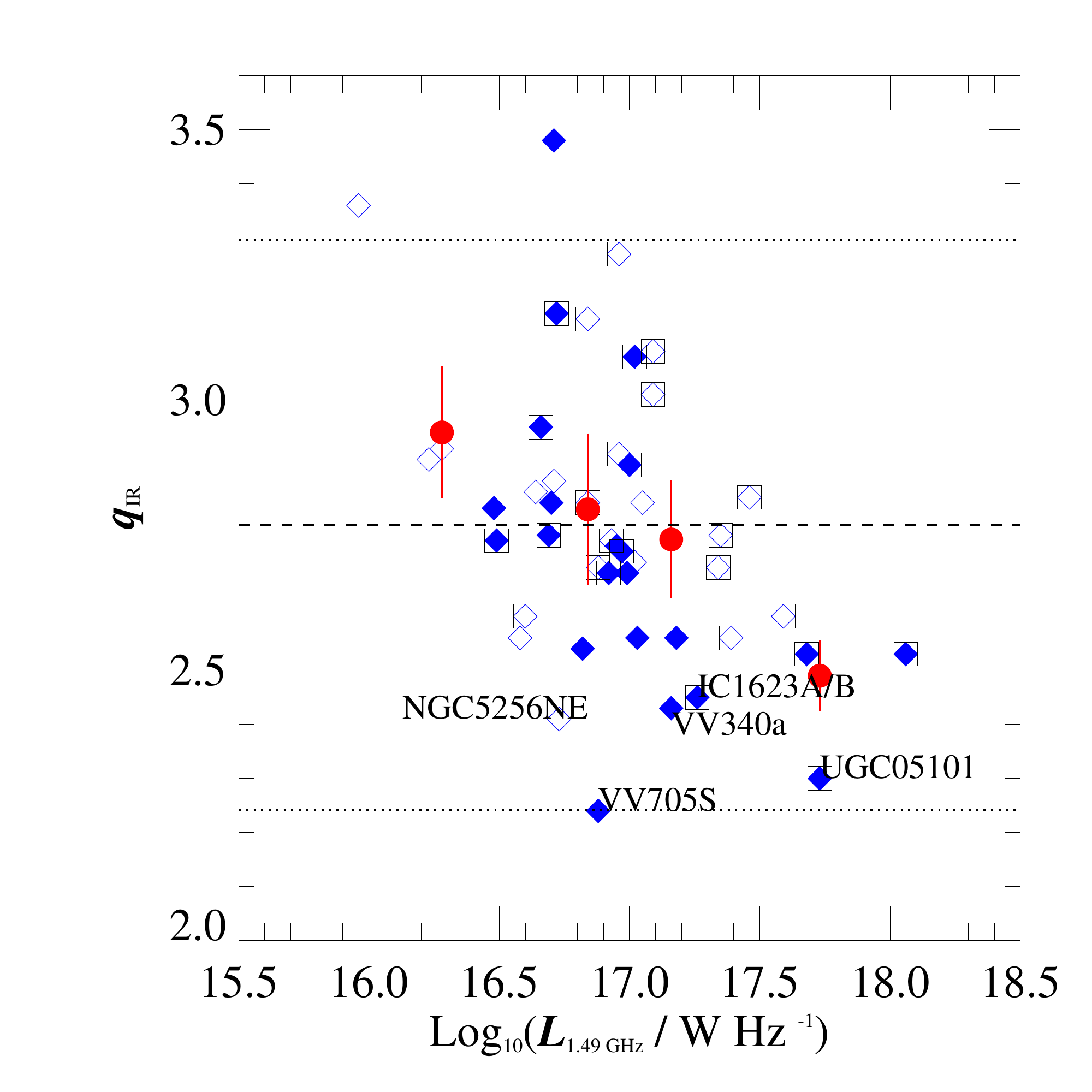}}
   \caption{($Left$) $q_{\rm IR}$ vs infrared luminosity $L_{\rm IR/8-1000 \mu m}$ in units of solar luminosity. ($Right$) $q_{\rm IR}$ versus radio luminosity at 1.49 GHz (see Table~\ref{data}). In both figures, the blue diamonds represent the 46 individual objects in our sample, where objects identified as radio-AGN based on the $\alpha$-maps are shown as filled blue diamonds (see Table~\ref{table:agn}). Black squares denote objects with mid-IR fraction above zero \citep{petric11}, thus they have a contribution to their mid-IR flux from an AGN. Red filled circles show the mean value of $q_{\rm IR}$ for each luminosity bin and the error bars are the standard deviation; note that the median luminosity is calculated instead of the mean to avoid edge effects due to bin selection. The dashed horizontal line shows the mean $q_{\rm IR}$ for the sample with value of 2.75 $\pm$  0.27; the {\bf2} $\sigma$ error is shown by the horizontal dotted lines. Objects with $q_{\rm IR}$ values below the 1 $\sigma$ error lines are marked with their name. 
      }
              \label{qir}%
    \end{figure*}

In Fig.~\ref{qir} we explore this relation for the 46 individual LIRGs in our sample. On the left we show the parameter $q_{\rm IR}$ as a function of the total infrared luminosity from 8-1000 $\rm \mu$m, and on the right with respect to the radio luminosity at 1.49 GHz. Higher resolution mid-IR observations \citep[see][]{tanio10b, tanio11, stierwalt13}, enabled us to measure total infrared luminosities for each galaxy of the system separately \citep{tanio13}, and calculate the value of $q_{\rm IR}$ for individual galaxies, which could not be done by \cite{condon91} due to lack of available observations. We note that the mean $q_{\rm IR}$ = 2.75 $\pm$ 0.27 we find, agrees with the average value reported in \cite{bell03} of 2.64 $\pm$ 0.26 dex for total infrared luminosity; the latter sample includes a selection of spiral, normal and irregular galaxies, as well as SBs and ULIRGs, that have a wide range in luminosities.

Objects with values of $q_{\rm IR}$ less than 2 $\sigma$ from the mean have excess radio emission making them radio-louder than the rest. From Fig.~\ref{qir} we see that only one galaxy (VV 705 S) lies just below the 2 $\sigma$ uncertainty line. VV 705 S is classified as a radio-AGN based on our radio analysis, while based on the mid-IR and optical diagnostics presented in Sec.~\ref{sec:mid-IR}, it is a SB (see Table~\ref{table:agn}). This diagram thus, presents additional evidence for an object containing an AGN. On the other hand, the rest of the objects in our sample are located above the 2 $\sigma$ uncertainty line, the bottom line in the figure, and as a result there is no strong evidence from this figure for the existence of an AGN due to excess radio emission above what expected from a SB.

\subsection{Comparison to mid-infrared and optical AGN diagnostics}
\label{sec:mid-IR}

LIRGs have been explored in detail in the mid-IR and one of the AGN diagnostics used extensively is the ratio of high to low ionisation lines (e.g. [Ne V]/[Ne II] and [O IV]/[Ne II]), as it is a powerful tool to identify AGN \citep[see e.g.][]{armus07, petric11}. The presence of the [NeV] line more, and to less extend that of the [O IV] line, in the mid-infrared spectrum of an object indicates an AGN \citep[e.g.][]{petric11}, since the ionisation potential of the [Ne V] line is too high to be produced by stars; for [O IV] the ionisation potential can be similar to that coming from OB stars. Furthermore, the presence of an AGN in the centre of the LIRG may cause the PAH features to disappear, albeit on sub-kpc scales, due to strong radiation fields. As a result the PAH features are weak or absent in the central regions of AGN \cite[e.g.][]{genzel98,armus07}. Empirically \cite[see][]{stierwalt13, murphy13}, an equivalent width (EQW) less than 0.27 $\rm \mu m$ in the 6.2 $\rm \mu m$ PAH feature suggests that an AGN is contributing significantly to the MIR emission of the LIRG.

\begin{figure}[!ht]  
    \resizebox{\hsize}{!}
            {\includegraphics{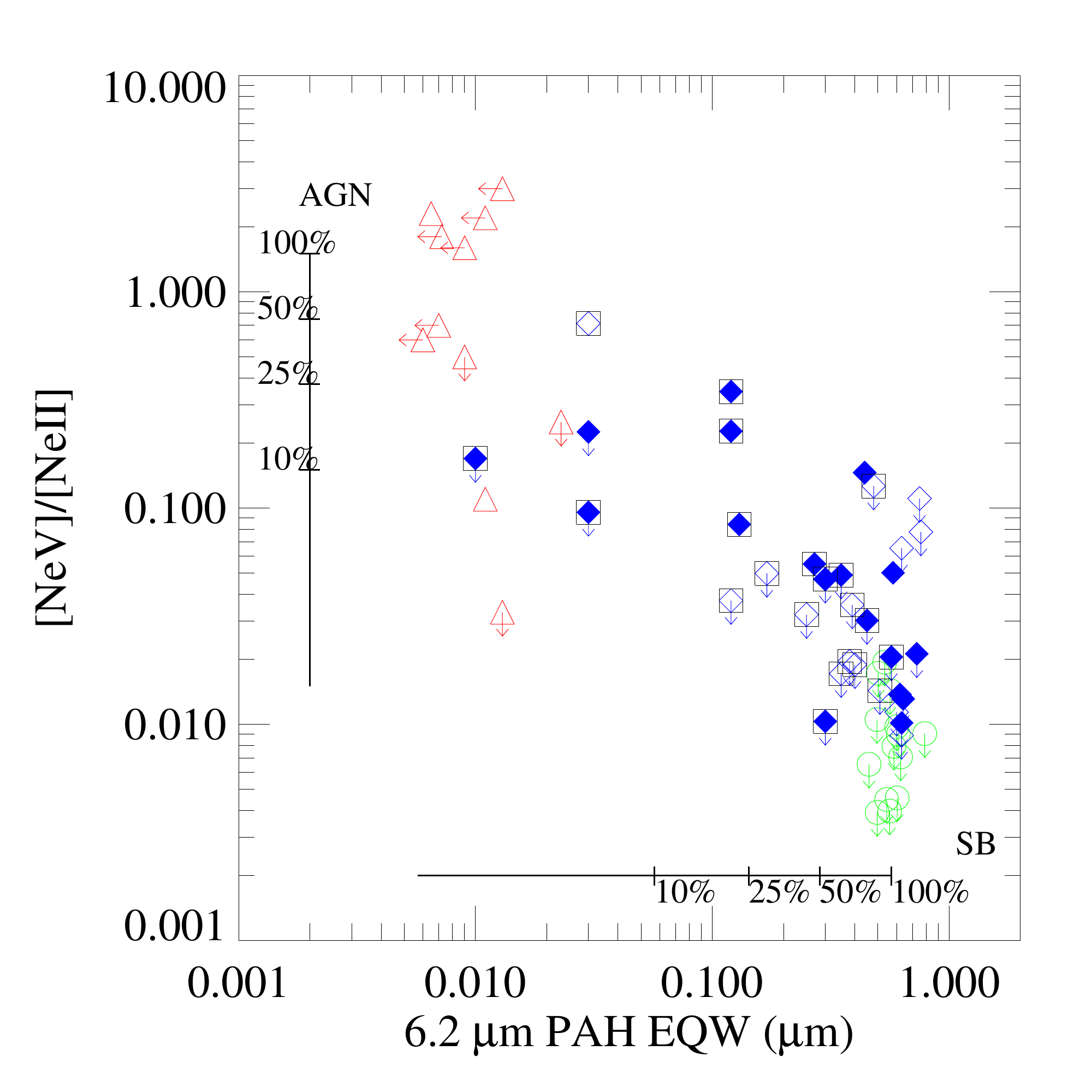}
            }
                \resizebox{\hsize}{!}{
             \includegraphics{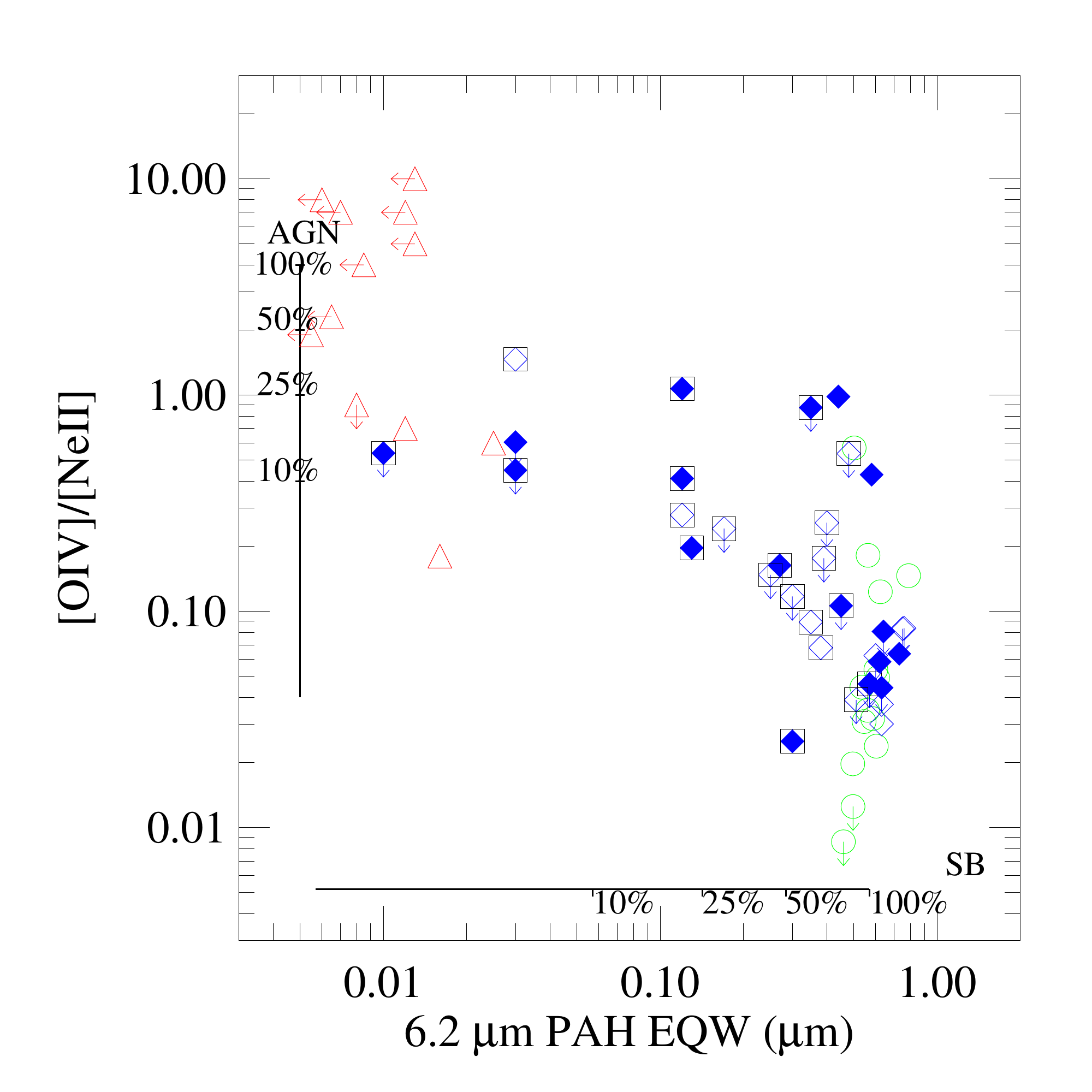}}
   \caption{[NeV]/[NeII] ($top$) and [OIV]/[NeII] ($bottom$) line ratios \citep{inami13} for the 46 objects in our sample (blue diamonds) versus the equivalent width (EQW) of the 6.2 $\rm \mu m$ PAH feature \citep[EQW values from][]{stierwalt13}; filled blue diamonds show objects identified as radio-AGN based on our radio diagnostic method, while black squares denote objects with mid-IR fraction above zero \citep{petric11}. For comparison we show a sample of AGN (red open triangles) and a sample of SB (including LIRGs; green open circles) taken from \cite{armus07}. The horizontal line gives the SB contribution, where the 100\% is the mean value of the 6.2 $\rm \mu$m PAH EQW. The vertical line shows the AGN contribution, where as 100\% we take the mean value of the [NeV]/[NeII] ratio. 
    }
              \label{nevpah}%
    \end{figure}

Following \citet{armus07} and \citet{petric11} we plot the [Ne V]/[Ne II] and [O IV]/[Ne II] line ratios \citep{inami13} to the 6.2 $\rm \mu m$ PAH equivalent width \citep[EQW;][]{stierwalt13} in Fig.~\ref{nevpah}. Our sample of LIRGs is shown in blue diamonds, where objects classified as radio-AGN based on our analysis are marked with filled blue diamonds, and objects with AGN fraction (the contribution of the AGN to the total bolometric luminosity) above zero \citep{petric11} are marked with black squares. For comparison, we overplot a sample of known AGN shown as open red triangles, and a sample of known SB shown as open green circles; the data\footnote{The AGNs used for comparison are I Zw 1, NGC 1275, Mrk 3, PG 0804+761, PG 1119+120, NGC 4151, PG 1211+143, 3C 273, Cen A, Mrk 279, PG 1351+640, Mrk 841, and PG 2130+099 \citep{weedman05}. The comparison starburst galaxies are NGC 660, NGC 1222, IC 342, NGC 1614, NGC 2146, NGC 3256, NGC 3310, NGC 4088, NGC 4385, NGC 4676, NGC 4818, NGC 7252, and NGC 7714 \citep{brandl06, bs09}.} are taken from \cite{armus07}. As it was discussed in detail in Section 3.3 of \cite{armus07}, average values from a set of "classical AGN" from
\cite{weedman05} as well as well known nearly "pure starbust" systems from \cite{brandl06} and \cite{bs09} were used to define the extreme high and low values of the [NeV]/[NeII] flux ratio and the 6.2 $\rm \mu m$ PAH EQW. Following that, it was assumed that the 100\% contribution of starburst is associated with an EQW  of $\sim$ 0.6 $\rm \mu m$ for the 6.2 $\rm \mu m$ PAH feature. This was decreasing linearly to the lowest possible measured value of 0.006 $\rm \mu m$ which marked the 0\% contribution and it was associated with the strongest AGN. Similar logic was used for the [NeV]/[NeII]  and [OIV]/[NeII] ratios presented in the Y-axes of Fig.~\ref{nevpah}. Following \cite{armus07}, the horizontal and vertical lines in  Fig.~\ref{nevpah} mark the fractional AGN and SB contributions to the [NeV]/[NeII] or [OIV]/[NeII] and 6.2 $\rm \mu m$ PAH EQW, respectively: the 100\% level is taken as the average value of the [Ne V]/[Ne II] or the [OI V]/[Ne II] ratio and 6.2 $\rm \mu m$ PAH EQW among the AGNs and SBs used for comparison, respectively. The 50\%, 25\% and 10\% levels are also shown.

None of our objects has a strong AGN contribution based on the [Ne V] detection, but there are some objects that have more than 10\% AGN contribution in their infrared luminosity. We further see that some objects (22 out of the 46), although there are not identified as AGN based on this diagnostic, are classified as radio-AGN based on the $\alpha$-maps (e.g. IC 1623A/B).

From the [Ne V] data, 8 LIRGs have a [NeV] detection and thus have an AGN contribution, 27 LIRGs have upper limits, and 7 don't have an entry in Table~\ref{table:agn}, since there is no high resolution spectrum available. It is known that the [OIV] 25.89$\mu$m emission line can be produced both by a weak AGN as well as from shocks near massive stars \citep[i.e.][]{lutz98,bs09}. However, we do include it for reasons of completeness.

Based on the EQW of the 6.2 $\mu \rm m$ PAH feature, 12 of our LIRGs are mid-IR AGN, i.e. have EQW less than 0.27 $\mu \rm m$. Cross-correlating the results from [Ne V] detection and 6.2 $\mu \rm m$ EQW we find 4 LIRGs that are identified as AGN based on those two mid-IR diagnostics (NGC 2623, UGC 05101, UGC 08696 and IC 5298), which are also radio-AGN.

The 6.2 $\mu \rm m$ EQW diagnostic gives us 12 AGN, 13 SB and 12 AGN/SB, where objects below 0.27 $\mu \rm m$ are AGN, objects above 0.54 $\mu \rm m$ are SB and objects with EQW in-between those values are AGN/SB. From the rest 9 objects without a EQW entry in Table~\ref{table:agn}, one system (ESO 550-IG025 N \& S) was not observed, and for the rest the archival SL staring mode of the observations was not centred on the galaxy nucleus \citep[IRAS F01417+1651, IRASF03359+1523, UGC 06436 SE, NGC 5256 NE \& C, IRASF17132+5313 NE \& SW][]{stierwalt13}. From the 13 objects that are classified as SB based on the 6.2 $\mu \rm m$ EQW, 2 have a [Ne V] detection (UGC 08387 and VV340a N; see notes on the objects in Appendix~\ref{sec:notes}), and thus are mid-IR AGN based on [NeV]. Furthermore, 6 of these 13 SB are classified as radio-AGN: UGC 02369 S, UGC 04881 SW, MCG+07-23-019, VV 705 S, NGC 6090 and Mrk 0331 N. There are also two objects in the sample, ESO550-IG025 S and IRAS F17132+5313 NE, that are identified as radio-AGN but do not have mid-IR data. In total, there are 8 objects out of the 46 that are classified as radio-AGN but are not identified as such based on the mid-IR AGN diagnostics presented here.

\begin{figure*}[!ht]  

    \resizebox{\hsize}{!}
            {\includegraphics{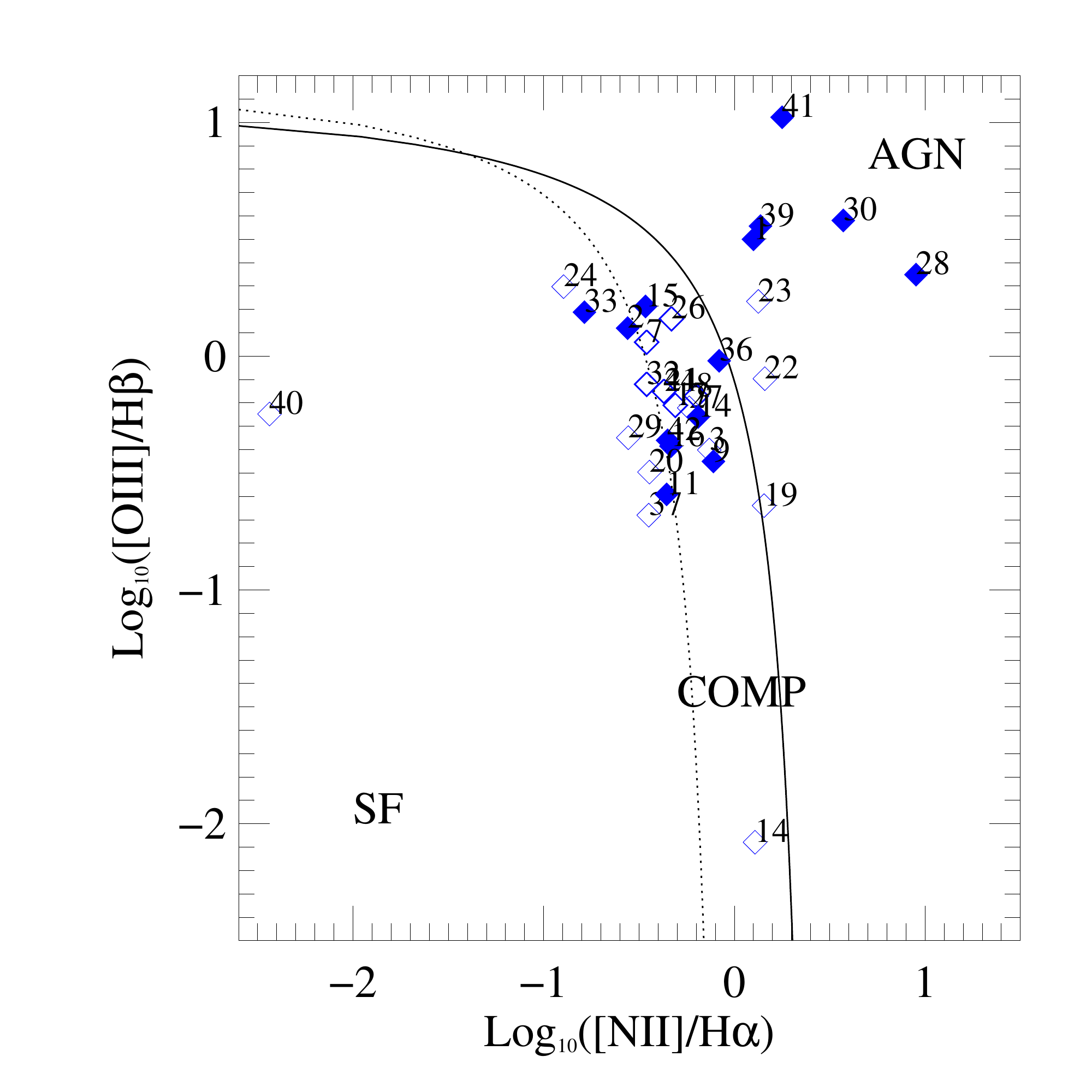}
            \includegraphics{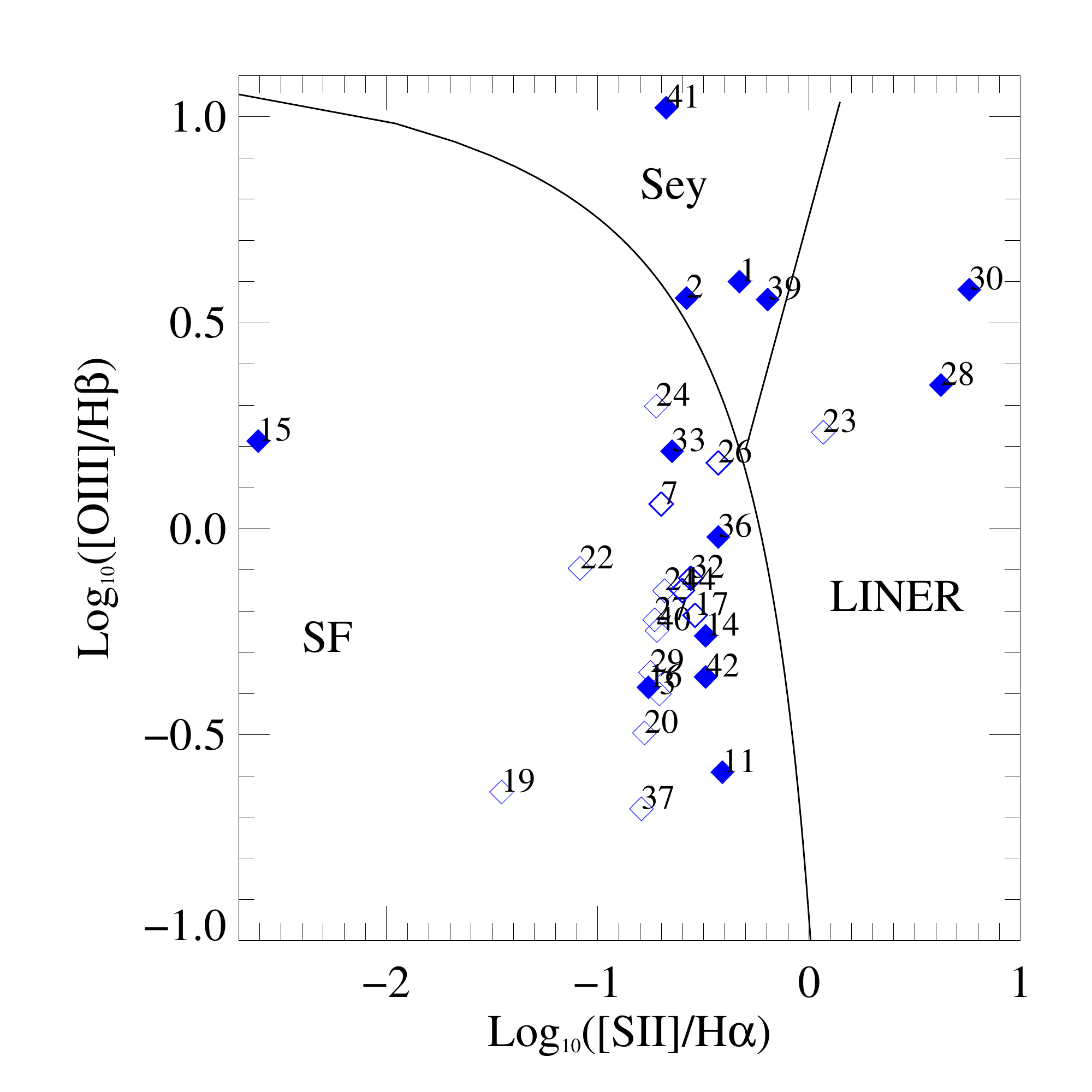}
             \includegraphics{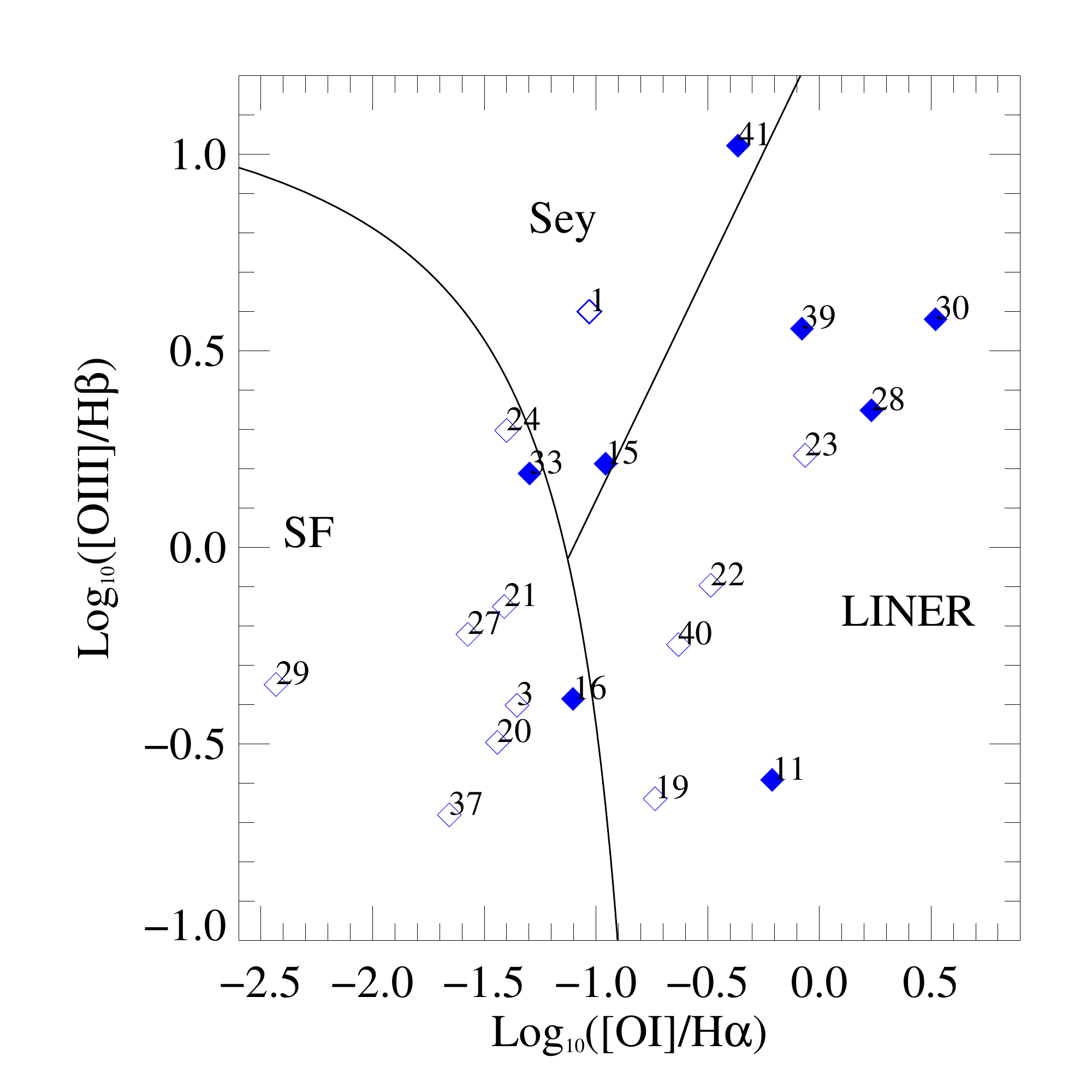}}
   \caption{BPT diagrams showing emission line ratios [OIII]/H$\beta$ versus [NII]/H$\alpha$ ($left$), [SII]/H$\alpha$ ($middle$) and [OI]/H$\alpha$ ($right$). In all figures, the blue diamonds represent the 46 individual objects in our sample, where objects identified as radio-AGN based on the $\alpha$-maps are shown as filled blue diamonds (see Table~\ref{table:agn}). Numbers next to the symbols represent the objects in our sample as given in Table~\ref{data}. On the $left$ graph, the solid and dotted lines separate the SF galaxies from the AGN and mark the area where composites (COMP) are located \citep{kauffmann03}.  The solid lines on the $middle$ and $right$ graphs are from \cite{kewley06} and separate the SF from Seyfert (Sey) and LINERs.   }
              \label{bpt}%
    \end{figure*}

We also obtained optical classifications for 44 out of the 46 objects in our sample (see Table~\ref{table:agn}) either from the literature as compiled in the NASA/IPAC Extragalactic Database (NED; 10 out of 46 objects), or from \cite{vega08} and \cite{mazzarella12} for 30 out of 46 objects, or by performing an analysis of our objects using the BPT diagram. For the latter we use the SDSS DR7 \citep{abazajian09} catalogue. After cross-correlating it to our radio RA and Dec, we obtained fluxes from optical spectroscopy for 22 out of 46 objects. We then used the standard BPT diagram \citep{kauffmann03}, [O III]/H$\beta$ versus [N II]/H$\alpha$ to separate star-forming (SF) objects, from composite (COMP) and AGN, and the [O III]/H$\beta$ versus [S II]/H$\alpha$ and [O III]/H$\beta$ versus [OI]/H$\alpha$ diagrams to separate Seyferts from LINERs \citep{kewley06}. We present these diagrams in Fig.~\ref{bpt}. From the 22 objects,  7 are SF objects, 7 COMP, 2 Seyfert and 4 LINER, while two  (IRASF01364-1042 and IRASF08572+3915) cannot be classified as the coverage of the optical lines is incomplete. We also use the optical line ratios from the IRAS Bright Galaxy Survey \citep{kim95}, and find optical classifications using the BPT diagram for 13 out of 46 objects. In Table~\ref{table:agn} we also report the classification given by \cite{veilleux95}. Note that we get multiple classifications for the same object depending on the search, literature or the BPT diagram. From the 8 radio-AGN that were not classified as AGN based on the published mid-IR diagnostics, only one is a Seyfert 2 (or HII region; Mrk 0331), while the rest are not classified as AGN based on the optical diagnostics presented here. 

To summarise the results from the AGN diagnostics: a) from the [Ne V] line we get 8 AGN while the rest are upper limits or no detections; b) from the 6.2 $\mu \rm m$ PAH we get 12 AGN, 13 SB and 12 composite objects; c) while from the optical diagnostics we get 12 Seyferts, 7 LINERs, 6 SF, 16 composites and 3 HII regions. We conclude that 3 objects out of the 46 in our sample ($\sim$ 6\%; UGC 02369 S, MCG+07-23-019, VV 705 S) were not identified as AGN based on the mid-IR and optical AGN diagnostics presented here, but are classified as radio-AGN based on our radio analysis. Also, there is no indication from the X-ray study of \cite{iwasawa11} that these objects contain an AGN; MCG +07-23-019 is not included in their study and there is not indication that it is an AGN based on diagnostics other than our radio analysis.

We need to mention the high fraction of radio-AGN in our sample ($\sim$ 45\%). From the 21 radio-AGN, only 10 out of 46 ($\sim$ 21\%) are mid-IR AGN based on both [NeV] detection and PAH EQW. The high radio-AGN fraction found in our study can be the result of the luminosity cut introduced by \cite{condon91}, where log$(L_{\rm IR} / L_{\odot}) \geq$ 11.25. This might introduce a bias towards excess radio emission. Furthermore, the high AGN fraction in the radio can be the result of the lower resolution of the PAH feature compared to our radio observations. As a result the observed PAH emission can suffer from blending from a circumnuclear SB and an AGN. In the case of enhanced star-formation in the circumnuclear region detecting the destruction of PAHs due to the X-ray hard-UV photos from an AGN is challenging since the PAH EQW from the surrounding regions will be large. Based on the 6.2 $\mu \rm m$ PAH EQW we have $\sim$ 26\% of objects containing an AGN, while based on the [Ne V] detection this drops to $\sim$ 17\%. The latter is in good agreement with the \cite{petric11} estimate that there is evidence for an AGN in the mid-infrared spectra in 18\% of the GOALS sample. Nevertheless, the results we present demonstrate how radio observations can be used towards unveiling AGN that are heavily obscured even at mid-IR wavelengths.

\subsection{Estimating the compact area using the flat radio-spectral-index values}

\label{sec:radiosize}

Based on \cite{condon91} and \cite{murphy13}, the value of the radio spectral index calculated between 1.49 and 8.44 GHz is connected to the compactness of the starburst, i.e. the flatter the radio spectral index the more compact the emitting area. If a flat radio spectral index is associated with compactness, then by calculating the area where the radio spectral index is flat we can estimate the size of the compact starburst. Here we calculate the flat area in kpc$^{2}$ from the $\alpha$-maps. This corresponds to the area where the radio spectral index is $\leq$ 0.5. The values are given in Table~\ref{table:radiosizes}, while errors were calculated by adding or subtracting the error per pixel value in the error $\alpha$-maps. Objects that have unresolved (COM) radio structures at both 1.48 and 8.44 GHz (see Table~\ref{data}) are placed in brackets. These are IRAS F01417+1651, IRAS F08572+3915, IRAS F10173+0828, IRAS F12112+0305 (NE), IRAS F12112+0305 (SW), UGC 08058, IRAS F14348-1447 (NE), IRAS F14348-1447 (SW), IRAS F15250+3608.  

We also want to explore how the size of the emitting area in the mid-IR compares to the size of the area which corresponds to a flat radio spectral index. In Table~\ref{table:radiosizes} we give the size of the area corresponding to flat radio spectral index, i.e. $\alpha \le$ 0.5, in kpc$^{2}$. The mid-IR sizes are calculated from the 13.2 $\rm \mu m$ continuum in the IRS spectrum of each object \citep{tanio10b}. The spatial profiles of each source along the slit were extracted and fitted with a Gaussian. The stellar profiles were also fitted with a Gaussian and scaled to the galaxy profile. Finally the stellar PSF was subtracted to give the intrinsic size of emission. In order to compare to the flat area, we assume a circular mid-IR area in kpc$^{2}$ (Table~\ref{table:radiosizes}). Ideally we want to compare the flat area determined above, with the mid-IR area as measured from mid-IR imaging, but these measurements are not available at the time of writing. Since the mid-IR and flat areas, as given in Table~\ref{table:radiosizes}, were not calculated in a similar manner, we cannot perform a one-to-one comparison.

In total, areas where the radio spectral index is flat range between $\sim$ 0.6-120 kpc$^{2}$, with a mean of 45$\pm$24 kpc$^{2}$. Excluding unresolved and barely resolved objects at 1.49 GHz the range is $\sim$ 5-100 kpc$^{2}$. The mid-IR areas for the sample presented here range between $\sim$ 0.6-45 kpc$^{2}$ with a mean of 9$\pm$11 kpc$^{2}$. In general the mid-IR areas are much smaller, but this may be partly attributed to the different methods of calculating the mid-IR and radio areas. Radio-area measurements are based on a 2D image, while mid-IR-area measurements are based on a 1D cut extrapolated to 2D assuming azimuthal symmetry \citep{tanio10b}; the latter would give roughly the same measurement for an object that is symmetric. Also, we have to note that most mid-IR areas are upper limits (see Table~\ref{table:radiosizes}). Thus we will refrain from drawing any strong conclusions from this analysis.

\begin{table}
\caption{Range of flat-alpha radio area and mid-IR area per radio class}             
\label{tab:ranges}      
\centering                          
\begin{tabular}{l l l }        
\hline\hline                 
\multicolumn{1}{c}{radio-class} & \multicolumn{1}{c}{radio area} & \multicolumn{1}{c}{mid-IR area}\\    
 & \multicolumn{1}{c}{(kpc$^{2}$)} & \multicolumn{1}{c}{(kpc$^{2}$)}\\  
\hline                        
\multicolumn{1}{c}{(1)} & \multicolumn{1}{c}{(2)}  & \multicolumn{1}{c}{(3)}  \\
\hline
AGN & 5-100 (44$\pm$26) & 0.6-44 (11$\pm$11)\\
SB    & 5-83 ~~(45$\pm$25)  & 0.6-45 (13$\pm$17) \\
AGN/SB & 16-65 (46$\pm$15) & 0.5-26 (5$\pm$7) \\
 \hline                                   
\end{tabular}
\tablefoot{Range of area in the radio where the radio spectral index is flat and of mid-IR area in kpc$^{2}$ grouped by radio class, excluding unresolved sources. Values in parenthesis give the mean and standard deviation.
}
\end{table}

In order to compare the objects in our sample we group them by radio classification, as shown in Table~\ref{tab:ranges}. These results suggest that the areas where the radio spectral index is flat are not statistically different for radio-AGN and radio-SB. This also applies to the mid-IR areas, as both radio-AGN and radio-SB display comparable ranges. Still, the radio area that corresponds to flat radio spectral index is larger than the mid-IR area when comparing objects grouped by their radio classification, as in Table~\ref{tab:ranges}. 

In the radio, the area that the radio spectral index is flat corresponds to the area over which the SB is very compact and the opacity is $>$ 1 at GHz frequencies \citep[e.g.][]{murphy13}, or in the case of an AGN the area where there is freshly ejected material. On the other hand, in the mid-IR we measure the area over which the SB or AGN is heating the dust to about 500 K. These results suggest that the sphere of influence of the AGN or the SB is larger in the radio than in the mid-IR.

\begin{figure}
           {\includegraphics[width=\hsize]{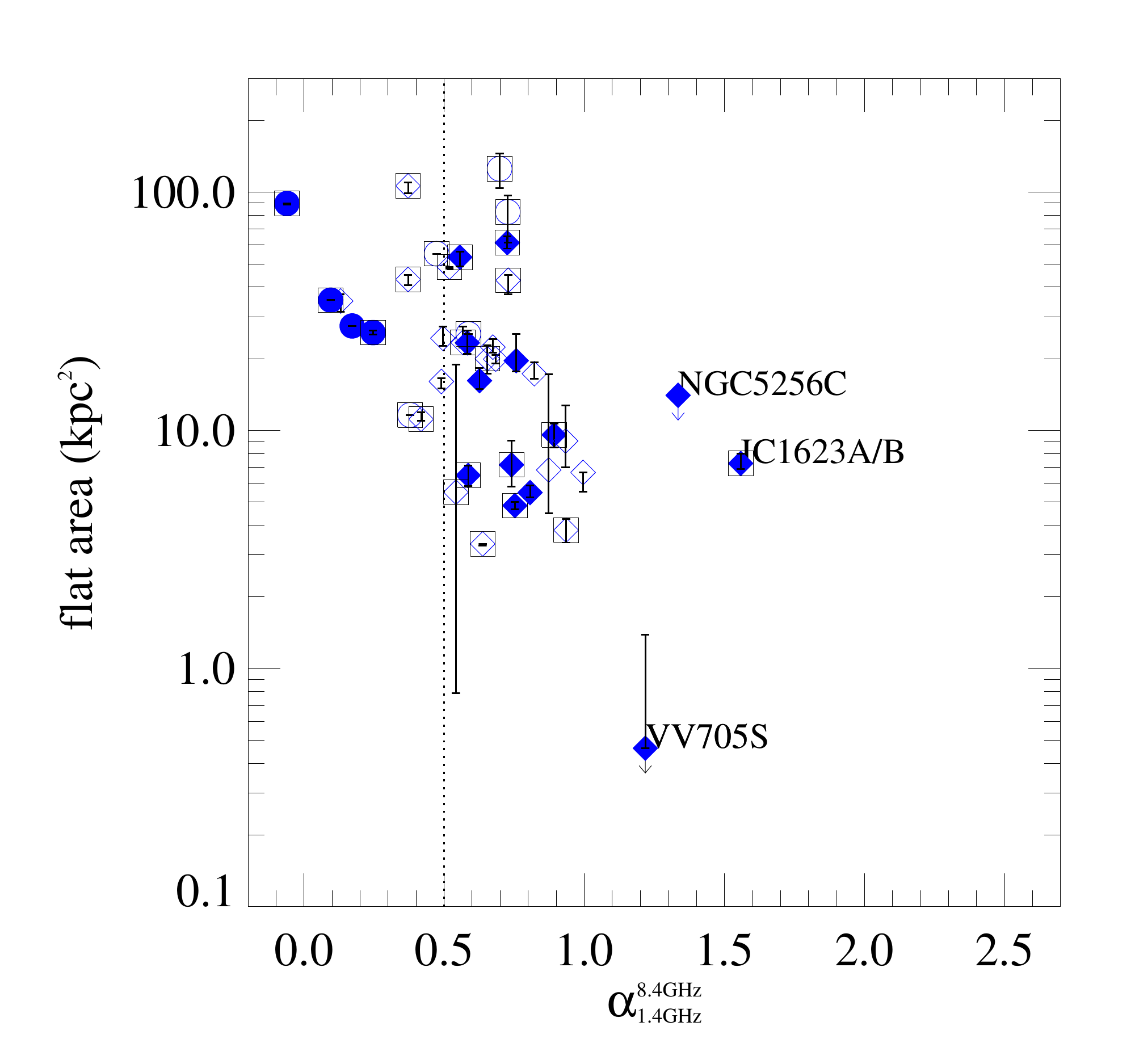}
            }   
            \resizebox{\hsize}{!}
           {\includegraphics[width=\hsize]{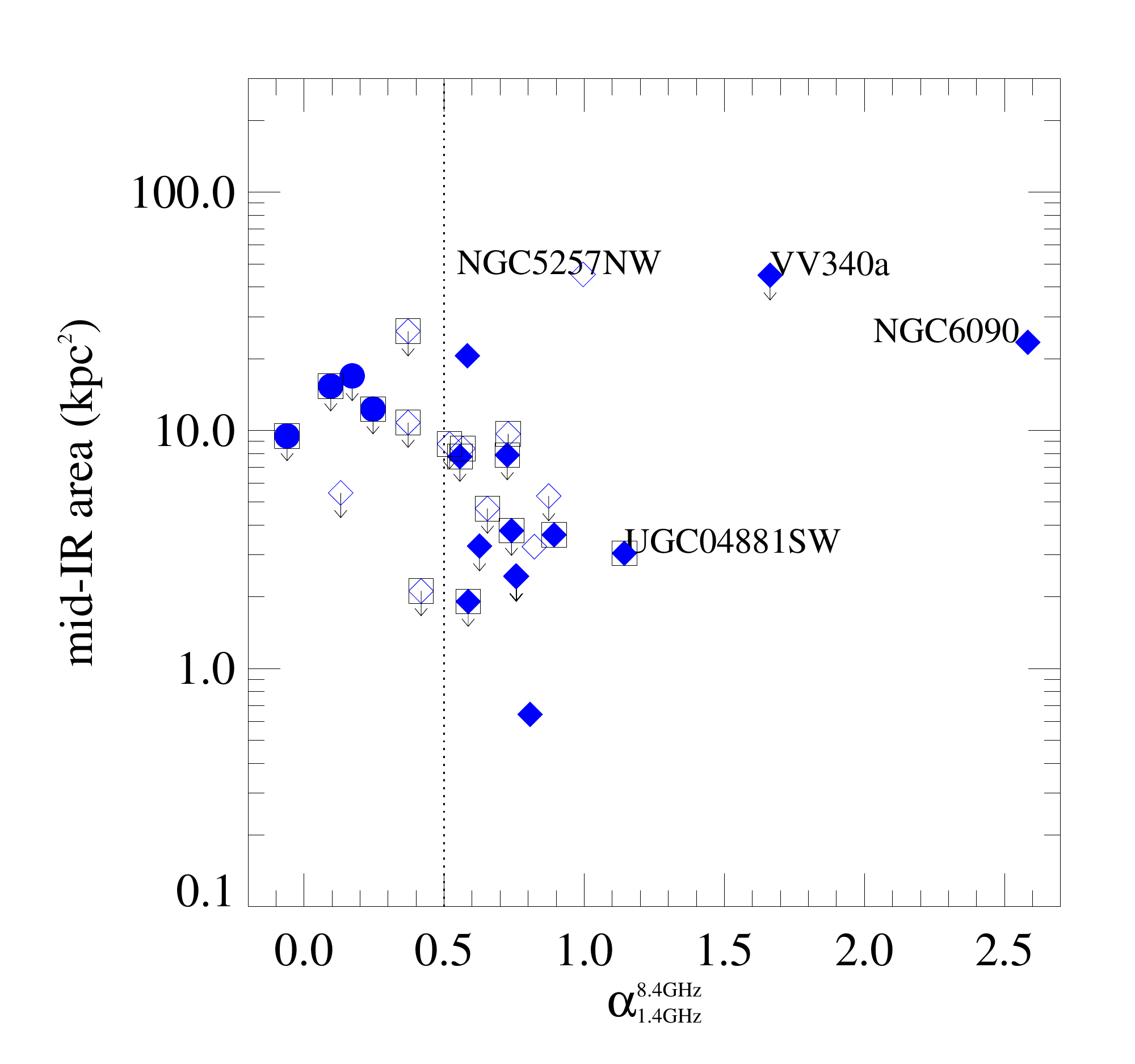}
    }
     
    \caption{($Top$) Size of area corresponding to flat radio spectral index (in kpc$^{2}$) versus integrated radio spectral index calculated between 1.49 \& 8.44 GHz. The dotted horizontal line marks the flat to steep division in radio spectral index. The Spearman $\rho$ = -0.255 with 10\% probability that the correlation is not present. ($Bottom$) Mid-IR area in kpc$^{2}$ of the core of mid-IR emission, assuming emission from a circular area (see Section~\ref{sec:radiosize}), versus integrated radio spectral index (see Table~\ref{data}). The Spearman $\rho$ = -0.189 with 21\% probability that the correlation is not present. In both plots, the dotted vertical line marks the flat to steep division in the radio spectral index. Symbols as in Fig.~\ref{qir}; unresolved objects are marked with circles instead of diamonds.
   }
            \label{ir-radio_size}%
 \end{figure}

In Fig.~\ref{ir-radio_size} we plot the flat area with the radio spectral index ($top$), and the mid-IR size versus the radio spectral index ($bottom$). We see a decrease of radio size with increasing radio spectral index meaning the steeper the radio spectral index, the smaller the size of the flat area, which is expected. The object with the smallest flat-$\alpha$ size is VV705 (S), and is an upper limit. We calculate the Spearman rank correlation coefficient using censoring for the upper limits with the available packages in IRAF. We find a negative correlation of -0.255 with 10\% probability that a correlation is not present. Similarly, we find a negative correlation in Fig.~\ref{ir-radio_size}-$bottom$, with $\rho$ = -0.189 and 21\% probability that a correlation is not present. We see that the mid-IR area decreases as the integrated radio spectral index steepens. There are three objects that do not follow this trend (NGC 5257 NW, VV 340a N and NGC 6090), as they have large mid-IR size and very steep radio spectral indices. NGC 6090 is very diffuse at 8.44 GHz with low flux density, giving a high radio spectral index between 1.49 and 8.44 GHz. Our results are limited due to the small number statistics and the large amount of upper limits in the mid-IR areas. We present this figure as a general comparison to the flat areas, but we will refrain from making a general statement about it, also because of the way the mid-IR areas were calculated.

\subsection{IR8 parameter for our sample} 

Using our sample we explore the ratio of the total infrared luminosity $L_{\rm IR}$ to the rest-frame luminosity at 8 microns $L_{8 \rm \mu m}$ (IR8), which is a star-formation compactness indicator \citep{daddi10a, elbaz11}. Star-forming galaxies follow the infrared main sequence, defined by IR8 $\approx$ 4.9 with a scatter of $\sigma \approx$ 2.5 dex \citep{elbaz11, magdis13}. Galaxies with the most intense compact starbursts are located above the main sequence \cite{elbaz11}, while objects below the main sequence are likely AGN-dominated. We find that three of the objects in our sample fall inside the main sequence (NGC 5257 NW, VV 340a N and IRAS F08572+3915), while the rest have IR8 values above the main sequence for star-forming galaxies, i.e. IR8 = 7.4, placing them in the compact starburst group. Yet, part of the 8 $\mu \rm m$ luminosity can be the result of a dust-obscured AGN, hidden by the SB. To reveal a possible contribution from a dusty AGN we correct for starburstiness R$_{\rm SB}$ \citep{elbaz11}, which is the ratio of the specific star-formation-rate (sSFR) to the specific star-formation-rate of a galaxy of the main sequence at the same redshift as our object (sSFR$_{\rm MS}$). We calculate the sSFR using the relation sSFR [yr$^{-1}$]= $L_{\rm IR}$ / $M^{\star} \times$ 1.72 $\times 10^{-10}$ \citep{kennicutt98}, where  $L_{\rm IR}$ the infrared luminosity as in Table~\ref{data} and $M^{\star}$ the stellar mass \citep{howell10, tanio13}, and the sSFR$_{\rm MS}$ [Gyr $^{-1}$] = 26 $\times~ t^{-2.2}$ \citep{elbaz11}, where $t$ is the cosmic time elapsed since the Big Bang in Gyr calculated using at the redshift of each object.

We correct the IR8 values of the object in our sample for starburstiness and find that 2 objects fall below 1 $\sigma$ from the mean, i.e. IR8/R$_{\rm SB}$ = 2.2: NGC5257 NW is a radio-SB and VV340a N is a radio-AGN. This correction suggests these objects may have a hidden by dust AGN, that has not been revealed by infrared or radio observations. From this we see that even our radio diagnostic can be challenging in cases of objects that are very dust-obscured, or in case of resolution limitations. Finally, we compare the corrected and uncorrected for starburstiness IR8 values to the integrated radio spectral index and the luminosity at 1.49 GHz, but find no correlation.

\subsection{A closer look at the dispersion in the values of $\alpha$}
\label{sec:adisp}

\begin{figure}
    \resizebox{\hsize}{!}
            {\includegraphics{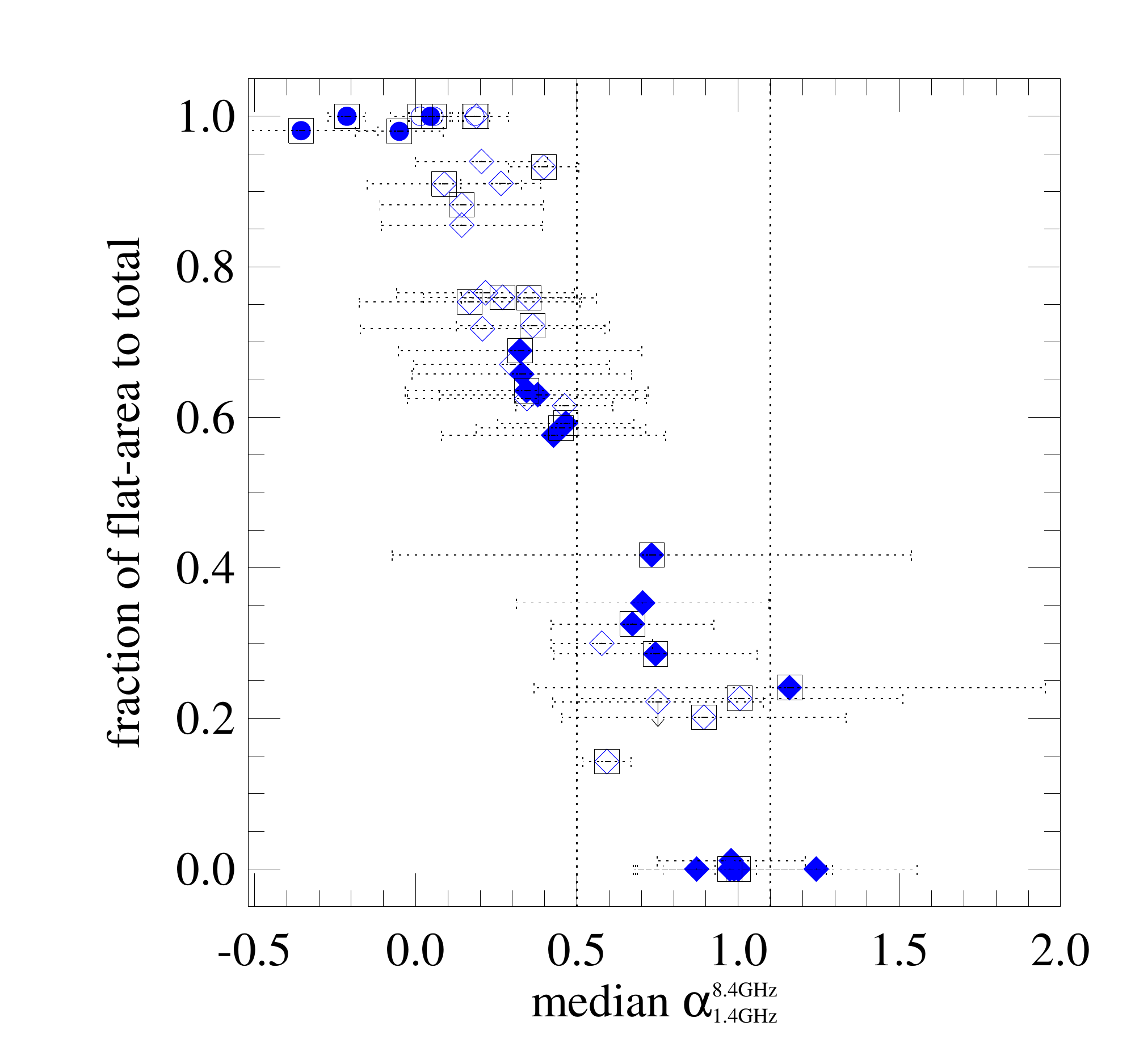}
            }

     \caption{
     The ratio of size of the area corresponding to flat radio spectral index (in kpc$^{2}$) as in Table~\ref{table:radiosizes}, to the total radio size as in the $\alpha$-maps versus median radio spectral index calculated from the $\alpha$-maps. The dispersion (standard deviation) around the median is shown with horizontal dotter error-bars. Symbols: blue open diamonds denote the objects of our sample, blue filled diamonds denote objects that are classified as radio-AGN based on the $\alpha$-maps, and black squares denote objects that have an AGN contribution in their mid-IR luminosity \citep{petric11}; unresolved objects are marked with circles instead of diamonds. The vertical dotted lines at 0.5 and 1.1 mark the flat-steep division and the limit above where we don't expect steepening from compact starburst, respectively.
   }
              \label{adispersion}%
    \end{figure}

Here we present the ratio of the flat area to the total area ($f_{f}$) as given in the $\alpha$-maps, with the resolution of the 1.49 GHz maps, for fluxes above 3 $\sigma$ (Fig.~\ref{adispersion}). We see a decrease in the fraction $f_{f}$ as the radio spectral index steepens. This is expected as the area where the radio spectral index is flat gets smaller as the median radio spectral index in the $\alpha$-map gets steeper (larger values). Unresolved sources that display low median values of $\alpha$ have large fractions of $f_{f}$  where we believe the AGN dominates for objects where median $\alpha <$ 0 (see Sec.~\ref{sec:radioAGNclass} for an explanation of an inverted radio spectral index). The vertical lines mark the flat-to-steep division in $\alpha$ and the limit above where we do not expect to see a steepening of the radio spectral index in compact starbursts ($\alpha \sim$ 1.1; see Sec.~\ref{sec:radioAGNclass}). 

Examining in detail the $\alpha$-histograms we observe that in several objects, the radio spectral index reaches values above  $\alpha$ = 1.1, and we will refer to this extension in the histograms as long steep tails. These long steep tails extend further than the dispersion of the median $\alpha$ shown by the error bars in Fig.~\ref{adispersion}. Thus examining the histogram by eye and not depending solely on the results of Fig.~\ref{adispersion} is important for the interpretation of the data. We attribute these long steep tails to the existence of an AGN, as discussed in Sec.~\ref{sec:radioAGNclass}.

We also compare the median values of $\alpha$ to the flat area and the brightness temperature Tb, but we do not find an obvious correlation. Just a trend for higher brightness temperatures and flat areas to correspond to flatter mean $\alpha$. 

\section{Conclusions}
\label{sec:conc}

We present results from the study of radio continuum observations of 35 LIRG systems, or 46 individual galaxies, from the GOALS sample, observed at both 1.49 \& 8.44 GHz with the VLA \citep{condon90, condon91}. Our objective is to identify the AGN component in these systems using radio observations that are largely unaffected by dust obscuration. We use the highest resolution maps available at the time of the study, down to 1$\farcs$5 resolution at 1.49 GHz and 0$\farcs$25 resolution at 8.44 GHz, in order to probe the central kpc of these LIRGs. These objects are selected by \cite{condon91} to be brighter in the infrared than log$(L_{\rm IR} / L_{\odot}) \geq$ 11.25, a luminosity cut that might be preferential towards selecting AGN. 

We find that distinguishing between AGN and SB is not straight-forward, even in the radio that is a dust free probe of emission. The main reason for that is the resolution of our observations. Although they allow us to probe the central kpc of these dusty LIRGs and observe through the dust, even through the starburst core itself \citep[see][]{murphy13}, still the variations in the values of the radio spectral index can be the result of either the AGN or the SB, or a combination of both. \cite{murphy13} claim that the dust obscuration in these sources occurs in the vicinity of the compact starburst itself, and not by extended dust located in the foreground galaxy disks. Since in LIRG nuclei SBs and AGN can co-exist, and since both give similar signatures in the radio, observations with higher resolution are needed in order to separate the contribution of each component in greater accuracy than provided in this study. 

We further investigate whether objects with low $q_{\rm IR}$ values, i.e. brighter at radio than at IR wavelengths, contain an AGN based on the radio-spectral-index maps and compare our results to optical and the following mid-IR AGN diagnostic tools: (a) the [Ne V] emission line and (b) the equivalent width of the 6.2 $\rm \mu$m PAH feature. Finally, we give the size of the area corresponding to flat radio spectral index, which is connected to compactness, and thus gives the size of the compact starburst at low-GHz radio wavelengths. 

In summary, our results are:
   \begin{enumerate}
      \item Out of the 46 individual objects, we find 21 ($\sim$ 45\%) radio-AGN based on the radio analysis, 9 ($\sim$ 20\%) radio-SB and 16 ($\sim$ 35\%) AGN/SB.
      \item 8 out of the 46 ($\sim$ 17\%) of the objects in our sample are radio-AGN, but were not identified as AGN in the mid-IR. 
      \item 3 out of the 46 ($\sim$ 6\%) of the objects in our sample are radio-AGN, but were not identified as AGN at both optical and mid-IR wavelengths.
      \item We estimate the size of the area in the radio of a compact starburst having a mean of 45 kpc$^{2}$.
       \end{enumerate}

In order to separate the AGN and SB components more accurately than presented here, one needs higher-resolution observations, e.g. with the EVLA, in order to resolve in greater detail the radio structures associated with the centre of these LIRGs.

\begin{acknowledgements}
      We would like to thank the anonymous referee whose careful reading and detailed comments greatly improved the manuscript. EV is funded by the Action ''Supporting Postdoctoral Researchers'' of the Operational Program ''Education and Lifelong Learning'' (Action's Beneficiary: General Secretariat for Research and Technology), and is co-financed by the European Social Fund (ESF) and the Greek State. EV would like to acknowledge Hanae Inami and Jim Condon for most useful conversations, and Grigoris Maravelias, Thomas Robitaille and Laure Ciesla for their most helpful assistance in Python. VC would like to acknowledge partial support from the EU FP7 Grant PIRSES-GA-2012-316788.
\end{acknowledgements}

\bibliographystyle{aa} 


%
\begin{table*}
\caption{Comparison of radio and mid-IR AGN diagnostics}             
\label{table:agn}      
\centering          
\begin{tabular}{l l r r l l}     
\hline\hline       
                
\multicolumn{1}{c}{Name} & \multicolumn{1}{c}{radio} & 
\multicolumn{1}{c}{[Ne V]/} & \multicolumn{1}{c}{[O IV]/} & \multicolumn{1}{c}{6.2 $\rm \mu m$ PAH EQW ($\rm \mu m$)} & 
\multicolumn{1}{c}{optical} \\ 
           
 & \multicolumn{1}{c}{classification} & 
\multicolumn{1}{c}{[Ne II]}&\multicolumn{1}{c}{[Ne II]}  & \multicolumn{1}{c}{(and classification)} & \multicolumn{1}{c}{classification} \\
 \hline
NGC 0034 (S)& AGN (6)&  <  0.03 &  < 0.11 &   0.45 (AGN/SB)& Sey2$^{\rm V08}$ / Sey$^{\rm BPT (r)}$ / Sey2$^{\rm VX95}$\\
IC 1623A/B & AGN (6)&  <  0.01 &    0.03 &   0.30 (AGN/SB) & SF$^{\rm BPT (r)}$ / HII$^{\rm VX95}$  \\
CGCG 436-030 & AGN/SB (1) &  <  0.02 &    0.09 &   0.35 (AGN/SB) & COMP$^{\rm BPT (s)}$ / LINER, HII$^{\rm V08}$\\
IRAS F01364-1042 & AGN/SB (1)&  <  0.04 &  <  0.18 &   0.39 (AGN/SB) &  LINER$^{\rm V08, NED}$\\
IRAS F01417+1651 & SB (2)&  --- &  --- & ---  & LINER, HII$^{\rm V08}$ \\
UGC 02369 (S) & AGN (5)&  < 0.02 &  <  0.05 &   0.57 (SB) & HII$^{\rm V08~(t)}$ \\
IRAS F03359+1523 & AGN/SB (1)&  <  0.01 &  <  0.07 & --- & COMP$^{\rm BPT (r)}$ / HII$^{\rm VX95}$\\
ESO 550-IG025 (N) & AGN/SB (1)& --- & --- & --- & COMP$^{\rm BPT (r)}$ / LINER$^{\rm VX95}$ \\
ESO 550-IG025 (S) & AGN (3)& --- & --- & --- & COMP$^{\rm BPT (r)}$ / LINER$^{\rm VX95}$ \\
IRAS F05189-2524 & SB (2) &    0.71 &    1.46 &   0.03 (AGN) & Sey1$^{\rm NED}$ / Sey2$^{\rm V08}$\\
NGC 2623 & AGN (1)&    0.06 &    0.16 &   0.27 (AGN) & COMP$^{\rm BPT (s)}$ / LINER, Sey2$^{\rm V08}$\\
IRAS F08572+3915 & AGN (4)&  <  0.23 &  <  0.61 &   0.03 (AGN) &  LINER, Sey2$^{\rm V08}$\\
UGC 04881 (NE) & AGN/SB (1)&  <  0.02 &  <  0.26 &   0.40 (AGN/SB) & COMP$^{\rm BPT (s)}$ / HII$^{\rm VX95}$ / HII$^{\rm V08(t)}$\\
UGC 04881 (SW) & AGN (3)& --- & --- & 0.61 (SB) & COMP$^{\rm BPT (r)}$ / HII$^{\rm VX95}$ / HII$^{\rm V08~(t)}$ \\
UGC 05101 & AGN (1)&    0.08 &    0.20 &   0.13 (AGN) & SF$^{\rm BPT (s)}$ / Sey1$^{\rm NED}$ / LINER, Sey1.5$^{\rm V08}$\\
IRAS F10173+0828 & AGN (4)&  <  0.05 &  <  0.88 &   0.35 (AGN/SB) & COMP$^{\rm BPT (s)}$\\
IRAS F10565+2448 & SB (2)&  <  0.01 &  <  0.04 &   0.51 (AGN/SB) & COMP$^{\rm BPT (r)}$ / HII$^{\rm V08}$\\
MCG +07-23-019 & AGN (1)&  <  0.01 &  <  0.08 &   0.64 (SB) & HII$^{\rm V08}$ \\
UGC 06436 (NW) & AGN/SB (1)&  <  0.01 & <  0.06 & 0.60 (SB) & COMP$^{\rm BPT (s)}$ / LINER, HII$^{\rm V08(t)}$ / SB$^{\rm NED(t)}$\\
UGC 06436 (SE) & SB (1)& --- & --- & --- & SF$^{\rm BPT (s)}$ / LINER,HII$^{\rm V08(t)}$ / SB$^{\rm NED(t)}$\\
NGC 3690 (E) & AGN/SB (5)&  <  0.02 &    0.07 &   0.38 (AGN/SB) & COMP$^{\rm BPT (s)}$\\
NGC 3690 (W) & AGN/SB (5)&  <  0.04 &    0.28 &   0.12 (AGN) & LINER$^{\rm BPT (s)}$ \\
IRAS F12112+0305 (NE) & SB (2)& --- &  <  0.12 &   0.30 (AGN/SB) & LINER$^{\rm BPT (s)}$ / HII$^{\rm V08}$\\
IRAS F12112+0305 (SW) & AGN/SB (4)& <  0.05 & --- & 0.30 (AGN/SB) & SF$^{\rm BPT (s)}$ / HII$^{\rm V08}$\\
UGC 08058 & AGN (1)&  <  0.17 &  <  0.54 &   0.01 (AGN) & Sey1$^{\rm V08, NED}$\\
VV 250a (NW) & AGN/SB (1)&  <  0.01 &    0.03 &   0.63 (SB) & COMP$^{\rm BPT (r)}$ / HII$^{\rm VX95}$\\
VV 250a (SE) & SB (2)&  <  0.08 &  <  0.08 &   0.76 (SB) & COMP$^{\rm BPT (s)}$\\
UGC 08387 & AGN (1)&    0.01 &    0.06 &   0.62 (SB) & LINER$^{\rm BPT (s)}$ / LINER, HII$^{\rm V08}$\\
NGC 5256 (NE) & SB (1)& --- & --- & --- & SF$^{\rm BPT (s)}$ / LINER$^{\rm M12}$\\
NGC 5256 (SW) & AGN (1)& 0.15 & --- & 0.44 (AGN/SB) & LINER$^{\rm BPT (s)}$ / Sey2$^{\rm M12, NED}$\\
NGC 5256 (C) & AGN/SB (1)& --- & 0.98 & --- & ---\\
NGC 5257 (NW) & SB (2)&  <  0.07 &    0.04 &   0.63 (SB) & SF$^{\rm BPT (r)}$ / HII$^{\rm VX95}$ \\
UGC 08696 & AGN (1)&    0.23 &    1.07 &   0.12 (AGN) & SF$^{\rm BPT (s)}$ / Sey2$^{\rm V08, NED}$ / LINER$^{\rm V08}$\\
IRAS F14348-1447 (NE) & AGN/SB (1)& --- &  <  0.15 &   0.25 (AGN) & LINER$^{\rm V08}$ \\
IRAS F14348-1447 (SW)& AGN/SB (1)& <  0.03 & --- & 0.25 (AGN) & LINER$^{\rm V08}$ \\
VV 340a (N) & AGN (3) &    0.05 &    0.43 &   0.58 (SB) & COMP$^{\rm BPT (r)}$ / LINER$^{\rm VX95}$ \\
VV 705 (N) & AGN/SB (1)&  <  0.11 &  <  0.08 &   0.75 (SB) &SF$^{\rm BPT (s)}$ / HII$^{\rm V08~(t)}$\\
VV 705 (S) & AGN (3)& --- & --- & 0.75 (SB) & HII$^{\rm V08~(t)}$ \\
IRAS F15250+3608 & AGN (4)&  <  0.10 &  <  0.45 &   0.03 (AGN) & Sey, LINER$^{\rm BPT (s)}$ / LINER$^{\rm V08}$\\
UGC 09913 & AGN/SB (1)&  <  0.05 &  <  0.24 &   0.17 (AGN) & prob. Sey$^{\rm NED}$ / SF$^{\rm BPT (s)}$ / HII, LINER$^{\rm V08}$\\
NGC 6090 & AGN (3)&  <  0.02 &    0.06 &   0.73 (SB) & Sey$^{\rm BPT (s)}$\\
IRAS F17132+5313 (NE) & AGN (1)&  <  0.01 &  <  0.04 & --- & COMP$^{\rm BPT (r)}$ / HII$^{\rm VX95}$  \\
IRAS F17132+5313 (SW) & SB (2)& --- & --- & --- &---  \\
IRAS F22491-1808 & AGN /SB(1)&  <  0.13 &  <  0.54 &   0.48 (AGN/SB) & COMP$^{\rm BPT (r)}$ / HII$^{\rm VX95}$\  \\
IC 5298 & AGN (1)&    0.35 &    0.41 &   0.12 (AGN) & Sey2$^{\rm V08, NED}$ / LINER, HII$^{\rm V08}$\\
Mrk 0331 (N)& AGN (1)&  <  0.01 &  <  0.04 &   0.63 (SB) & HII, Sey2$^{\rm V08}$ \\
\hline                  
\end{tabular}
\tablefoot{ Columns: (1) Name of the galaxy; systems are marked with the relative position to each other (e.g.\ N, S, E, W). (2) Classification based on radio spectral index maps and histograms. In parenthesis we give the histogram category. (3) Ratio of [Ne V] to the [Ne II] emission line: if the [Ne V] line is detected then the galaxy contains an AGN; if detection is an upper limit then it is a possible AGN; and if there is no high-resolution spectrum for the object, then we mark with '---'. (4) Ratio of the [O IV] to the [Ne II] emission line; if there is no high-resolution spectrum for the object, then we mark with '---'. (5) EQW in microns of the 6.2 $\rm \mu m$ PAH feature. AGN or SB classification are based on empirical values as follows: values below 0.27 $\rm \mu m$ suggest the presence of an AGN; values above 0.54 $\rm \mu m$ denote a SB; and values in-between suggest the presence of both AGN and SB \cite[see][]{stierwalt13, murphy13}. (6) Optical classification based on optical spectroscopy; reference is given in each case. For classification based on the BPT diagram (s stands for SLOAN data and r for RBGS data; \citep{kim95}), SF stands for object in the star-forming region, COMP stands for objects between star-forming and AGN regions in the BPT \cite[e.g.][]{kewley01, kauffmann03, kewley06}; Sey stands for Seyfert with subcategories 1, 1.5 and 2. References: 'NED' for objects drawn from the NASA/IPAC Extragalactic Database (http://ned.ipac.caltech.edu); 'BPT' from the BPT diagram; 'M12' from \cite{mazzarella12}; 'V08' from \cite{vega08}; 'VX95' from \cite{veilleux95}; (t) stands for reference for the galaxy system.
}
\end{table*}
%

%
\begin{table*}
\caption{Size of area corresponding to flat radio spectral index, it's fraction to total, and mid-IR area}             
\label{table:radiosizes}      
\centering          
\begin{tabular}{l l c r }    
\hline\hline       
                
\multicolumn{1}{c}{Name} & \multicolumn{1}{c}{flat area} &  fraction of flat area &
\multicolumn{1}{c}{mid-IR} \\
 && to total area&\multicolumn{1}{c}{area} \\
\hline
& \multicolumn{1}{c}{(kpc$^{2}$)} & in $\alpha$-map&\multicolumn{1}{c}{(kpc$^{2}$)}  \\
 \hline
NGC 0034 (S)&  4.84$^{+0.17}_{-0.17}$&0.4 &--- \\
IC 1623A/B &  7.27$^{+0.73}_{-0.37}$& 0.2&--- \\\ 
CGCG 436-030 & 20.01$^{+2.71}_{-2.71}$ & 0.72&  (4.7) \\
IRAS F01364-1042 & 43.04$^{+1.90}_{-2.38}$ & 0.91& (10.7) \\ 
IRAS F01417+1651 & [11.61] & 1.00&--- \\
UGC 02369 (S)& 9.58$^{+1.09}_{-1.09}$ & 0.28& 3.6 \\
IRAS F03359+1523 & 16.05$^{+0.53}_{-1.06}$ & 0.7&--- \\
ESO 550-IG025 (N) & 24.38$^{+2.87}_{-1.77}$ & 0.7&--- \\ 
ESO 550-IG025 (S)&--- & 0.00 & --- \\
IRAS F05189-2524 & 48.53$^{+0.00}_{-0.81}$& 1.00 &  (8.7) \\
NGC 2623 &  6.47$^{+0.62}_{-0.62}$ & 0.68&  (1.9) \\
IRAS F08572+3915 & [27.45] &1.00 & (16.9) \\ 
UGC 04881 (NE) & 23.42$^{+3.92}_{-1.18}$&0.75 &  (8.3) \\
UGC 04881 (SW) &---&0.00 &  3.05 \\
UGC 05101 & 53.39$^{+2.87}_{-4.60}$ & 0.59&  (7.7) \\
IRAS F10173+0828 & [25.74]$^{+0.53}_{-0.53}$&0.98 & (12.2) \\
IRAS F10565+2448 & 42.68$^{+2.20}_{-5.29}$ & 0.93&  (9.6) \\ 
MCG +07-23-019 & 23.35$^{+2.96}_{-2.31}$ & 0.65& 20.5 \\
UGC 06436 (NW)& 17.30$^{+1.99}_{-0.85}$ & 0.67&  3.2 \\ 
UGC 06436 (SE)&  9.03$^{+3.72}_{-2.01}$ & 0.61&--- \\ 
NGC 3690 (E)&  3.34$_{-0.05}$ & 0.22&--- \\
NGC 3690 (W)&  3.81$^{+0.43}_{-0.43}$& 0.20 &--- \\
IRAS F12112+0305 (NE)& [55.21] & 1.00&--- \\ 
IRAS F12112+0305 (SW)& [25.40] & 1.00&--- \\ 
UGC 08058 & [89.72]$_{-1.17}$ &  0.98&(9.48) \\ 
VV 250a (NW)&  6.82$^{+10.37}_{-2.33}$ & 0.30&  (5.3) \\
VV 250a (SE)& 34.94$^{+2.28}_{-3.42}$ & 0.91&  (5.4) \\ 
UGC 08387 & 16.18$^{+2.06}_{-1.27}$ & 0.57&  (3.2) \\ 
NGC 5256 (NE)& 19.91$^{+0.84}_{-0.84}$ & 0.85&--- \\ 
NGC 5256 (SW)& 19.62$^{+5.76}_{-1.92}$ & 0.63&  (2.4) \\
NGC 5256 (C)& (27.59)& 0.22 &  --- \\
NGC 5257 (NW)&  6.66$_{-1.14}$ & 0.94& 45.1 \\
UGC 08696 & 61.38$^{+3.81}_{-3.33}$ & 0.63&  (7.8) \\
IRAS F14348-1447 (NE) & [82.69]$^{+14.26}_{-21.38}$ & 0.58&--- \\
IRAS F14348-1447 (SW) & [125.46]$^{+19.96}_{-21.38}$ & 0.75&--- \\
VV 340a (N)&--- & 0.00& 44.84 \\
VV 705 (N)& 22.37$^{+1.86}_{-1.12}$ & 0.62&--- \\
VV 705 (S)&  (0.46)$^{+0.93}$&0.01 &--- \\
IRAS F15250+3608 & [35.24] & 1.00& (15.3) \\
UGC 09913 & 11.15$^{+0.77}_{-0.19}$&0.75 &  (2.1) \\
NGC 6090 &---&0.00 & 23.43 \\
IRAS F17132+5313 (NE)&  ---& 0.00 &--- \\
IRAS F17132+5313 (SW)&  5.50$^{+13.37}_{-4.72}$&0.14 &--- \\
IRAS F22491-1808 &106.31$^{+3.57}_{-7.15}$ & 0.88& (26.1) \\ 
IC 5298 &  7.17 $^{+1.89}_{-1.38}$ & 0.32&  (3.7) \\
Mrk 0331 (N)&  5.48$^{+0.39}_{-0.24}$ & 0.35&  0.6 \\
\hline                  
\end{tabular}
\tablefoot{ Columns: (1) Name of the galaxy; systems are marked with the relative position to each other (e.g.\ N, S, E, W). (2) Square kpc of the area corresponding to flat radio spectral index ($\alpha \leq$ 0.5) in the $\alpha$-maps. The errors were calculated by taking into account the error per pixel as shown in the error $\alpha$-maps. Values in parenthesis are upper limits, while square brackets denote unresolved sources. (3) Fraction of the flat area to the total area marked by the $\alpha$-map (Fig.~\ref{amaps}). (4) Size of area of mid-IR emission at 13.2 $\rm \mu m$ in kpc$^{2}$; the calculation method is described in \cite{tanio10b}; values in parenthesis denote upper limits, meaning the measured size is $<$ 10\% larger than the instrumental size (PSF) so an intrinsic size for the emitting region couldn't be calculated. The reported upper limits are on the measured size (not intrinsic), i.e. 1.1$\times$PSF size. For objects marked with '---' there is no available measurement. Note: IRAS F12112+0305 and IRAS F14348-1447 are double radio sources but their radio components are not completely separated. IRAS F17132+5313 1has two radio components, separated, but the infrared source is barely separated. NGC 5256 (SW) and (C) are barely separated at 1.49 GHz (see $\alpha$-map). Here we give the flat area of the SW radio component which is associated with the most luminous infrared source. The total flat area of NGC 5256 (SW) and (C) is 47.19$^{+9.52}_{-9.18}$ kpc$^{2}$. We put a limit on NGC5256 (C) because the area is partly overlapping with the southernmost component.
}

\end{table*}

\appendix

\section{Notes on the objects}
\label{sec:notes}

\noindent
{\bf NGC 0034} This is a galaxy pair, but only the south companion has been observed at both frequencies used in our analysis. The centre of radio emission at 1.49 GHz does not coincide with the HST bulge position \citep{haan11}, but is $\sim$ 0.8 arcsec away. The centre of the 8.44-GHz emission is matching the HST bulge position. The maximum flux pixel in the 1.49-GHz map is not at the same R.A. and Dec and the maximum flux pixel in the 8.44-GHz map. This offset is of the order of $\sim$ 1 arcsec, which is larger than the pixel size of the 1.49-GHz image (0.4 arcsec). We believe that the shift in the core emission is real and can happen for example when a jet component is blended with the core in a lower resolution observation \citep{paragi00}. It has been reported \citep{lobanov98}, that this shift is dependent on observing frequency due to a change in synchrotron self-absorption opacity \citep[see][]{paragi00}, and the observed brightness peak gets closer to the emission position as we go to higher frequencies. This shift is imprinted in the $\alpha$-map, where we see a transition of the values of $\alpha$ from flat to steep as we go from SE to NW. Furthermore, there is no indication from the study of \cite{inami13} that there is shocked line emission based on IRS spectroscopy of this object. \cite{prouton04} estimate a 40\% contribution from the AGN to the total infrared luminosity, and based on the optical diagnostics presented in this paper this objects is associated with a Seyfert 2 (Table~\ref{table:agn}). We classify this object as a radio-AGN because of the shift, and because the radio spectral index in the $\alpha$-map exceeds out theoretical value of 1.1.

\noindent
{\bf IC 1623A/B} \cite{condon91} does not give a measurement for the flux density at 8.44 GHz of this source, but we provide a flux density above 3 $\sigma$ in Table~\ref{data} measured from the map. The source is large (24 arcsec long) at 1.49 GHz, but due to the small angular size of the 8.44 GHz map, only a part of the object is observed, hence the $\alpha$-map is small. The radio spectral index has {\bf a} wide range of values. The maximum flux pixel in the 1.49-GHz map is not at the same R.A. and Dec as the maximum flux pixel in the 8.44-GHz map. This offset is of the order of $\sim$ 1.5 arcsec, which is larger that the pixel size of the 1.49-GHz image (0.4 arcsec). The image absolute coordinate frames are correct within 0.1 arcsec (Jim Condon priv. comm.). Thus we believe the shift to be real. As a result, the flat radio-spectral-index area is offset from the radio centre at 1.49 GHz, but matches the peak brightness at 8.44 GHz. \cite{rich11} find shocked emission and outflow associated with the east galaxy, as a result of the merging process in this late type merging galaxy system. On the contrary, there is no indication from the study of \cite{inami13} that there is shocked line emission based on IRS spectroscopy of this object, and the FeII line present in the spectrum, a good indicator of shocked emission,  is probably not dominated by shocks. We classify this object as a radio-AGN because of the shift, and because the radio spectral index in the $\alpha$-map exceeds out theoretical value of 1.1.

\noindent
{\bf CGCG 436-020} The $\alpha$-map and the $\alpha$-histogram show a flat range of values ($\sim$ 0.3) in the centre, rapidly steepening outwards. The values of $\alpha$ barely exceed our theoretical cut value to separate AGN from SB of $\alpha$ = 1.1 if we take into account the uncertainties. We thus classify it as AGN/SB in the radio. We note, there is a HST bulge associated with the centre of radio emission \citep{haan11}, as additional evidence an AGN could precent. 

\noindent
{\bf IRAS F01364-1042} From the $\alpha$-map of this object we see that the centre of radio emission is associated with a very flat index value, which is slightly steepening outwards. Based on our classification, this steepening might be due to the AGN, but since it does not exceed much, within the uncertainties, our theoretical limit it can also be due to a SB located in an HII region. Then the CR electrons as they escape the SB are losing energy which is shown as steepening in the radio spectral index. We classify this object as AGN/SB. 

\noindent
{\bf IRAS F01417+1651} This is a COM radio source with flat range of $\alpha$ values and we classify it as a SB. To be an AGN, the brightness temperature should be above 10$^{5}$-10$^{6}$ K \cite[e.g.][]{clemens08, condon91}, but this object has a value of  $\sim 3\times10^{4}$ K. Note: the inverted values of $\alpha$ are located at the edges of the $\alpha$-maps, and because the errors are of the order of 0.2, we don't take them into consideration.

\noindent
{\bf UGC 02369 (S)} Only the south companion has been observed at 8.44 GHz. From the radio structure at 8.44 GHz we see a core with diffuse radio emission around it. There is no indication for a weak jet from the radio structure. There is no evidence that this is an AGN based on mid-IR AGN diagnostics. Yet, the $\alpha$-map gives very steep values, above 1.1, suggesting an AGN. Thus we classify this object as radio-AGN.

\noindent
{\bf IRAS F03359+1523} This radio source is associated with the east companion of a galaxy pair, while the HST bulge is associated with the west companion. It has an extended radio source at 1.49 GHz, that presents a core-jet-like feature at 8.44 GHz. The radio spectral index at the centre of the source is flat, steepening outwards with values of $\alpha$ exceeding 1.1, but due to the uncertainties we will classify this as AGN/SB.

\noindent
{\bf ESO 550-IG025 (N)} This is the north radio component of the galaxy pair. This object is brighter in the radio than the IR, unlike the south companion where we observe the opposite. This gives a low value of $q_{\rm IR}$, but not lower than the 1 $\sigma$ uncertainty (see Fig.~\ref{qir}). The north component presents a flat radio-spectral-index distribution steepening outwards. We have to note though that the brightness temperature is low for this object to have a powerful AGN ($\sim 10^{3}$ K; Table~\ref{tab:Tb}). Due to the uncertainties in the steep values of $\alpha$ we classify it as AGN/SB.

\noindent
{\bf ESO 550-IG025 (S)} This object is associated with the south companion of the galaxy pair, and the $\alpha$-map exhibits steep values that indicate synchrotron emission ($\alpha \sim$ 0.8), getting higher than values expected from a nuclear SB, i.e. $\alpha >$ 1.1. Thus we classify it as radio-AGN.

\noindent
{\bf IRAS F05189-2524} This object has a [NeV] detection making it a mid-IR AGN. Similarly the 6.2 $\mu \rm m$ PAH EQW suggests an AGN. The optical classification suggests a Seyfert galaxy, probably of type 2. Based on the $\alpha$-map and the $\alpha$-histogram, this is a radio-SB, since the $\alpha$-map gives values of $\alpha$ that are flat. The SB is also suggested by it's brightness temperature ($\sim 2\times10^{4}$ K; Table~\ref{tab:Tb}). The $\alpha$-map in this case does not help us identify the AGN in this ULIRG, although it could be a young AGN and we could be seeing emission from a jet that is being formed. Evidence for this could be the slightly elongated towards N-S radio structure at 8.44 GHz, but it can also be an artefact of the radio map.

\noindent
{\bf NGC 2623} The mid-IR diagnostics classify this objects as an AGN. In the optical it is a composite, having both AGN and SB contributions. The $\alpha$-map shows a flat distribution, steepening outwards with values of $\alpha >$ 1.1, within the uncertainties. Based on our classification this is a radio-AGN.

\noindent
{\bf IRAS F08572+3915} This object shows a compact core at 8.44 GHz with several diffuse point-like structures in the surrounding area. \cite{vega08} show that this object contains an AGN that dominates the bolometric luminosity of the galaxy. From the $\alpha$-map we see that the core emission gives an inverted $\alpha$, getting flatter outwards. Because values of $\alpha$ at the core of radio emission are inverted, suggesting emission from a face-on jet, we classify this object as a radio-AGN.

\noindent
{\bf UGC 04881 (NE)} The north-east radio component of a galaxy pair presents flat values of $\alpha$ that get steeper outwards, as seen by the $\alpha$-histogram (Fig.~\ref{amaps}-$Right$), but values of $\alpha$ barely exceed the value of 1.1, within the uncertainties, thus we classify it as AGN/SB. The mid-IR diagnostics suggest an AGN/SB, and in the optical the BPT diagram gives a composite including both AGN and SB. We see that the $\alpha$-map and $\alpha$-hist does not help us distinguish between the AGN and the SB.

\noindent
{\bf UGC 04881 (SW)} The $\alpha$-map of the SW radio component shows steep values of $\alpha \sim$ 0.8, indicating synchrotron emission, that get larger than 1.1. This steepening and spectral ageing suggest an AGN and not a nuclear SB, since the latter have short lifetimes. We note that the brightness temperature is low ($\sim 1.5\times10^{2}$ K; Table~\ref{tab:Tb}). Based on the 6.2 $\mu \rm m$ EQW diagnostic, this is a SB, while based on the BPT diagram it is a composite. We classify it as a radio-AGN.

\noindent
{\bf UGC 05101} The $\alpha$-map shows a flat distribution in the centre, steepening outwards with values of $\alpha >$ 1.1. Based on our classification this is a radio-AGN. This object is also an AGN based on the mid-IR diagnostics presented in the analysis, and a Seyfert galaxy in the optical.

\noindent
{\bf IRAS F10173+0828} The $\alpha$ values are negative suggesting self-absorption as expected in a face-on AGN with the jet pointing towards the observer. The brightness temperature is $\sim 8\times10^{4}$ K (Table~\ref{tab:Tb}), close to the value for AGN. We classify it as a radio-AGN.

\noindent
{\bf IRAS F10565+2448} This is a compact radio source at 1.49 GHz, but at 8.44 GHz it appears more structured. Based on the distribution of values of the radio spectral index, and the histogram, this is a radio-SB. The inverted values of $\alpha$ located at the edges of the $\alpha$-map suggest optically thick region at a distance of 2 arcsec from the centre of the radio source at 1.49 GHz. Having an optically thick region so far from the nucleus is rare, but has been observed before \cite[e.g.][]{inami10}. After checking the HST image of this galaxy we do not see any feature or excess emission at the position where the radio spectral index in inverted. We believe this is an area where the obscuration is either too high even for radio observations to penetrate or the inverted values are caused by the very diffuse radio emission at 8.44 GHz at the edges of this radio source. {\bf Since} the errors are of the order of 0.2, we don't take these values into consideration.

\noindent
{\bf MCG +07-23-019} This is an extended radio source at 1.49 GHz, and at 8.44 it presents a jet-like structure \citep[similar to FRI;][]{fr74}, whose centre is not associated with the centre of IR emission at 8 $\mu \rm m$ (map not shown here). The $\alpha$ values are flat in the centre of radio emission at 1.49 GHz, steepening outwards and exceeding the value of 1.1, which is also seen in the histogram. Based on our classification this is a radio-AGN.

\noindent
{\bf UGC 06436 (NW)} The north-west companion of this system exhibits flat values of $\alpha$, steepening outwards, as seen in the $\alpha$-map and histogram exceeding the value of 1.1, but due to the uncertainties  we classify it as AGN/SB. We note the low brightness temperature ($\sim 6\times10^{2}$ K; Table~\ref{tab:Tb}) that does not favour a powerful AGN.

\noindent
{\bf UGC 06436 (SE)} The $\alpha$-map of the south-east radio companion of this galaxy pair has flat values in the centre steepening slightly outwards, not exceeding the value of 1.1. We classify this galaxy as a radio-SB.

\noindent
{\bf NGC 3690 (E)} The east radio component of this interacting system has a complex radio structure, and the $\alpha$-map shows four discrete areas. The central radio component, the one associated with the bulge, has flat values of $\alpha$ in the centre steepening rapidly outwards and getting higher than 1.1 suggesting a radio-AGN. The other areas have the characteristics of SB regions. The wide range of $\alpha$ values in the histogram are due to the fact the radio source is composed of smaller radio components at 8.44 GHz. By adding up these separate $\alpha$-maps creates a histogram with wide range of values of $\alpha$. We classify this object as AGN/SB in the sense that it contains both AGN and SB. 

\noindent
{\bf NGC 3690 (W)} The west radio companion of this galaxy merger also has a complex radio structure, and in the $\alpha$-map we see three discrete areas. The NE area displays steep values of $\alpha$ getting larger than 1.1 going outwards, suggesting an AGN. The other two regions show a flat centre, steepening outwards slightly, probably from a SB. The wide range of $\alpha$ values in the histogram are  due to the fact the radio source is composed of smaller radio components at 8.44 GHz. By adding up these separate $\alpha$-maps creates a histogram with wide range of values of $\alpha$. We classify the west radio companion of the galaxy pair as AGN/SB in the sense that it contains both AGN and SB. 

\noindent 
{\bf IRAS F12112+0305 (NE)} This is a double radio source in a galaxy merger, where the NE component is associated with the NE merging galaxy. The distribution of $\alpha$ values goes from inverted to flat, but the spatial distribution is particular in the sense that from SW to NE $\alpha$ gets from inverted to flat values. Also, the centre of the NE radio component has highest values of $\alpha$ than the surroundings. We see that the NE radio component falls mainly in category 2 of histograms. The inverted values of $\alpha$ are located at the edges of the radio source, and due to the uncertainty we do not take them under consideration. We classify it as a radio-SB. For the NE radio component we have a [OIV] measurement, which does not on it's own confirm the presence of an AGN. In order to obtain measurements for the total infrared luminosity, we found the fractional contribution of the two components at 8 $\mu \rm m$ and applied this to the $L_{\rm IR}$ of the system. The values are given in Table~\ref{data}.

\noindent
{\bf IRAS F12112+0305 (SW)} The SW radio component of this galaxy pair is associated with the SW merging galaxy, and has a category 2 histogram with some inverted values at the edges of the source. Due to the uncertainty we will classify it as AGN/SB.

 \noindent
{\bf UGC 08058} The inverted radio spectral index and the range of variations suggest emission coming from a face-on AGN with the jets pointing towards the observer. We classify it as a radio-AGN. 

\noindent
{\bf VV 250a (NW)} This is the north-west companion of an interacting galaxy-system. The flat values of $\alpha$ in the centre of this source, steepening outwards are highly uncertain around the value of 1.1. Thus we cannot distinguish between an AGN or a SB. We classify it as AGN/SB. 

\noindent
{\bf VV 250a (SE)} Taking into account the errors in $\alpha$ and after looking at the histogram for this source, we classify it as category 2 and thus a SB due to the flat values of $\alpha$.

\noindent
{\bf UGC 08387} The map at 8.44 GHz reveals a jet-like radio structure, FRI-like \citep{fr74}. The centre of radio emission matches the HST bulge position. The $\alpha$-map shows a distribution of flat values which are steepening outwards, getting higher than 1.1. We classify this as a radio-AGN. Although the 6.2 $\mu \rm m$ PAH EQW suggests this is a SB, we get a [NeV] detection telling us this object contains an AGN. The question is, how is it possible for an object to have strong 6.2 $\mu \rm m$ PAH feature that was not destroyed by the AGN. One explanation could be that the PDR is located in a circumnuclear region that does not get influenced by the AGN, thus the PAH feature is not being affected.

\noindent
{\bf NGC 5256 (NE)} This is the north-east companion of a galaxy system, a compact radio source. Based on the distribution of values of the radio spectral index and the histogram, and the steepening in the values of alpha at the edges of the source ($\alpha <$ 1.1), we classify this as a radio-SB. Note that in the optical it is classified as LINER \citep{mazzarella12}. However, the presence of a compact, luminous hard X-ray source and evidence for a bi-conic ionisation cone in B- and I-band HST images aligned with corresponding radio continuum features provide strong evidence for an AGN-driven outflow from this nucleus \citep{mazzarella12}.

\noindent
{\bf NGC 5256 (SW) \& (C)} At 1.49 GHz the SW and C (for centre) radio components are barely separated. The C radio component at 13:38:17.5 +48:16:36 is not associated with the 8 $\mu \rm m$ source (map not shown here). The source is in-between the two galaxies of this interacting system, is believed to be a very fast and short-lived shock that effectively destroys most of the dust grains, sublimates the molecular gas, and inhibits star formation \citep{mazzarella12}. Based on our radio classification, and the steepening of $\alpha$ ($<$ 1.1), we classify it as a AGN/SB, in the sense that is associated with an AGN since it has the signature of a radio-AGN. The SW radio component shows flat distribution of values steepening outwards rapidly and exceeding the value of $\alpha$ = 1.1. Also the radio centre is associated with the bulge. We classify this one as radio-AGN. We have [Ne V] measurement for the SW radio component and [O IV] for the C radio component. The [NeV] emission from this nucleus is also consistent with the AGN classification. The fact that the PAH feature suggests an AGN/SB can be explained by a circumnuclear star-forming region that does not get affected by the AGN.

\noindent
{\bf NGC 5257 (NW)} Only the north-west companion of this interacting galaxy-system has been observed at 8.44 GHz. This radio source has a complex radio structure where the radio at 1.49 GHz traces the gas in the spiral arms of the NW galaxy. Only the nucleus was observed at 8.44 GHz due to the smaller field of view of the 8.44 GHz observations compared to the one at 1.49 GHz. In Fig.~\ref{app:amaps} we give just the radio source associated with the nucleus for presentation issues. The $\alpha$-map and histogram show a flat $\alpha$ distribution, and thus we classify it as a radio-SB. We note that there are some steep values at the edges of the central radio source and the explanation is cooling of CR electrons; yet the uncertainty is of the order of 0.2 for these values for $\alpha$.

\noindent
{\bf UGC 08696} At 8.44 GHz we see a complex system. There is a double radio source associated with the IR source at 8 $\mu \rm m$ (map not shown here), which has jet-like structure and has two additional features, the brightest of which is located north-east of the radio centre. The $\alpha$-map shows flattening in the centre, following the 8.44-GHz structure. The values get steeper towards the SE of the centre. Based on the $\alpha$-map and the histogram this is a radio-AGN, due to the steepening of $\alpha >$ 1.1.

\noindent 
{\bf IRAS F14348-1447 (NE)} This is the NE radio component of a double radio source, associated with a merging galaxy-system. The radio centre at 1.49 GHz is associated with the HST bulge. The radio sources are not completely separated, so the histograms created are imperfect. We see that the NE component matches category 1 of our classification, with $\alpha$ steepening above 1.1 but due to the uncertainties we classify it as AGN/SB. It is also a mid-IR AGN based on the PAH feature, and a LINER in the optical. In order to obtain measurements for the total infrared luminosity, we found the fractional contribution of the two components at 8 $\mu \rm m$ and applied this to the $L_{\rm IR}$ of the system. The values are given in Table~\ref{data}.

\noindent
{\bf IRAS F14348-1447 (SW)} This is the SW radio component of a double radio source, associated with a merging galaxy-system. 
The radio centre of the south component is not matching the one at 8 $\mu \rm m$ (map not shown here). The radio centre at 1.49 GHz is associated with the HST bulge. The radio sources are not completely separated, so the histograms created are imperfect. The SW radio component matches category 1, with steepening above $\alpha$ = 1.1, but due to the uncertainties we classify it as AGN/SB.

\noindent
{\bf VV 340a (N)} Only the north companion of the interacting system has been observed at both 1.49 \& 8.44 GHz. The x-shaped radio structure at 1.49 GHz suggests a double-double radio source or an X-shaped radio source \cite[e.g.][]{capetti02}, i.e. a restarted radio source associated with an AGN. We do not have further observations at 8.44 GHz to cover the whole area seen at 1.49 GHz. Based on our classification and the steep values of $\alpha$ ($>$ 1 going outwards), we classify it as radio-AGN. This object has a [NeV] detection which indicates the presence of an AGN, while the 6.2 $\mu \rm m$ EQW classifies it as a SB. Probably there is a circumnuclear PDR region that does not get affected by the AGN. Based on the BPT this is a SF object.

\noindent
{\bf VV 705 (N)} This is the north radio component of an interacting galaxy-system, showing flat $\alpha$ in the centre steepening outwards with $\alpha >$1 being highly uncertain. Thus we classify it as a AGN/SB. This object is classified as a SB based on the 6.2 $\mu \rm m$ EQW. The small radio contour between the north and south radio components has a steep distribution of values suggesting synchrotron emission, maybe as a result of the interaction between the galaxies. 

\noindent
{\bf VV 705 (S)} The south companion of this galaxy system shows values of $\alpha$ associated with synchrotron emission ($\alpha \sim$ 0.8), and steepening of the values outwards exceeding $\alpha$ = 1.1. Based on our classification this is a radio-AGN. This object is classified as a SB based on the 6.2 $\mu \rm m$ EQW. 

\noindent
{\bf IRAS F15250+3608} The $\alpha$-map of this compact radio source shows a negative distribution of values suggesting synchrotron self-absorption. A jet from an AGN pointing towards the observer is a possible explanation. The brightness temperature is of the order of 10$^{5}$ K (Table~\ref{tab:Tb}) which makes us confident to classify it as a radio-AGN.

\noindent
{\bf UGC 09913} At 8.44 GHz we see a double radio structure, with diffuse radio emission around it. The centre of radio emission is associated with inverted values of $\alpha$ suggesting synchrotron self-absorption, before emission becomes optically thin. The steepening of the radio spectral index from the centre outwards, as seen in the $\alpha$-map and histogram, can be either from an AGN or due to CR electrons in an HII region. Thus we classify it as AGN/SB.

\noindent
{\bf NGC 6090} This is a beautiful galaxy merger that displays a complex double radio structure at 1.49 GHz. Only the NE radio component has been observed at 8.44 GHz, and shows diffuse radio emission. The $\alpha$-map gives a distribution of steep values suggesting a radio-AGN. In the mid-IR it is classified as a SB, and as a Seyfert based on the BPT diagram.

\noindent 
{\bf IRAS F17132+5313 (NE)} This is the north-east radio components of a galaxy pair that shows a distribution of steep values of $\alpha$. 
We believe the NE component is a radio-AGN, due to the intense steepening of the radio spectral index further out from the core, suggesting synchrotron ageing from an AGN. We have [Ne V] and [O IV] measurement only for the NE component. In order to obtain measurements for the total infrared luminosity, we found the fractional contribution of the two components at 8 $\mu \rm m$ and applied this to the $L_{\rm IR}$ of the system. The values are given in Table~\ref{data}. \cite{condon91} do not give brightness temperature for the NE radio component.

\noindent
{\bf IRAS F17132+5313 (SW)} The south-west radio component shows flat values of $\alpha$ suggesting a compact SB. We do not have mid-IR nor optical classification for this object.

\noindent
{\bf IRAS F22491-1808} The HST bulge position agrees with the radio centre at 8.44 GHz but not at 1.49 GHz, but the distance between the radio centres is negligible. The $\alpha$-map gives a flat distribution of values, steepening outwards, but we cannot distinguish between AGN or SB due to the uncertainties. We classify this is as AGN/SB.

\noindent
{\bf IC 5298} This radio source resembles an FRI radio source \citep{fr74} at 8.44 GHz, and has a tail at 1.49 GHz. The distribution of $\alpha$ values, flat in the centre steepening outwards ($\alpha >$ 1.1), suggests a radio-AGN. We note that the uncertainties in $\alpha$, in particular the $\alpha- d\alpha$ value is 1.1, but we will keep the radio-AGN classification because there is also an indication from the radio structure. This object is also classified as an AGN based on the mid-IR diagnostics, and is a Seyfert 2 or LINER in the optical.

\noindent
{\bf Mrk 0331} This is the north companion of a galaxy system. At 8.44 GHz we see a ring-structure. The $\alpha$-map and histogram show a flat centre steepening outwards with values of $\alpha$ getting much higher than 1.1. We classify it as a radio-AGN. We note that in the mid-IR this is a SB, and a Seyfert 2 in the optical.

   \begin{figure*}[!ht]
    \resizebox{\hsize}{!}
            {\includegraphics{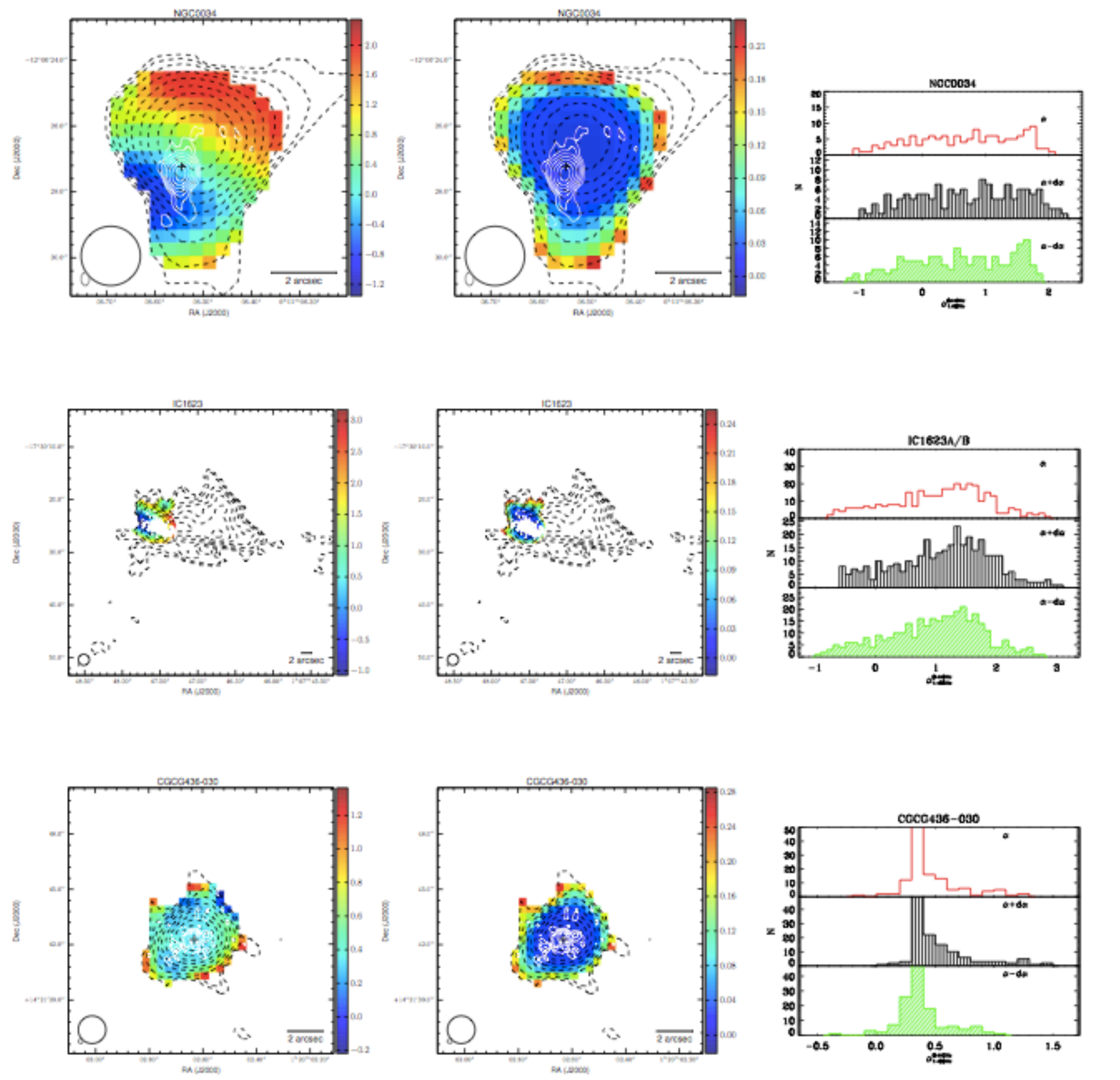}
          
            }

   \caption{Each row corresponds to one object or system from our sample. ($Left$) Radio-spectral-index maps ($\alpha$-maps) are represented with the colour-map; the colour bar on the right shows the radio spectral index values from low (flat $\alpha$) to high values (steep $\alpha$). Overlaid, the radio contours at 1.49 GHz (black dashed) and at 8.44 GHz (white solid; original unsmoothed maps), where the contour levels scale as log$_{10}$ of the flux density from the lowest flux value (3 $\sigma$) to the highest. On the bottom right we give a scale-bar of 2 arcsec, as visual aid. ($Middle$) The error map for the radio spectral index per pixel value. The colour bar on the right shows the range of values. Contours as on the $Left$. In both plots, on the bottom left corner we give the beam of the 1.49 GHz map as black circle, and the beam of the 8.44 GHz map as grey circle/ellipse. On the bottom right we give a scale-bar of 2 arcsec, as visual aid.
($Right$) Histograms of the $\alpha$-maps: ($Top$) Histograms of the radio spectral index, based on Fig.~\ref{app:amaps}-$Left$. ($Middle$) Histograms where the errors on $\alpha$ (Fig.~\ref{app:amaps}-$Middle$) have been added. ($Bottom$) Histograms where the errors on $\alpha$ have been subtracted.
   }
              \label{app:amaps}
    \end{figure*}
\addtocounter{figure}{-1}

%
   \begin{figure*}[!ht] 

            \resizebox{\hsize}{!}
            {\includegraphics{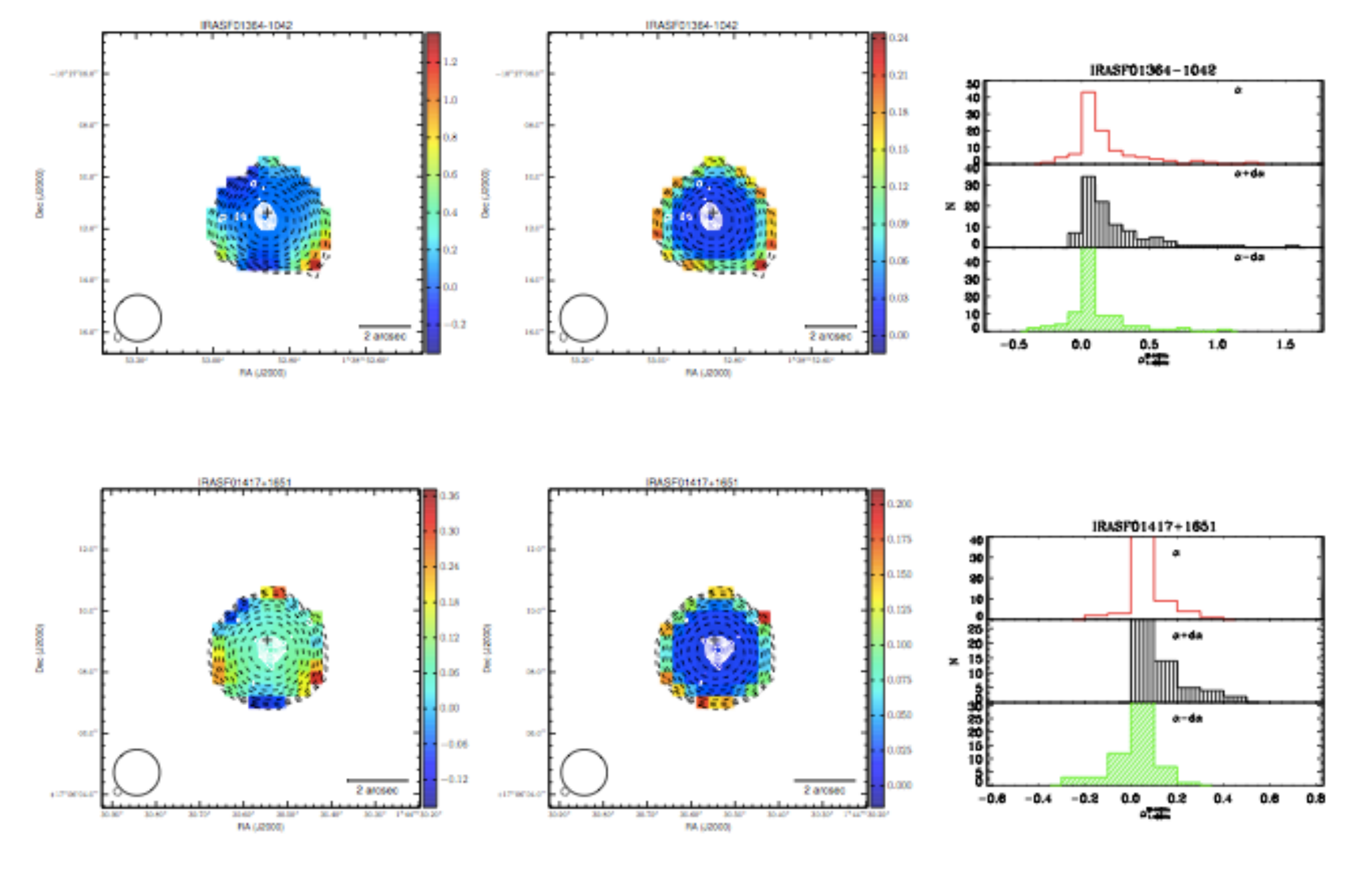}
         
    }

            \resizebox{\hsize}{!}
            {\includegraphics{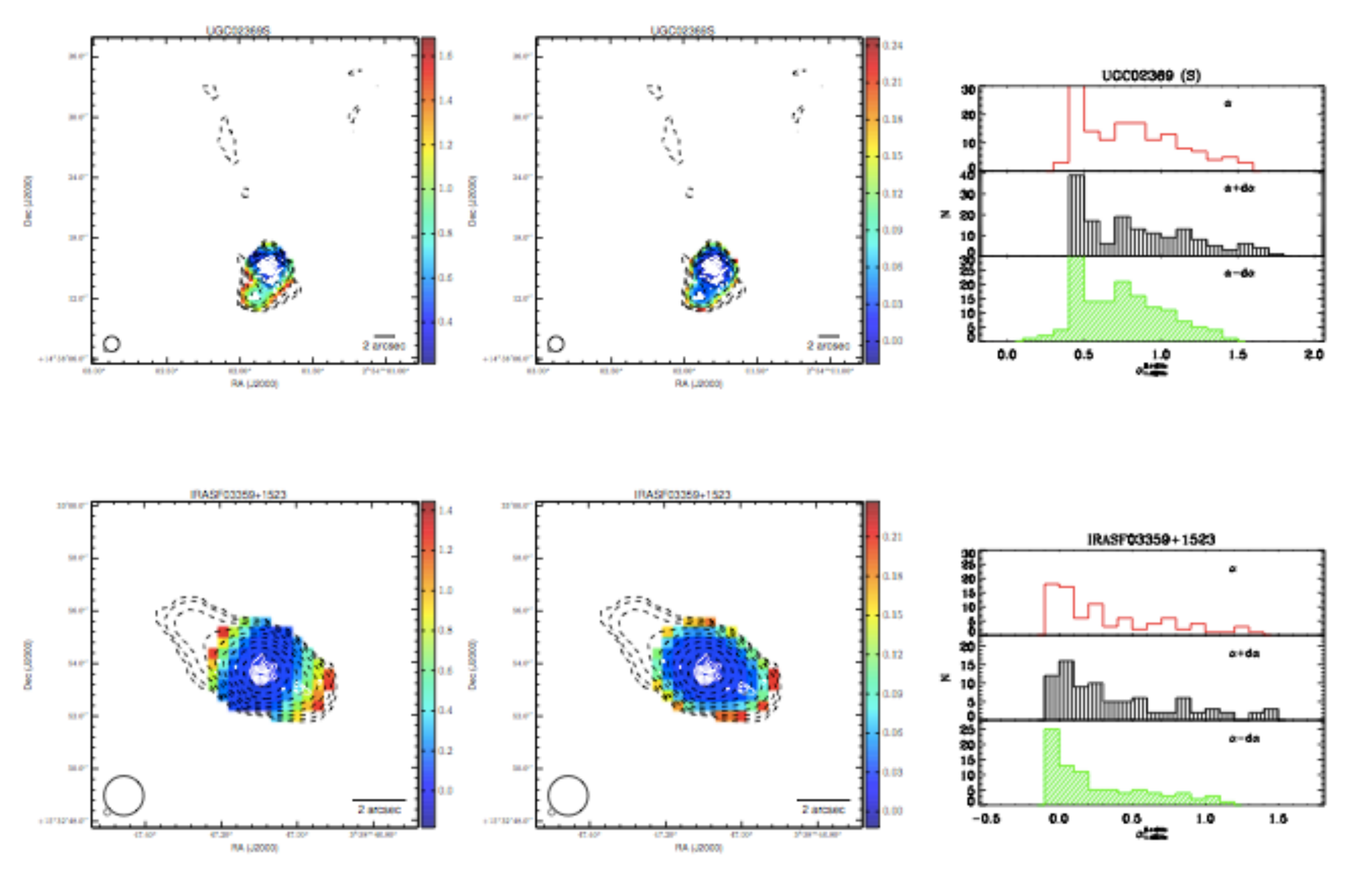}
          
            }

   \caption{(Continued)
   }
              \label{app:amaps}%
    \end{figure*}
\addtocounter{figure}{-1}

%
  %
   \begin{figure*}[!ht] 

               \resizebox{\hsize}{!}
            {\includegraphics{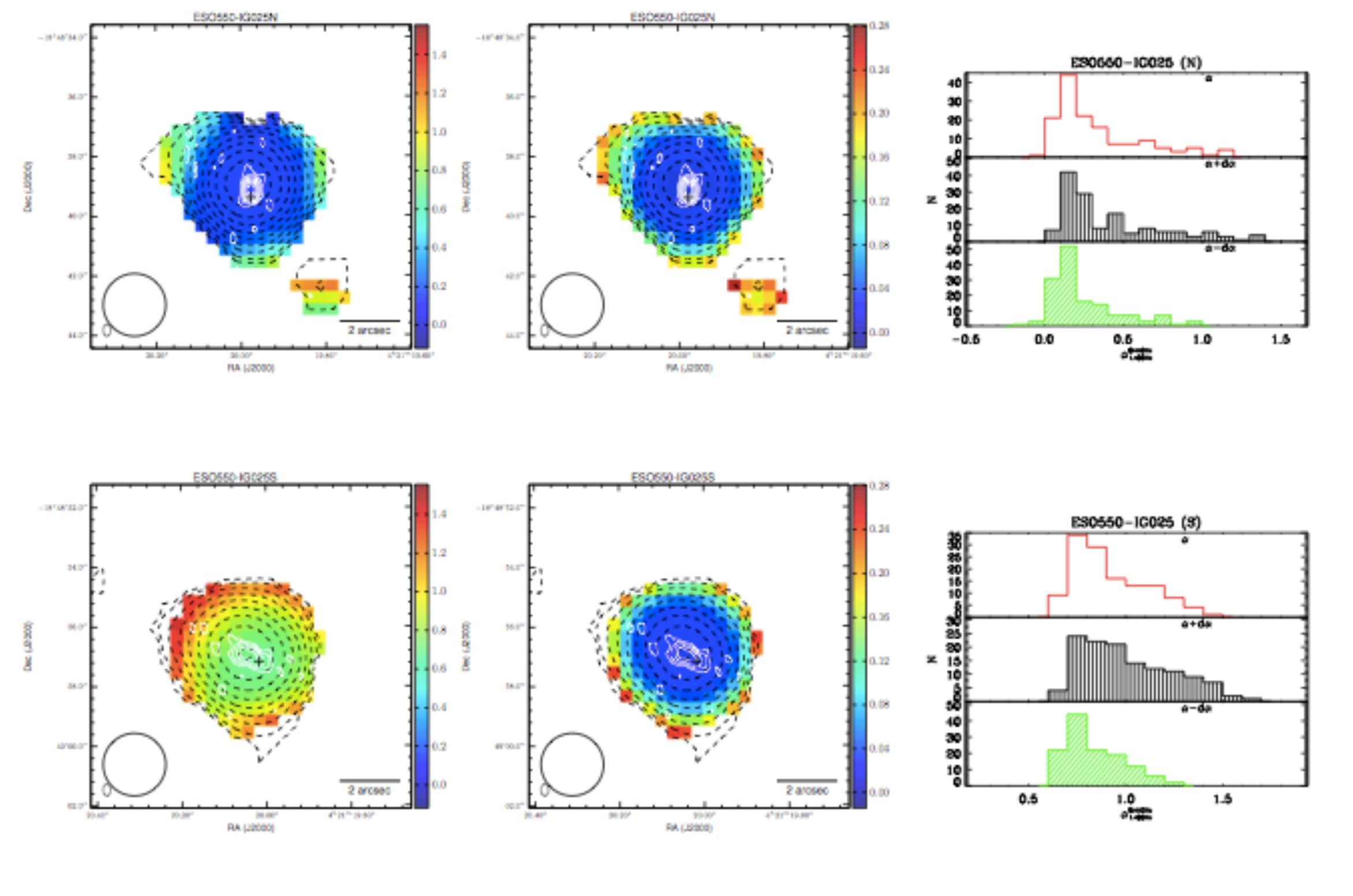}
    
}
                   
              \resizebox{\hsize}{!}
            {\includegraphics{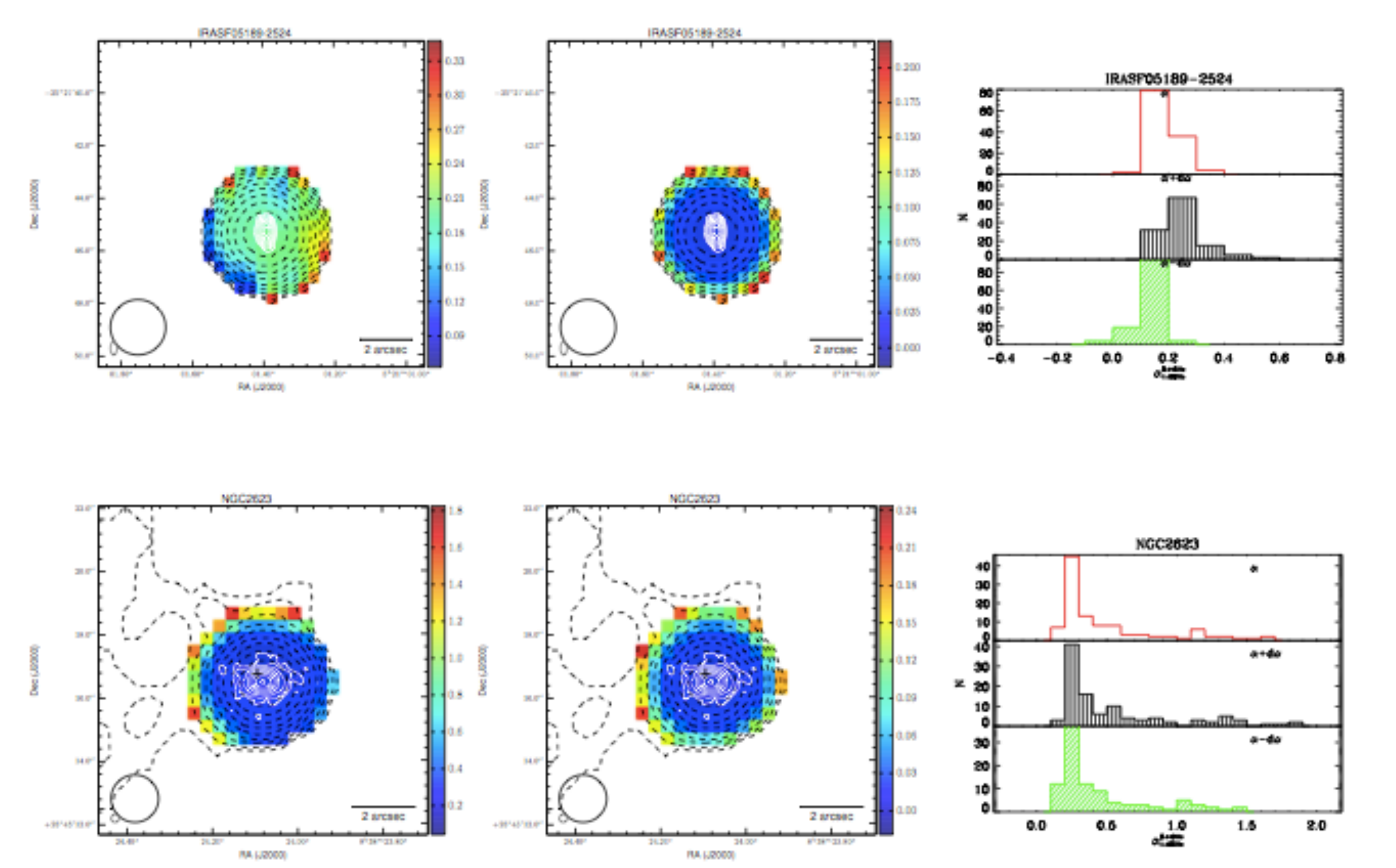}
    
            }

   \caption{(Continued)
   }
              \label{app:amaps}%
    \end{figure*}

\addtocounter{figure}{-1}

%
  %
   \begin{figure*}[!ht] 
                 \resizebox{\hsize}{!}
            {\includegraphics{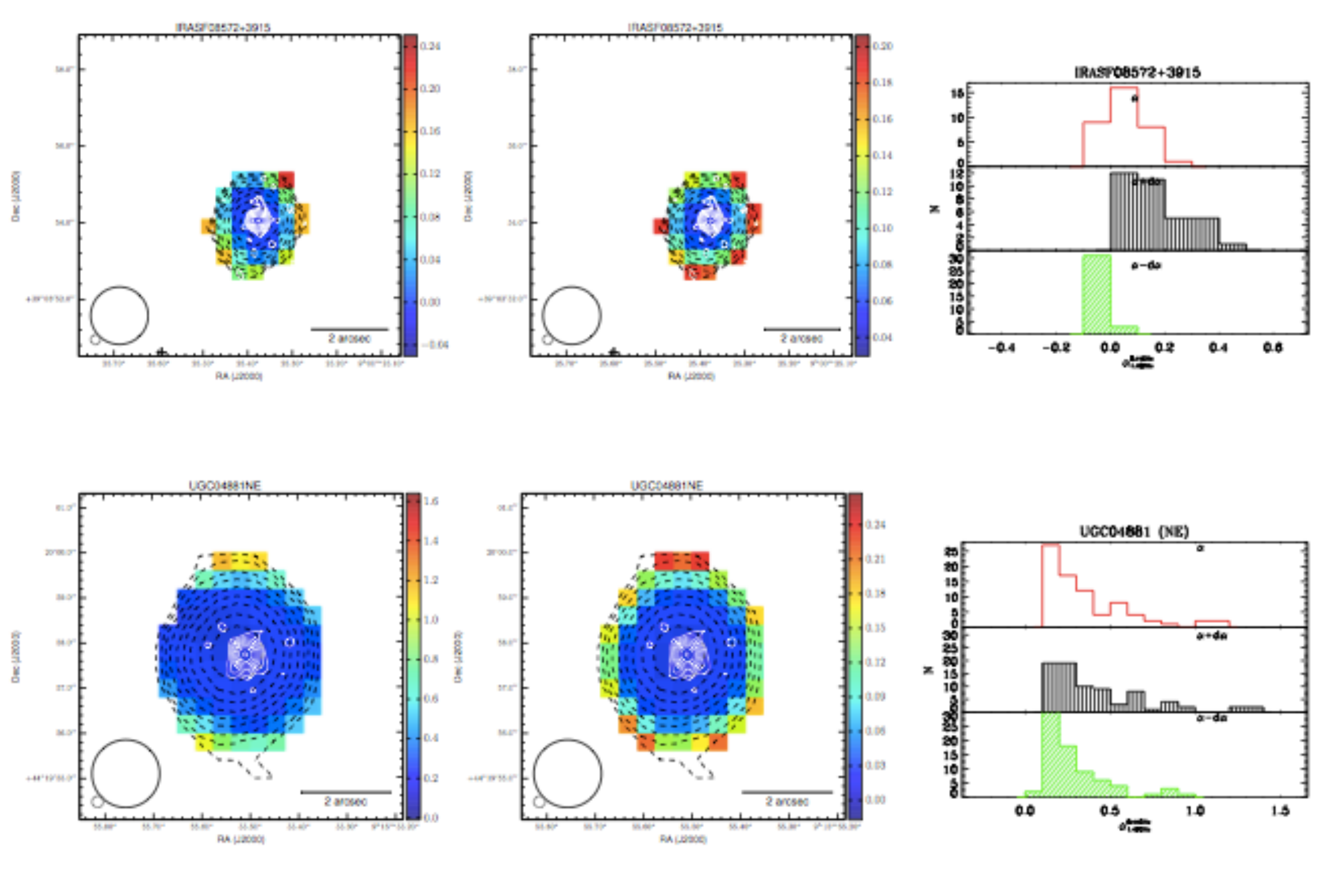}
                 }

                 \resizebox{\hsize}{!}
            {\includegraphics{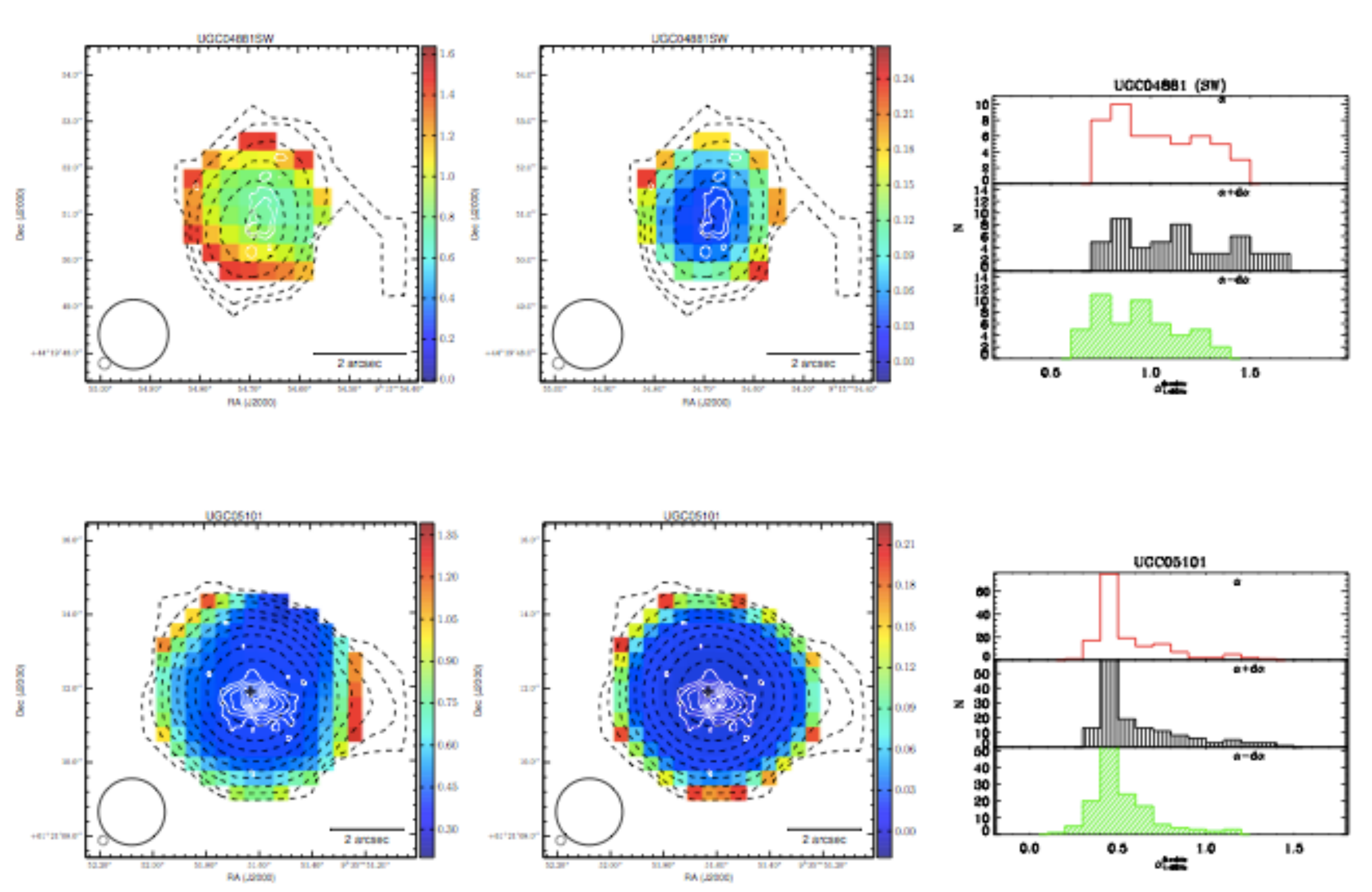}
       
            }

   \caption{(Continued)
   }
              \label{app:amaps}%
    \end{figure*}
\addtocounter{figure}{-1}

%
  %
   \begin{figure*}[!ht] 
                    \resizebox{\hsize}{!}
            {\includegraphics{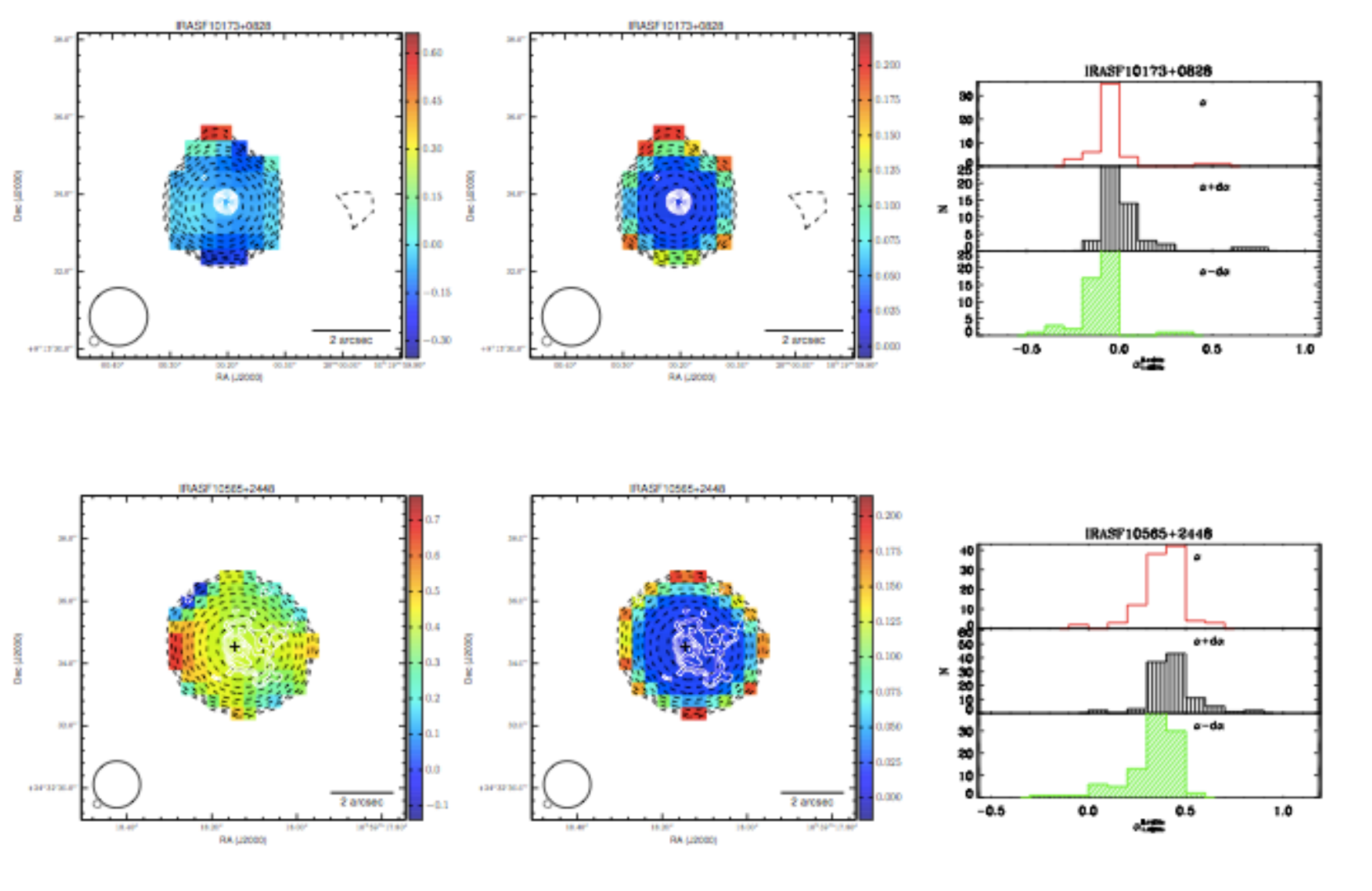}
            
            }  
     
               \resizebox{\hsize}{!}
            {\includegraphics{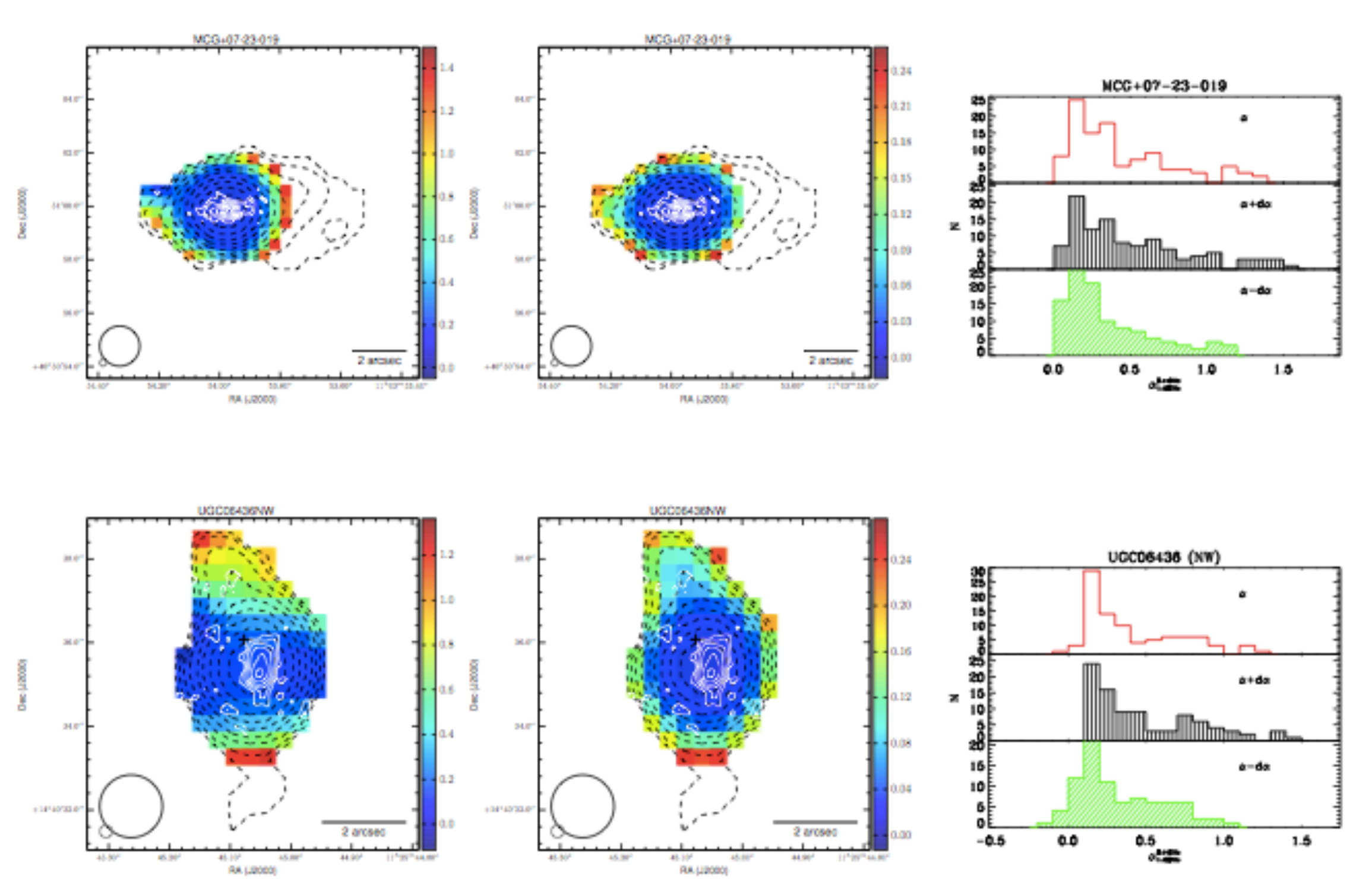}
  
            }

   \caption{(Continued)
   }
              \label{app:amaps}%
    \end{figure*}
\addtocounter{figure}{-1}

%
  %
   \begin{figure*}[!ht] 

         \resizebox{\hsize}{!}
            {\includegraphics{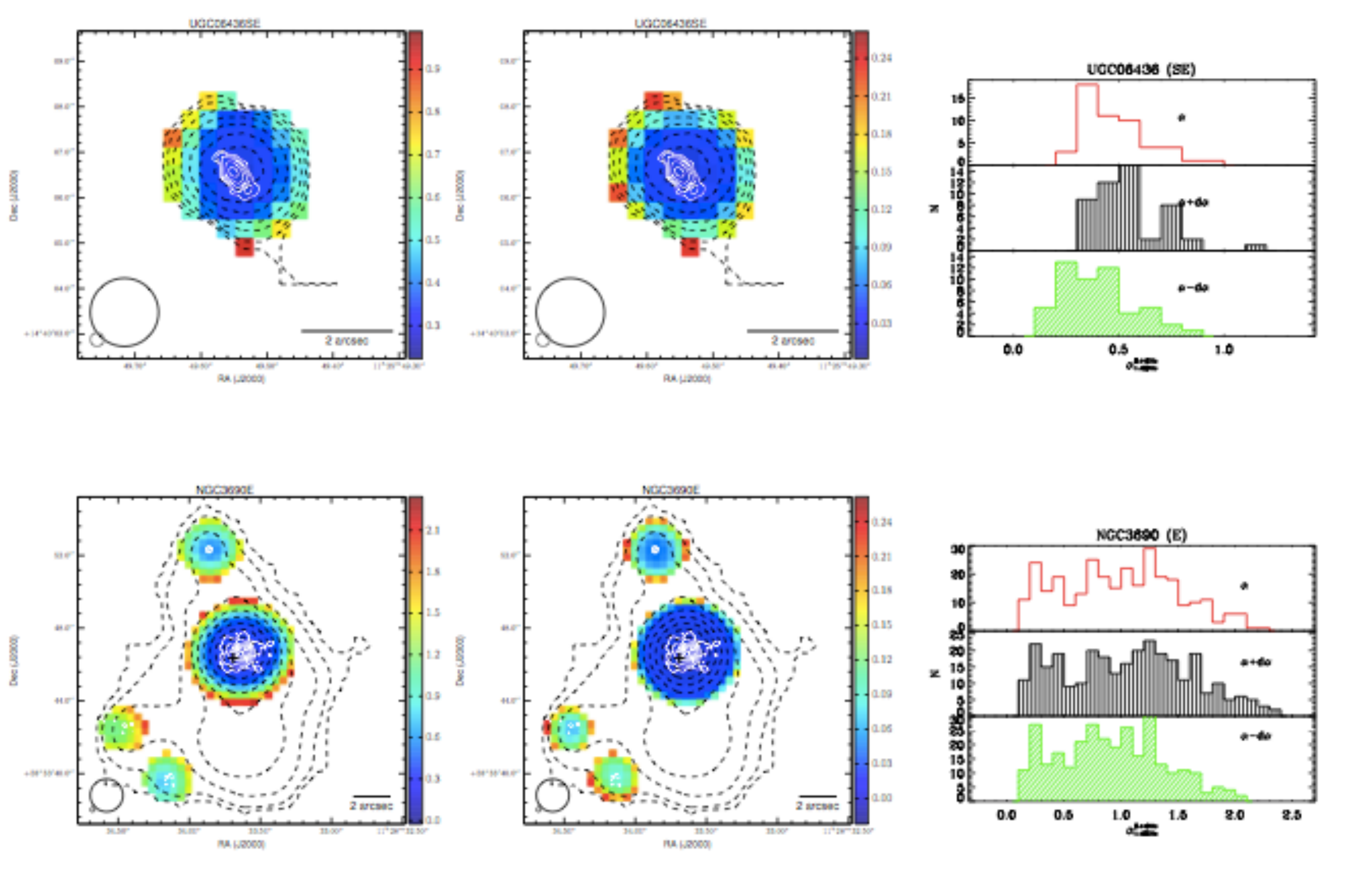}
       
            }
         
          \resizebox{\hsize}{!}
            {\includegraphics{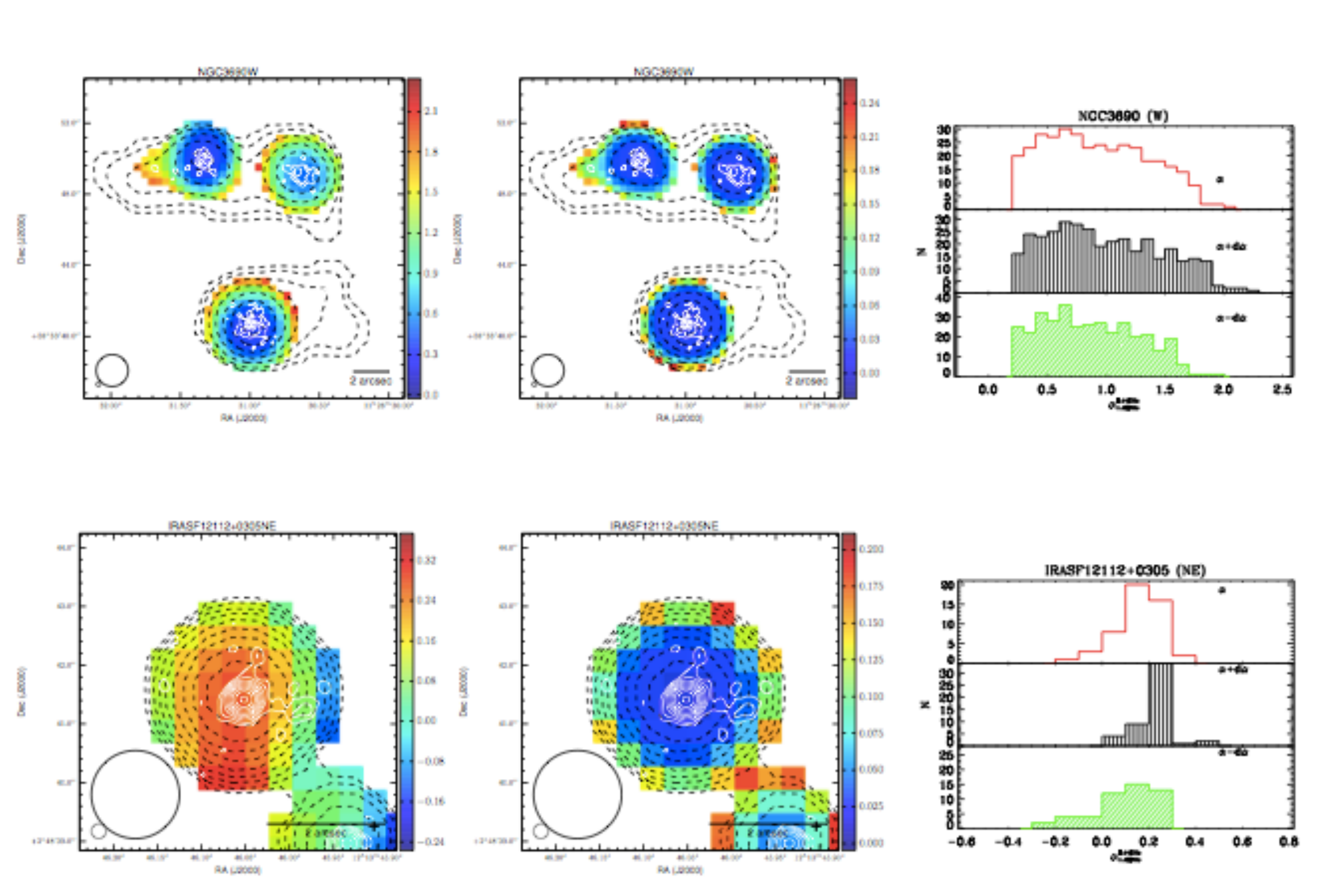}
          
            }
   \caption{(Continued) 
   }
              \label{app:amaps}%
    \end{figure*}

\addtocounter{figure}{-1}

%
  %
   \begin{figure*}[!ht] 

                      \resizebox{\hsize}{!}
            {\includegraphics{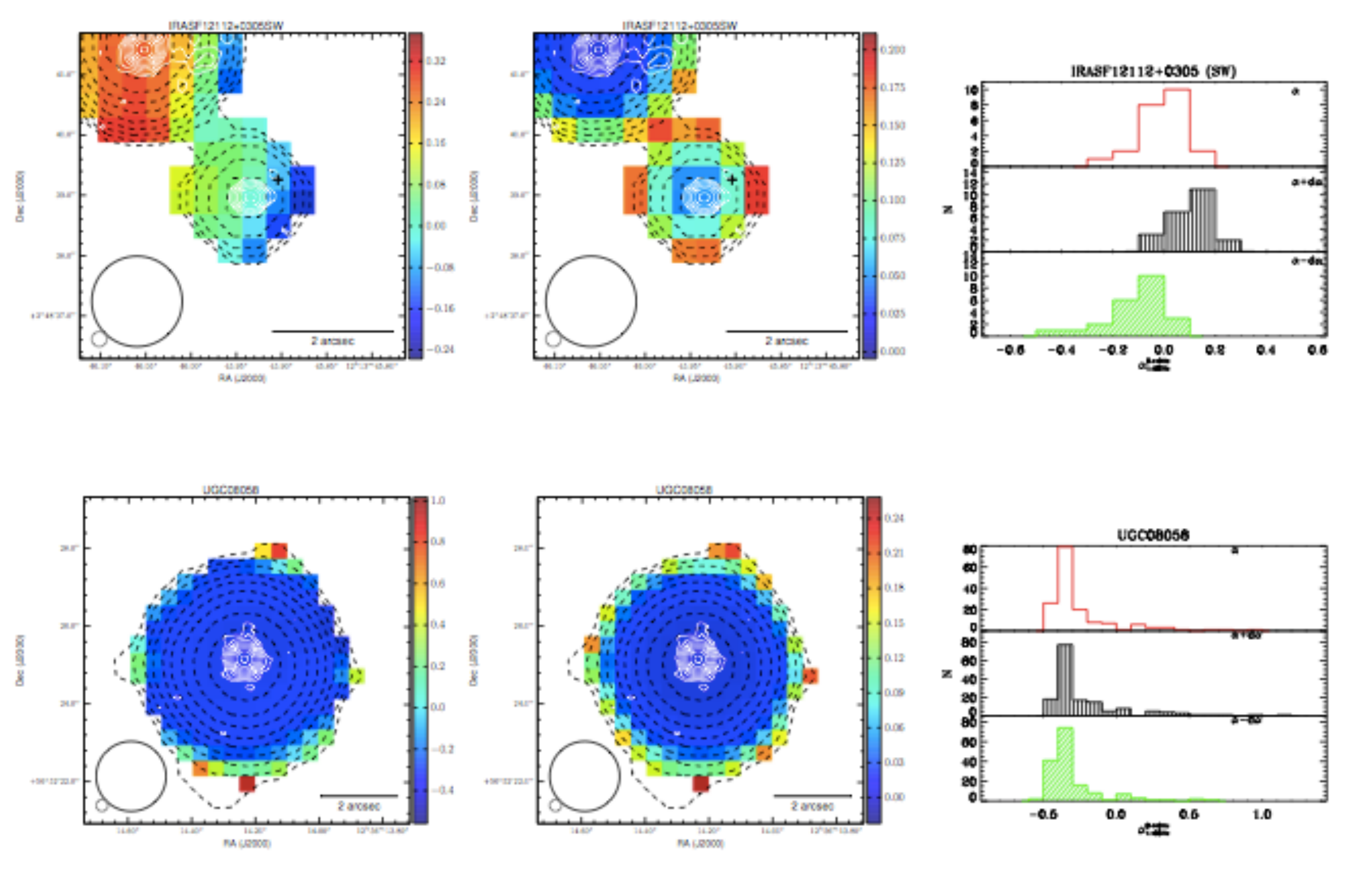}
      
            }   
                      \resizebox{\hsize}{!}
            {\includegraphics{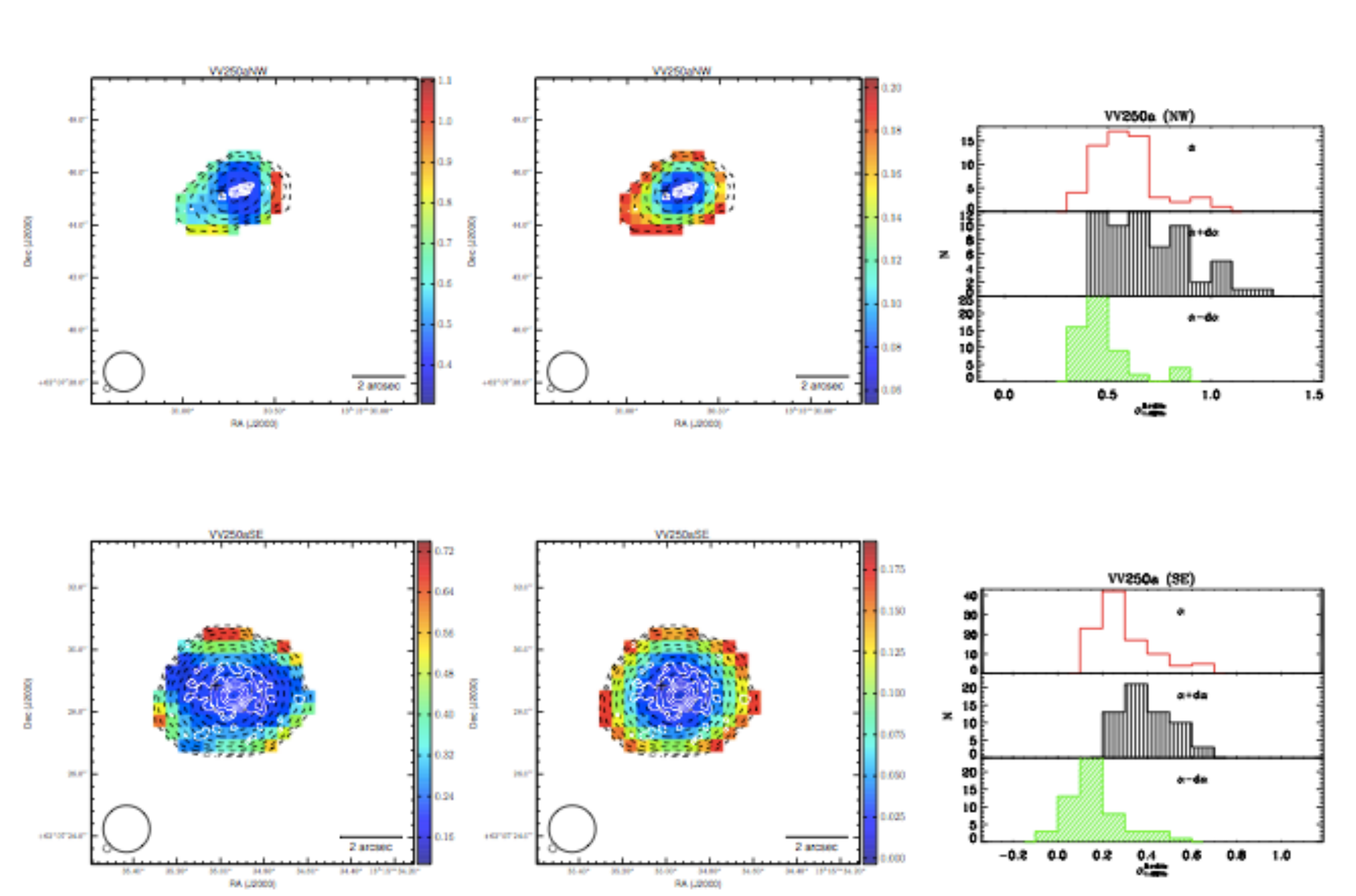}

            }

   \caption{(Continued) 
   }
              \label{app:amaps}%
    \end{figure*}

\addtocounter{figure}{-1}

%
  %
   \begin{figure*}[!ht] 

         \resizebox{\hsize}{!}
            {\includegraphics{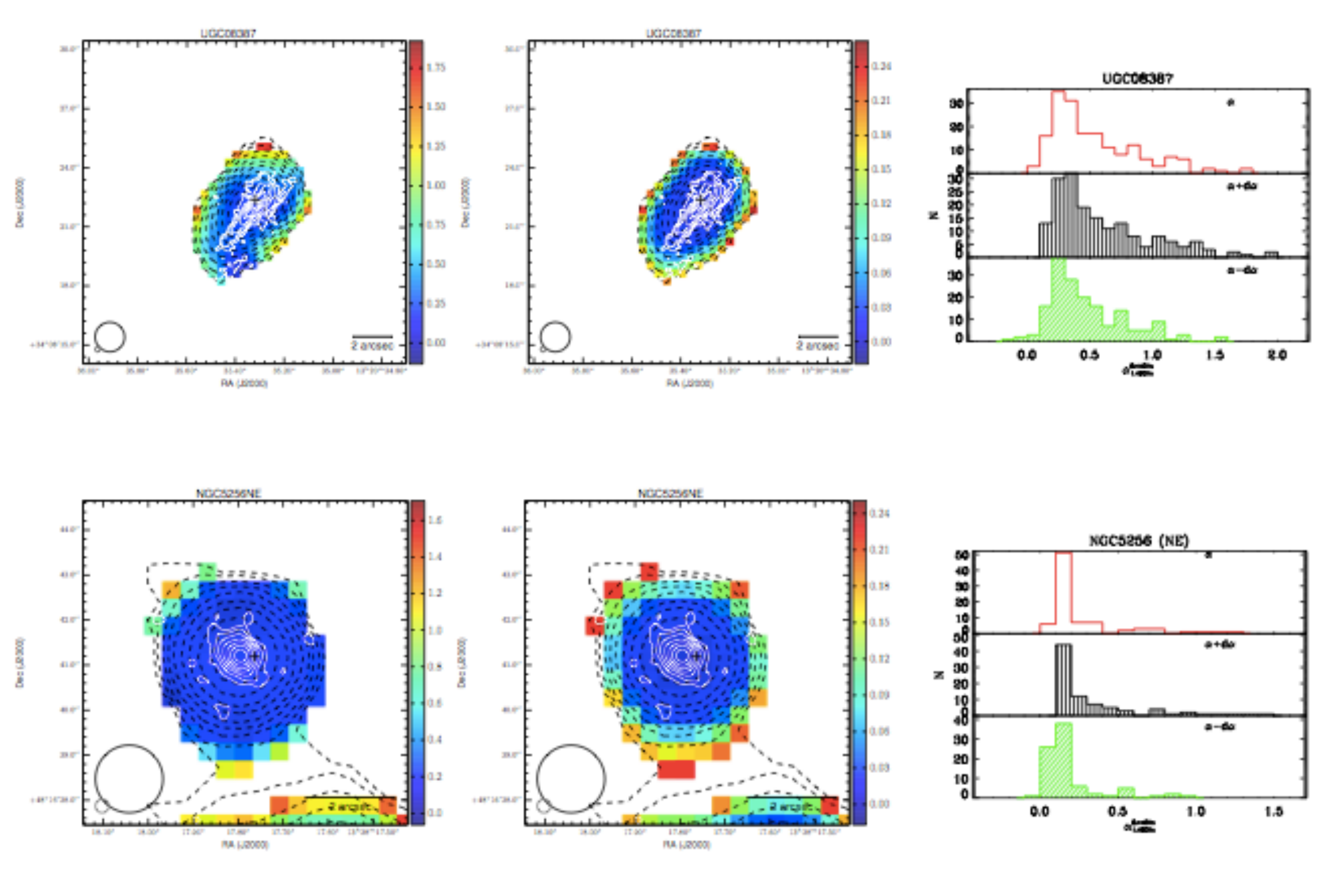}
          
            }

                \resizebox{\hsize}{!}
            {\includegraphics{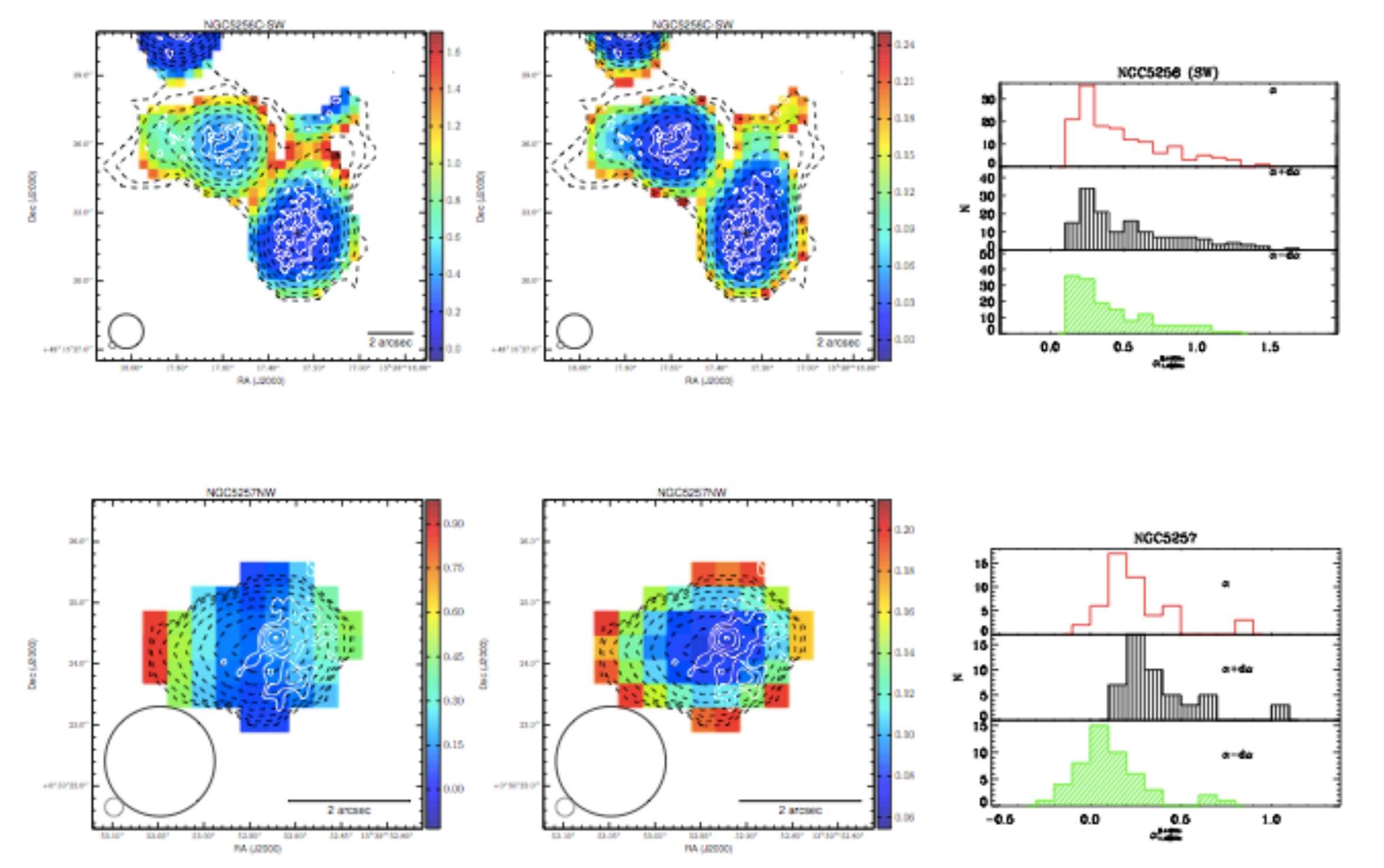}
      
            }
   \caption{(Continued) 
   }
              \label{app:amaps}%
    \end{figure*}

\addtocounter{figure}{-1}

%
  %
   \begin{figure*}[!ht] 
        \resizebox{\hsize}{!}
            {\includegraphics{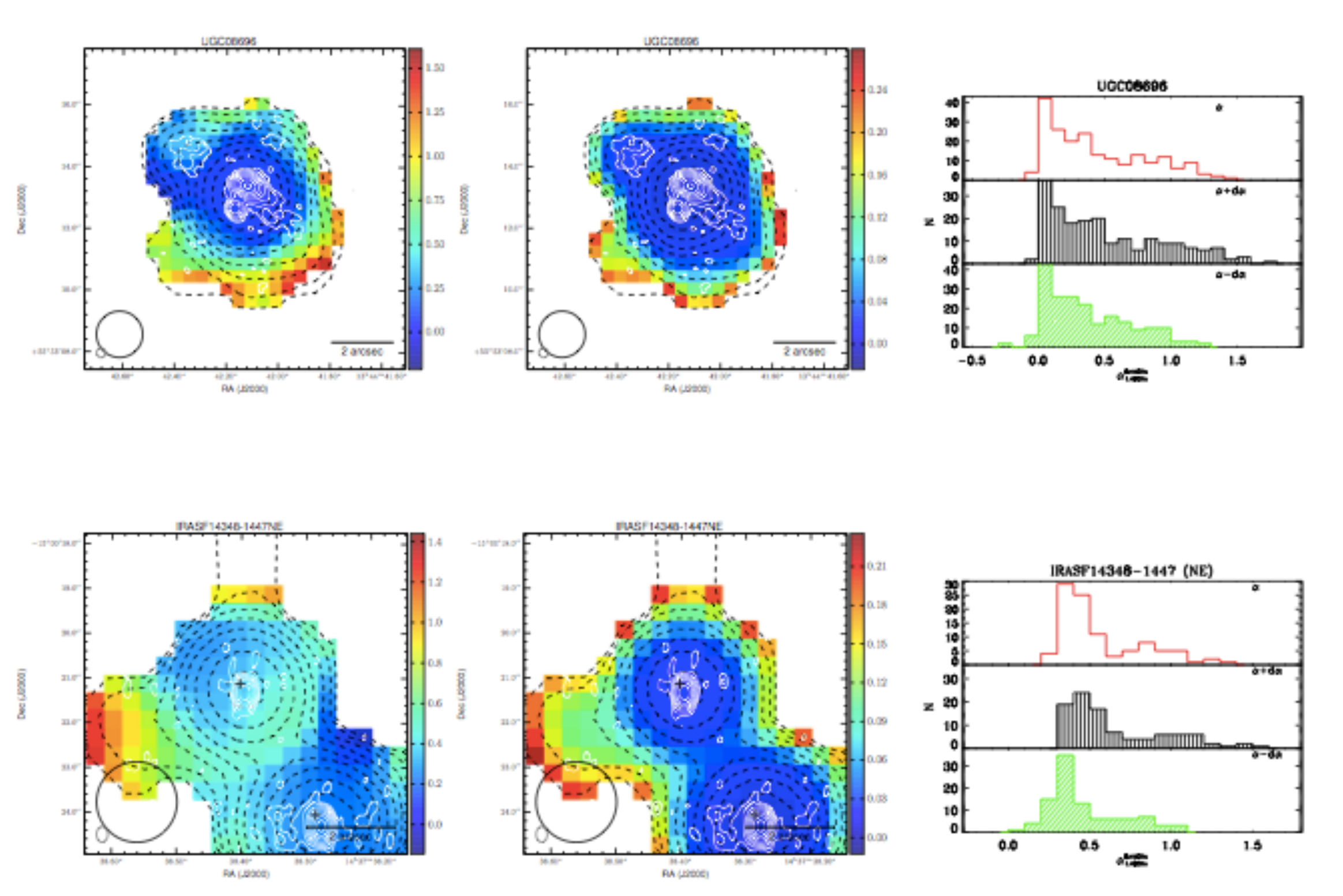}
           
            }
    
    \resizebox{\hsize}{!}
            {\includegraphics{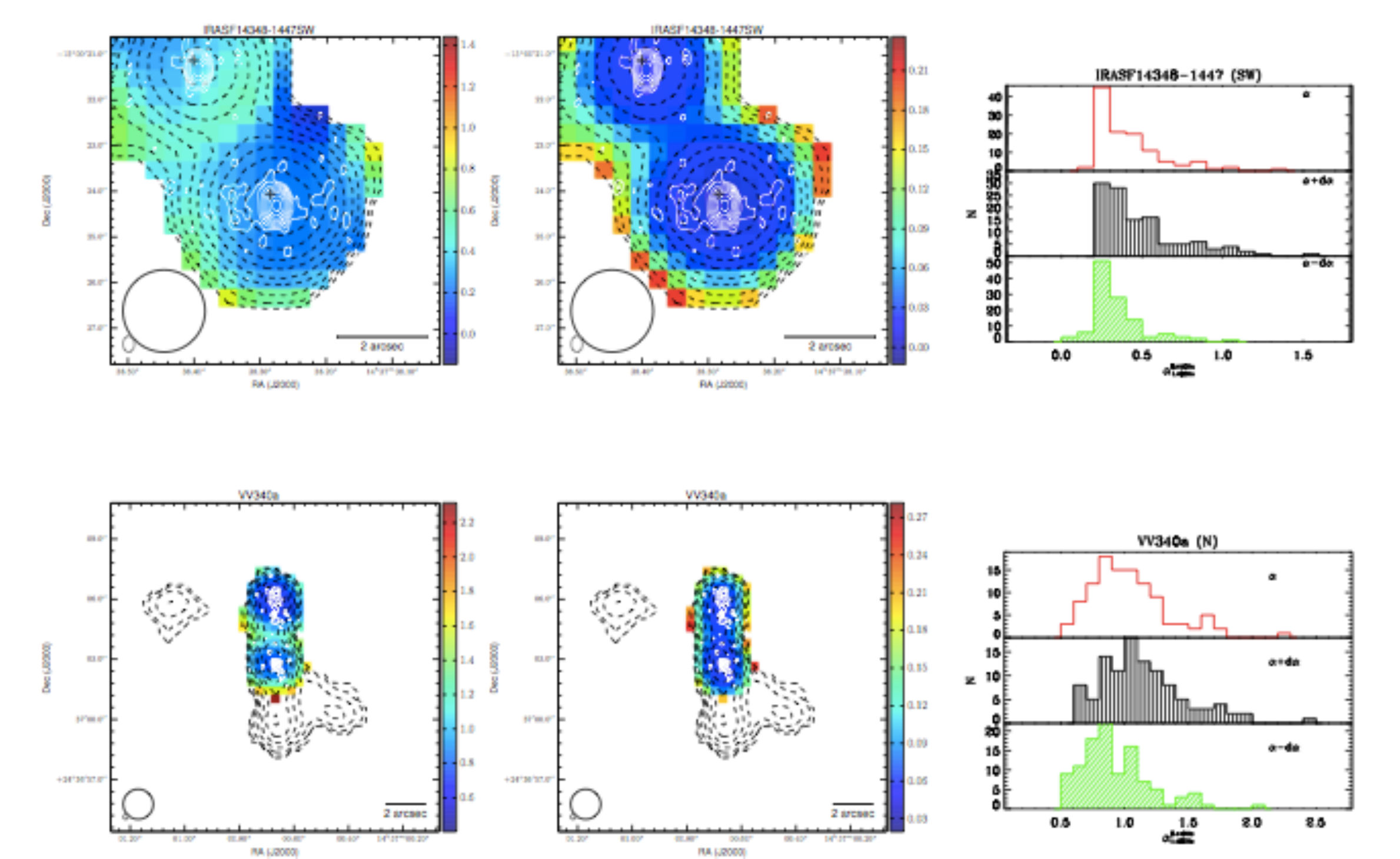}
                }

   \caption{(Continued) 
   }
              \label{app:amaps}%
    \end{figure*}

\addtocounter{figure}{-1}

%
  %
   \begin{figure*}[!ht] 
          \resizebox{\hsize}{!}
            {\includegraphics{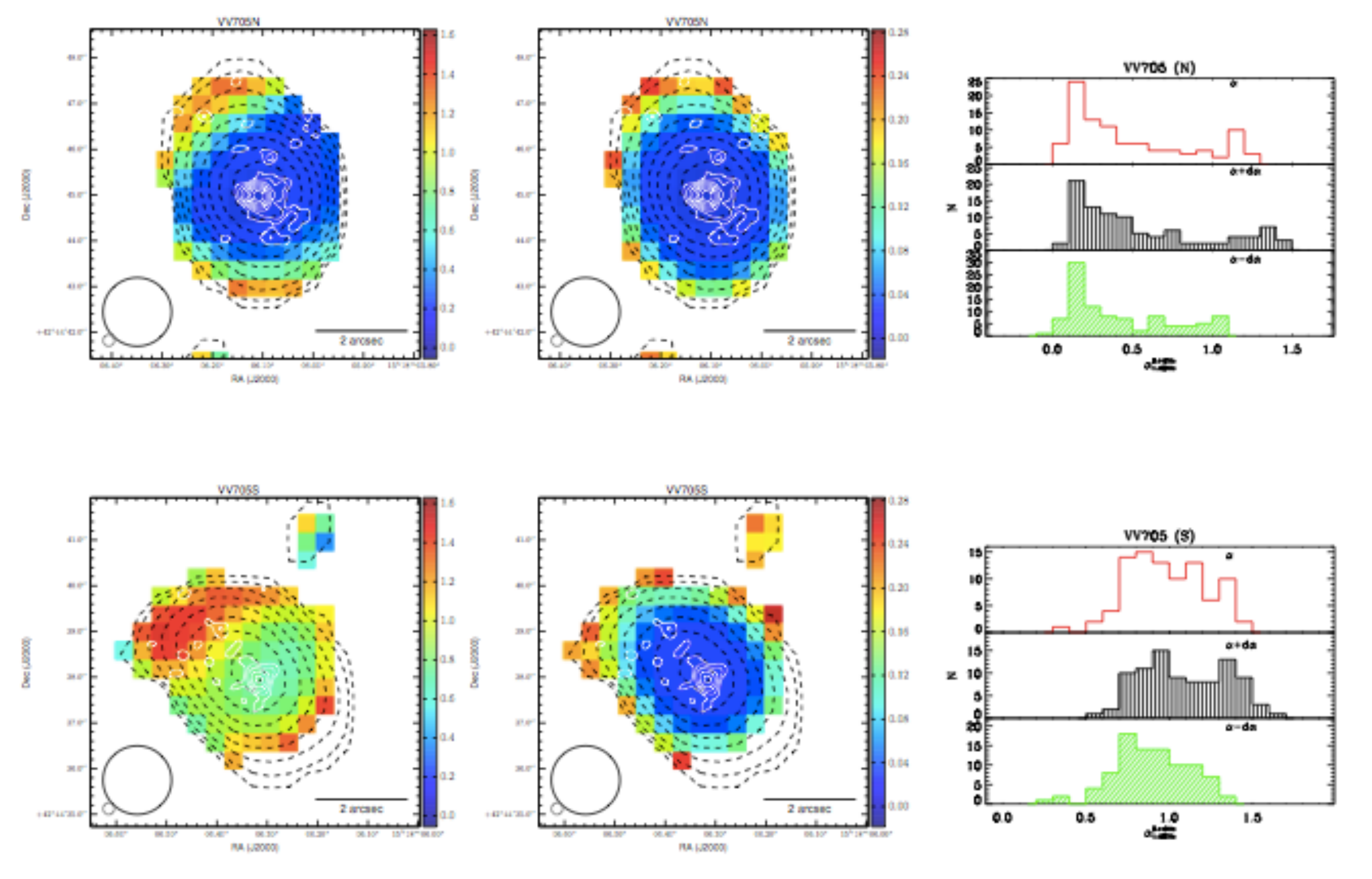}
         
            }
      
    \resizebox{\hsize}{!}
            {\includegraphics{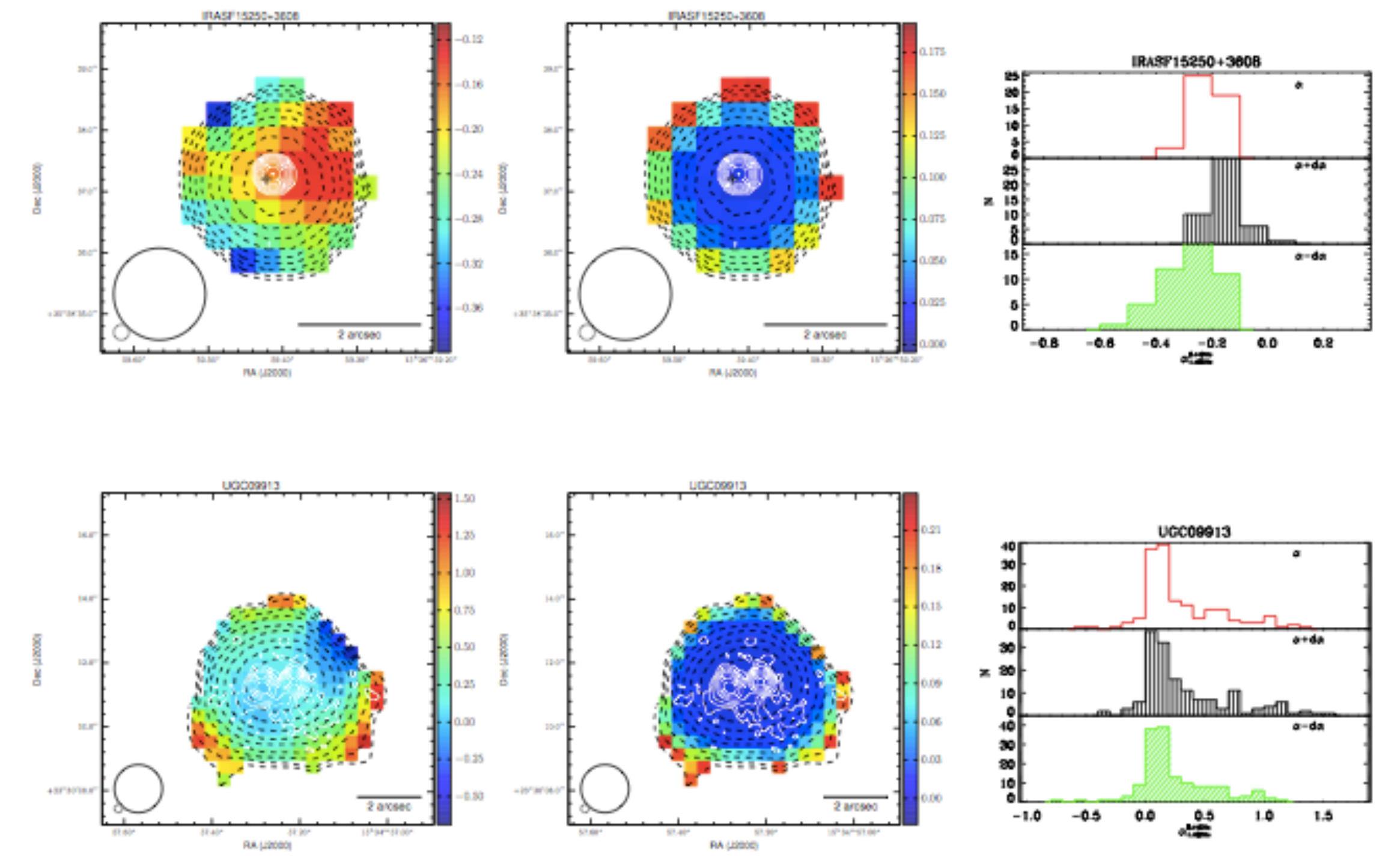}
       
            }

   \caption{(Continued) 
   }
              \label{app:amaps}%
    \end{figure*}

\addtocounter{figure}{-1}

%
  %
   \begin{figure*}[!ht] 
              \resizebox{\hsize}{!}
            {\includegraphics{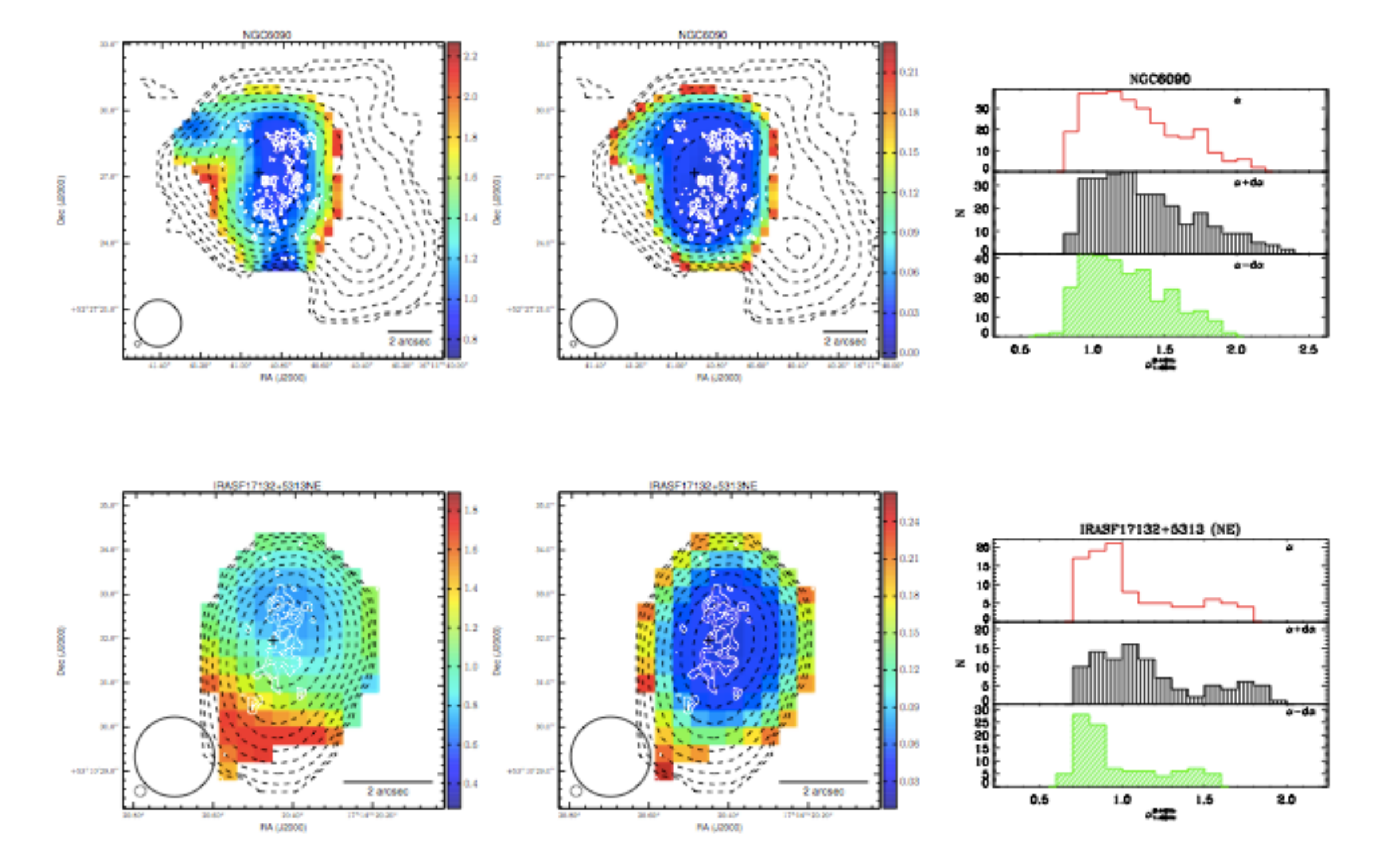}
              }

    \resizebox{\hsize}{!}
            {\includegraphics{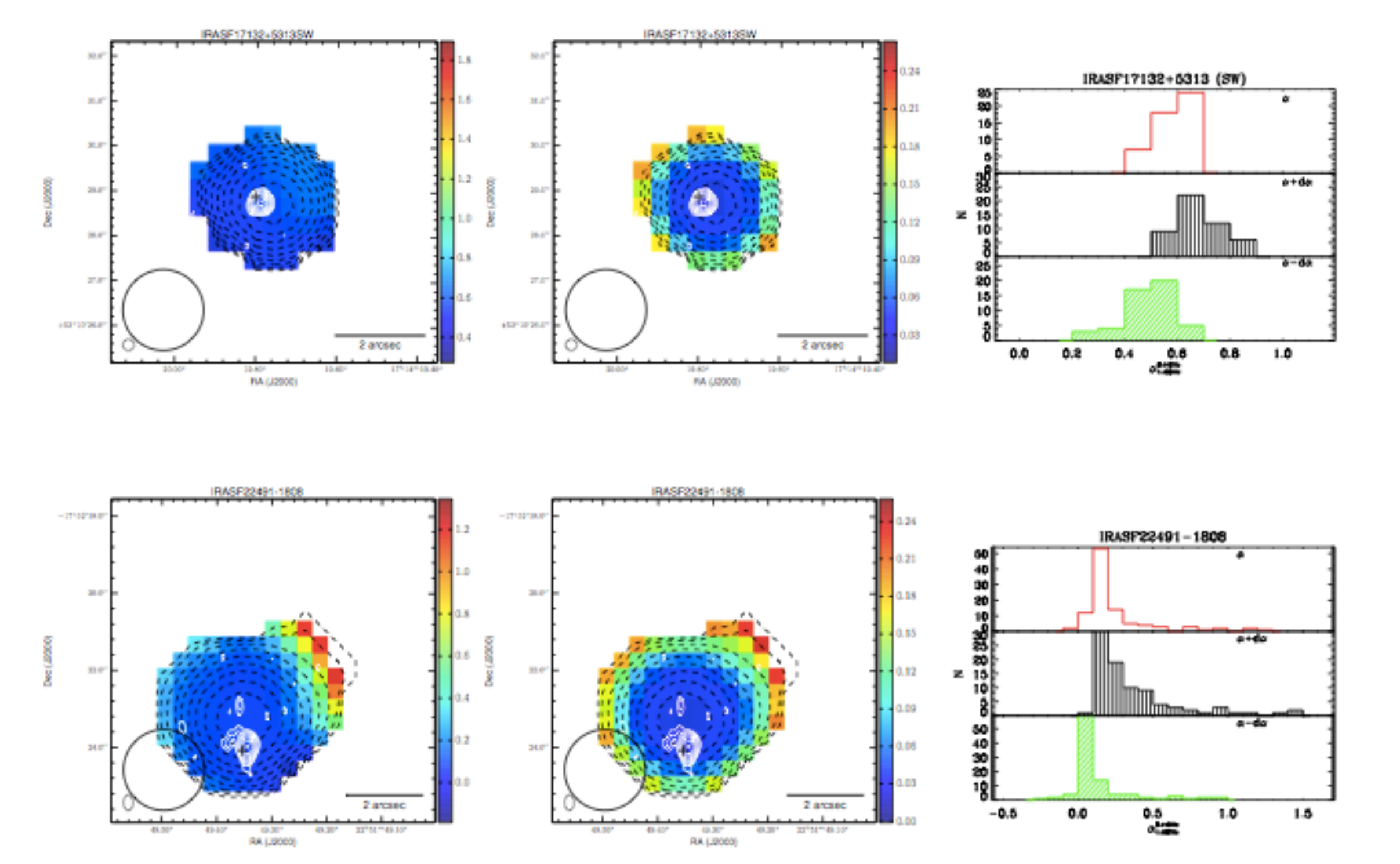}
  
            }

   \caption{(Continued) 
   }
              \label{app:amaps}%
    \end{figure*}

\addtocounter{figure}{-1}

%
  %
   \begin{figure*}[!ht] 
      \resizebox{\hsize}{!}
            {\includegraphics{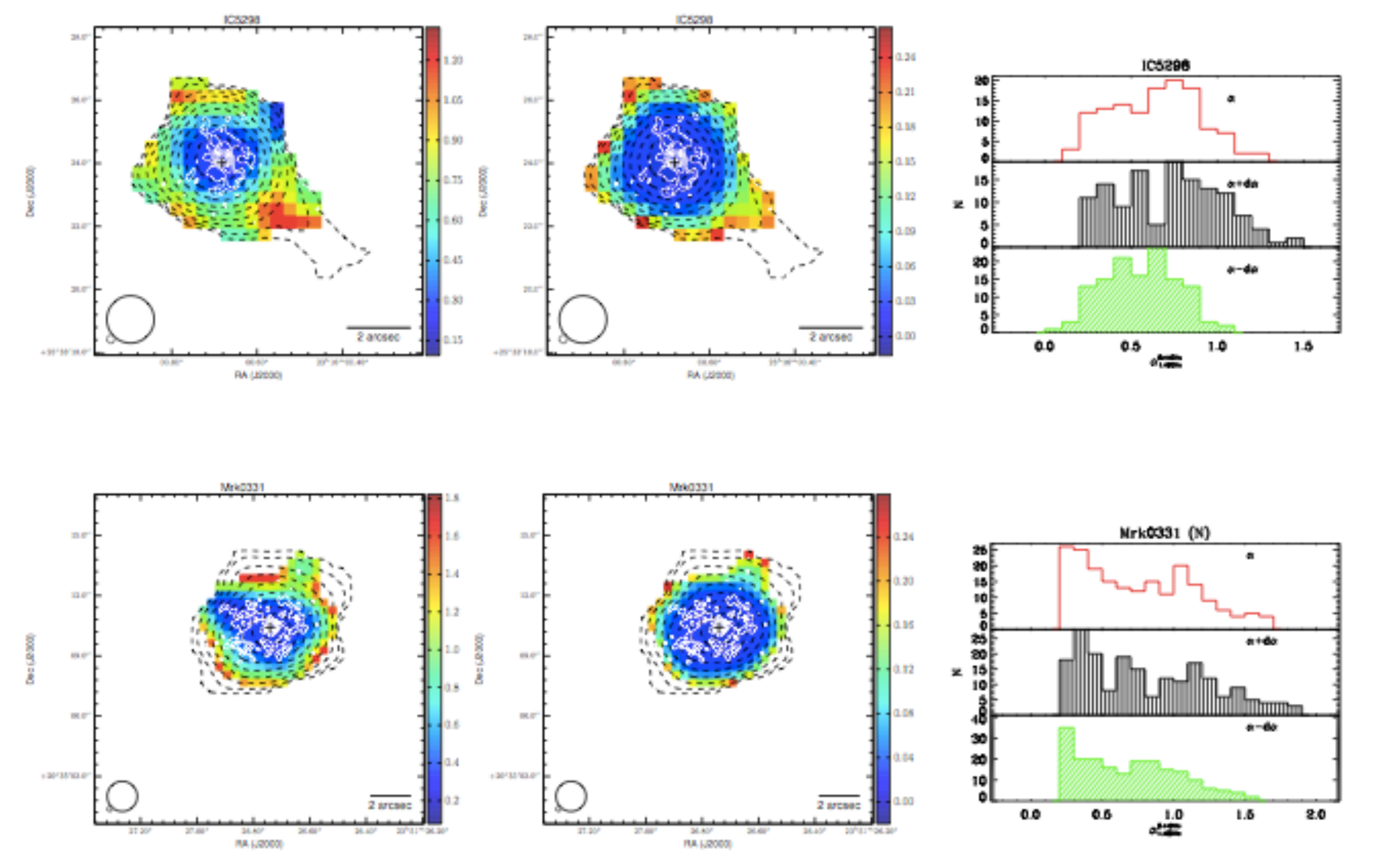}

            }

   \caption{(Continued) 
   }
              \label{app:amaps}%
    \end{figure*}

\end{document}